\documentclass[12pt]{article}
\usepackage{amsmath}
\usepackage{subcaption}
\usepackage{graphicx}
\usepackage{graphics}
\usepackage{enumerate}
\usepackage{natbib}
\usepackage{xcolor}
\usepackage{appendix}
\usepackage{url}
\usepackage{multirow}
\usepackage{comment}
\usepackage{amsthm} 

\usepackage{tablefootnote}
\usepackage{longtable}
\usepackage{threeparttable}
\usepackage{bm}
\usepackage[ruled,vlined]{algorithm2e}
\usepackage{caption}
\usepackage{placeins}
\usepackage{lscape}

\let\code=\texttt

\newcommand{\blind}{1}

\makeatletter
\newcommand{\pushright}[1]{\ifmeasuring@#1\else\omit\hfill$\displaystyle#1$\fi\ignorespaces}
\newcommand{\pushleft}[1]{\ifmeasuring@#1\else\omit$\displaystyle#1$\hfill\fi\ignorespaces}
\makeatother

\begin{document}

\def\spacingset#1{\renewcommand{\baselinestretch}%
{#1}\small\normalsize} \spacingset{1}

\if1\blind
{
  \title{\bf A Correlated Network Scale-up Model: Finding the Connection Between Subpopulations}
  \author{Ian Laga, Le Bao, and Xiaoyue Niu }
  \maketitle
} \fi

\if0\blind
{
  \bigskip
  \bigskip
  \bigskip
  \begin{center}
    {\LARGE\bf A Correlated Network Scale-up Model: Finding the Connection Between Subpopulations}
\end{center}
  \medskip
} \fi

\bigskip
\begin{abstract}
    Aggregated relational data (ARD), formed from ``How many X's do you know?'' questions, is a powerful tool for learning important network characteristics with incomplete network data. Compared to traditional survey methods, ARD is attractive as it does not require a sample from the target population and does not ask respondents to self-reveal their own status. This is helpful for studying hard-to-reach populations like female sex workers who may be hesitant to reveal their status. From December 2008 to February 2009, the Kiev International Institute of Sociology (KIIS) collected ARD from 10,866 respondents to estimate the size of HIV-related groups in Ukraine. To analyze this data, we propose a new ARD model which incorporates respondent and group covariates in a regression framework and includes a bias term that is correlated between groups. We also introduce a new scaling procedure utilizing the correlation structure to further reduce biases. The resulting size estimates of those most-at-risk of HIV infection can improve the HIV response efficiency in Ukraine. Additionally, the proposed model allows us to better understand two network features without the full network data: 1. What characteristics affect who respondents know, and 2. How is knowing someone from one group related to knowing people from other groups. These features can allow researchers to better recruit marginalized individuals into the prevention and treatment programs. Our proposed model and several existing NSUM models are implemented in the \code{networkscaleup} R package.
\end{abstract}

\noindent%
\vfill

\newpage

\spacingset{1.5} 
\section{Introduction}
\label{sec:intro} 

Hard-to-reach populations are groups of people that are not easily sampled by commonly used surveys, potentially due to their stigmatized status (e.g. female sex workers) or their infeasibility to be reached (e.g. people who committed suicide). There is a long history of developing methods to estimate the sizes of hard-to-reach populations, such as direct survey estimates, capture-recapture, and venue-based sampling, but still no method has emerged as the gold standard. UNAIDS/WHO outlined strengths and weaknesses of many of the methods \citep{unaids2010guidelines}. Direct estimates typically use random surveys of the general population and calculate what percent of respondents belong to the hard-to-reach population, but are inefficient or lead to biased results for small and hard-to-reach populations. For reasonable sample sizes, many of these surveys do not even reach the hard-to-reach populations, making it impossible to estimate the population size. Other methods require working with members of the hard-to-reach populations directly, which can lead to more accurate and precise estimates, but it is often difficult to directly contact populations that desire to stay hidden due to poor treatment and negative social stigma.

Originally motivated by estimating the size of people who have died in the 1985 Mexico City earthquake \citep{bernard1989estimating}, the network scale-up method (NSUM) avoids the need for samples from the hard-to-reach population entirely, making it more convenient to implement and shining a light on the scale of impossible-to-reach population. NSUM uses aggregated relational data (ARD), which contains the answers to surveys with questions of the form ``How many X's do you know?'' The ARD is collected from the general population, rather than from the target population. 

The basic premise of the NSUM is that the number of people that respondent $i$ knows in group $k$, denoted by $y_{ik}$, follows the scale-up equation given by
\begin{equation}
    \frac{y_{ik}}{d_i} = \frac{N_k}{N},
\end{equation}
where $d_i$ is the degree (or total number of people that respondent $i$ knows), $N_k$ is the size of group $k$, and $N$ is the total population size \citep{killworth1998estimation}. This model assumes that the probability that a member of respondent $i$'s social network belongs to group $k$ is proportional to the prevalence of $k$ in the general population. $N_k$ could be estimated directly given $y_{ik}$'s, $d_i$'s and $N$. However, $d_i$ is typically unknown and difficult to estimate directly, requiring the models to estimate both $d_i$ and $N_k$, either sequentially or simultaneously. In order to first estimate $d_i$, the ARD also includes questions about ``known population'' (e.g. people named John or postal workers, where the population sizes $N_k$ are known through census or other means). Note that we often use the terms ``group'' and ``subpopulation'' interchangeably. In this manuscript, we primarily use ``group'' when introducing the model, to recognize that in general not all NSUM applications are related to subpopulations. When referring to our specific application study, we prefer the term ``subpopulation.''

The most popular basic NSUM model was proposed by \cite{killworth1998estimation}. It assumes that the data come from a Binomial distribution given by
\begin{equation}
    y_{ik} \sim Binom \left(d_i, \frac{N_k}{N}\right).
\label{eq:basic}
\end{equation}
In order to estimate the unknown degrees and group sizes, the authors propose the following two-stage procedure: Stage 1 estimates the unknown $d_i$ as $\hat{d}_i$ by maximizing the likelihood 
\begin{equation*}
    L(d_i;\bm{y}) = \prod_{k=1}^{L} \binom{d_i}{y_{ik}} \left(\frac{N_k}{N} \right)^{y_{ik}}  \left(1 - \frac{N_k}{N} \right)^{d_i - y_{ik}},
\end{equation*}
with respect to $d_i$, where $L$ denotes the number of groups with known $N_k$ and the likelihood involves only the known populations. Stage 2 involves maximizing the likelihood which involves only the unknown $N_k$, denoted by $N_u$, i.e.
\begin{equation*}
    L(N_u;\bm{y}, \hat{\bm{d}}) = \prod_{i=1}^{n} \binom{\hat{d}_i}{y_{iu}} \left(\frac{N_u}{N} \right)^{y_{iu}}  \left(1 - \frac{N_u}{N} \right)^{\hat{d}_i - y_{iu}},
\end{equation*}
where $n$ is the number of respondents. This Stage 2 is repeated independently for each unknown group, i.e. there may be any number of $N_u$. There are other procedures to estimate the unknown $N_u$, although the general strategy of using responses corresponding to known $N_k$ to estimate $d_i$ and then back-estimating the unknown $N_u$ remains the same.

After being introduced to UNAIDS as a promising method to estimate most-at-risk people for HIV infection, many countries/cities have attempted to implement ARD surveys. One of the largest surveys is the 2009 Ukraine survey, in which the Kiev International Institute of Sociology (KIIS) collected ARD from 10,866 respondents aged 14 and above from December 2008 to February 2009 to estimate the size of 8 HIV-related subpopulations in Ukraine. The ultimate goal of the survey is to improve HIV response efficiency in Ukraine with the help of accurate size estimates. The authors relied on the NSUM to estimate these population sizes since existing methods like multiplier method and capture-recapture were too resource-intensive to obtain accurate estimates for all of Ukraine and would require studies in at least 60 settlements to obtain national size estimates \citep{paniotto2009estimating}.


Early frequentist models provided a solid foundation for quickly and easily estimating degrees and group sizes from ARD surveys (\cite{killworth1998social}, \cite{killworth1998estimation}). Recent Bayesian models have improved size estimates and answered important scientific questions about social networks. \cite{zheng2006many} included additional overdispersion in the model through a negative binomial overdispersion parameter, both better capturing the variability in the data than the existing methods and providing an estimate of the variation in respondents' propensities to know someone in each group. Later, \cite{maltiel2015estimating} aimed to model the NSUM biases (barrier effects, transmission error, and recall error) directly through the priors, estimating the strength of each bias within the groups. Most recently, \cite{teo2019estimating} included respondent covariates both about the respondent (e.g. age, gender) and how the respondent felt about each unknown group (e.g. what level of respect do you feel towards female sex workers) to adjust size estimates and study how these covariates influenced the number of people the respondents knew in each group. However, their model ignored the extra variability in the data and it resulted in small uncertainty intervals, similar to the \cite{killworth1998estimation} estimates. We refer readers to \cite{laga2021thirty} for a more complete review of the existing NSUM models and ARD properties.

Until now, all models assume that the response biases for a single participant is independent across all groups. However, we conjecture that this is not the case. \cite{zheng2006many} observed that the residuals from their model were correlated, and respondents who knew individuals who had suffered from one negative experience (e.g. suicide or rape) were more likely to know individuals who suffered from other negative experiences. We aim to properly model the correlation structure to further improve NSUM estimates and answer the sociological question of how different groups are related. 

In the Ukraine survey, information about the respondents' demographics and their acquaintance to multiple known and unknown subpopulations is collected. To better utilize all auxiliary information and learn more about the connections among subpopulations, we propose a new ARD model that accounts for overdispersion, decomposes the biases, and incorporates respondent characteristics, while also capturing the correlations between subpopulations. Our regression framework allows for more flexibility and ease-of-use than the existing approaches and provides quantitative measures of how covariates influence the number of people known in both known and unknown groups. The correlation estimates from our model provide insight into how social networks form and can hint at how different groups overlap in society. In addition, we propose various measures to assess the reliability of the model estimates. 

This paper is organized as follows. First, Section \ref{sec:background} describes the Ukraine dataset. We introduce our proposed NSUM models in Section \ref{sec:model}, along with a novel group size scaling method. The benefits and limitations of the models are discussed and our modeling choices are explained. We also show how the overall bias term in our model can be deconstructed into the three NSUM biases. We establish empirical properties of our model in Section \ref{sec:simulations}. We fit our proposed model to the Ukraine dataset in Section \ref{sec:application}. We discuss practical advice for future collection and analysis of ARD in Section \ref{sec:advice}. Final remarks and discussion are found in Section \ref{sec:discussion}.

\section{Ukraine Data}
\label{sec:background}

Of the Eastern European countries, Ukraine has the second highest rate of new HIV infections in the WHO European Region, motivating the study of key populations \citep{who2017hiv, paniotto2009estimating}. From December 2008 to February 2009, the Kiev International Institute of Sociology interviewed 10,866 respondents aged 14 and above, asking ``How many X's do you know?'' questions about 22 known groups and 8 unknown groups \citep{paniotto2009estimating}. 
We consider 4 of the 8 unknown subpopulations, women providing sexual services for payment over the last 12 months (FSW), men providing sexual services for payment over the last 12 months (MSW), men who have sex with men (MSM), and people injecting drugs over the last 12 months (IDUs), since these subpopulations belong to the World Health Organization's list of main key population groups vulnerable to HIV \citep{WHO2016consolidated}. Examples of the known groups include men aged 20-30, women who gave birth to a child in 2007, and men who served sentences in places of imprisonment in  2007.  Supplementary Table 1 lists the known and unknown groups and the sizes of the known groups.

In addition to the ARD ($\bm{Y}$), respondents were also asked demographic questions about their gender, age, education, nationality, profession, and whether they had access to the internet ($\bm{Z}$: individual characteristics), as well as ``what level of respect'' the respondent believed there was in Ukraine for each group on a 1-5 Likert scale, where 1 represents very low level of respect ($\bm{X}$: individual by group properties). After removing respondents with missing responses, the remaining sample has 9,241 respondents, which is 85.05\% of the original dataset\footnote{We recognize the importance of understanding and accounting for the significant amount of missing responses in our data. We examined several missing-data diagnostics and presented key findings in the Supplementary Material Section 4. We found that while there is a relationship between some of the covariates and the frequency of missing responses, this relationship is fairly weak and is subset to only a few of the subpopulations, most notably the subpopulations related to gender and age but not any of the unknown subpopulations. In general, we do not believe that removing the respondents with missing data significantly affects our inference. It may be of interest to explore sophisticated methods to handle missing data in ARD models.}. Furthermore, based on the accuracy of leave-one-out size estimates for the known groups, we keep only 11 of the 22 known groups, so for our analysis, $n = 9,241$ and $K = 15$.

\begin{figure}[!ht]
    \centering
    \begin{subfigure}{0.32\textwidth}
        \includegraphics[width=\textwidth]{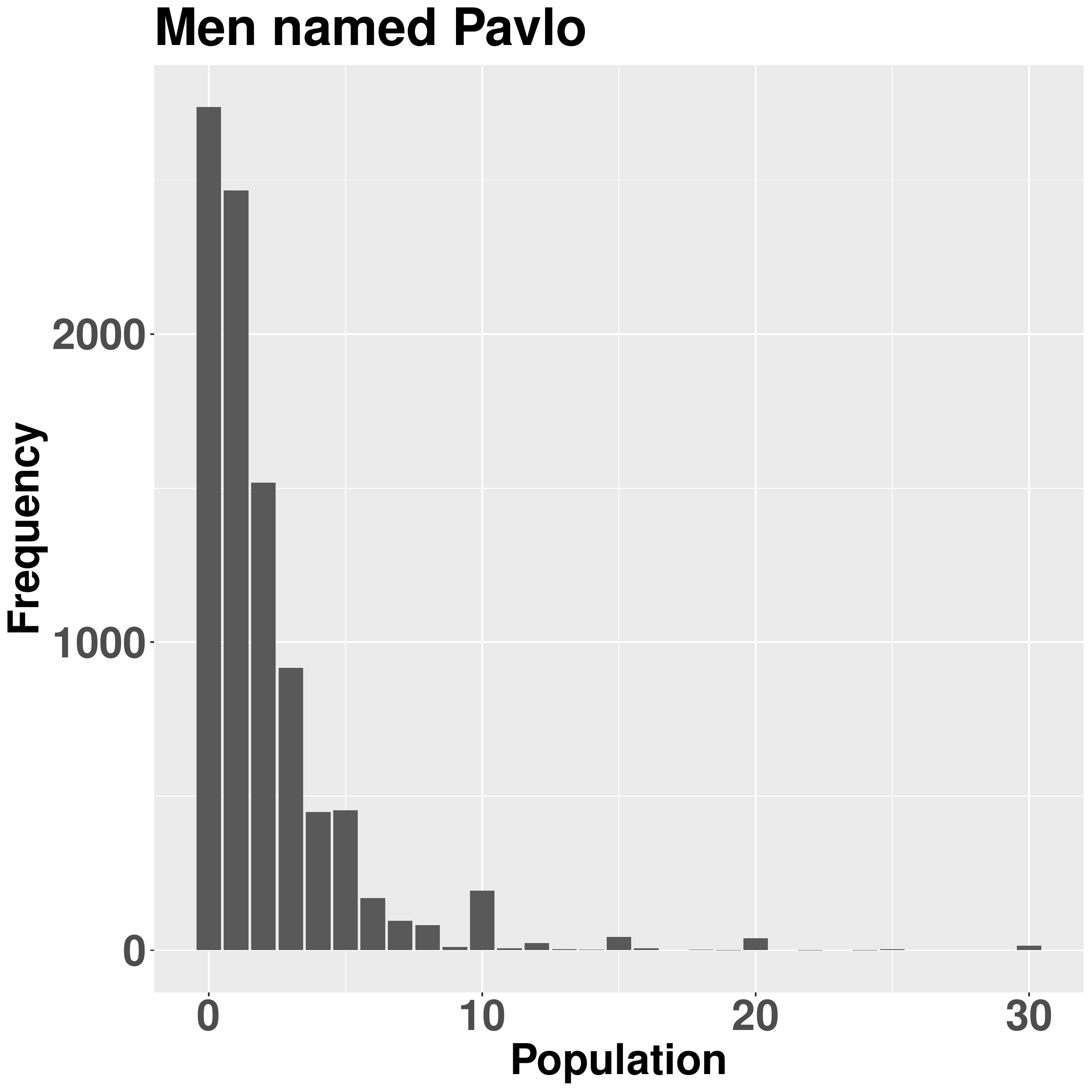}
        \caption{}
    \end{subfigure}
    \hspace*{\fill}
    \begin{subfigure}{0.32\textwidth}
        \includegraphics[width=\textwidth]{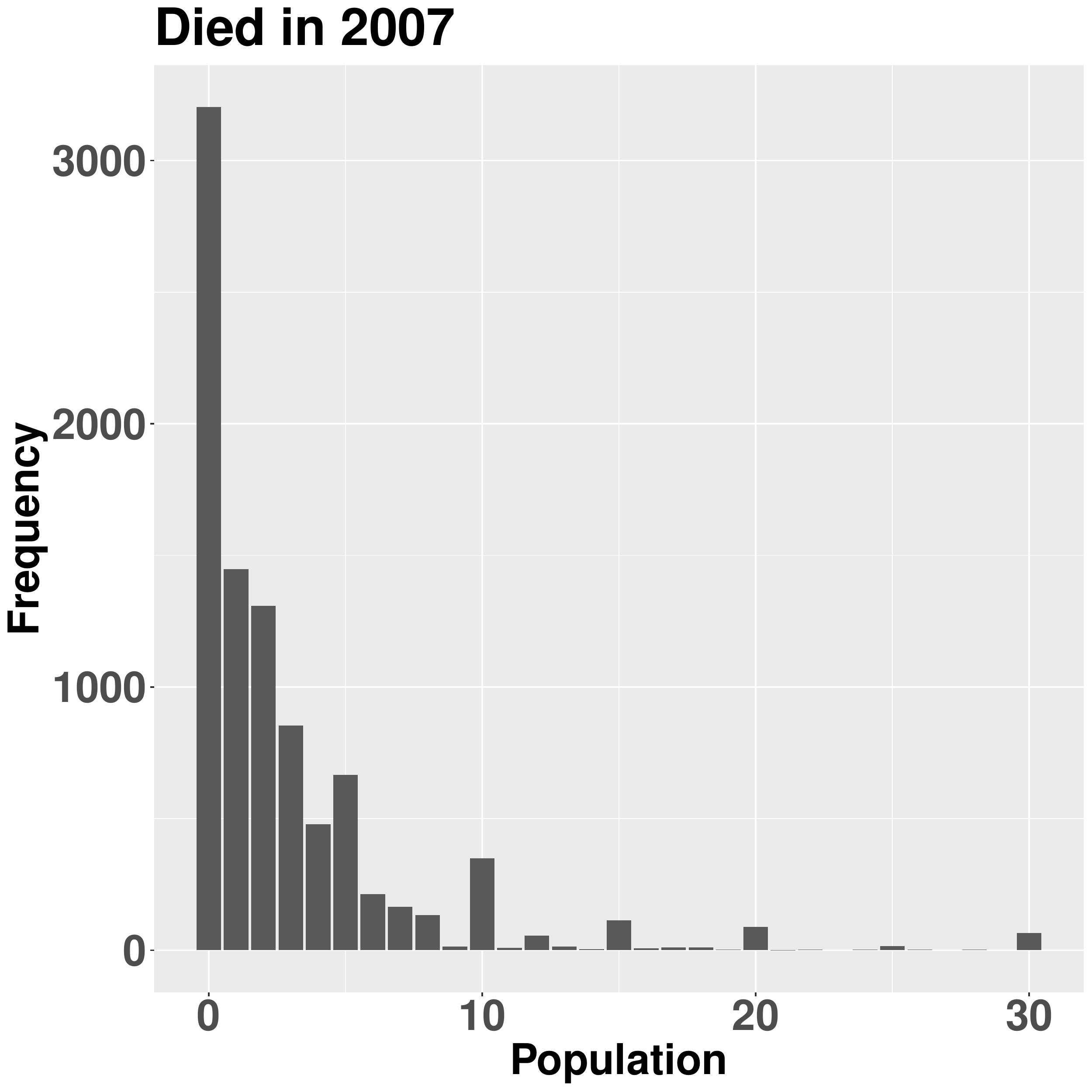}
        \caption{}
    \end{subfigure}
    \hspace*{\fill}
    \begin{subfigure}{0.32\textwidth}
        \includegraphics[width=\textwidth]{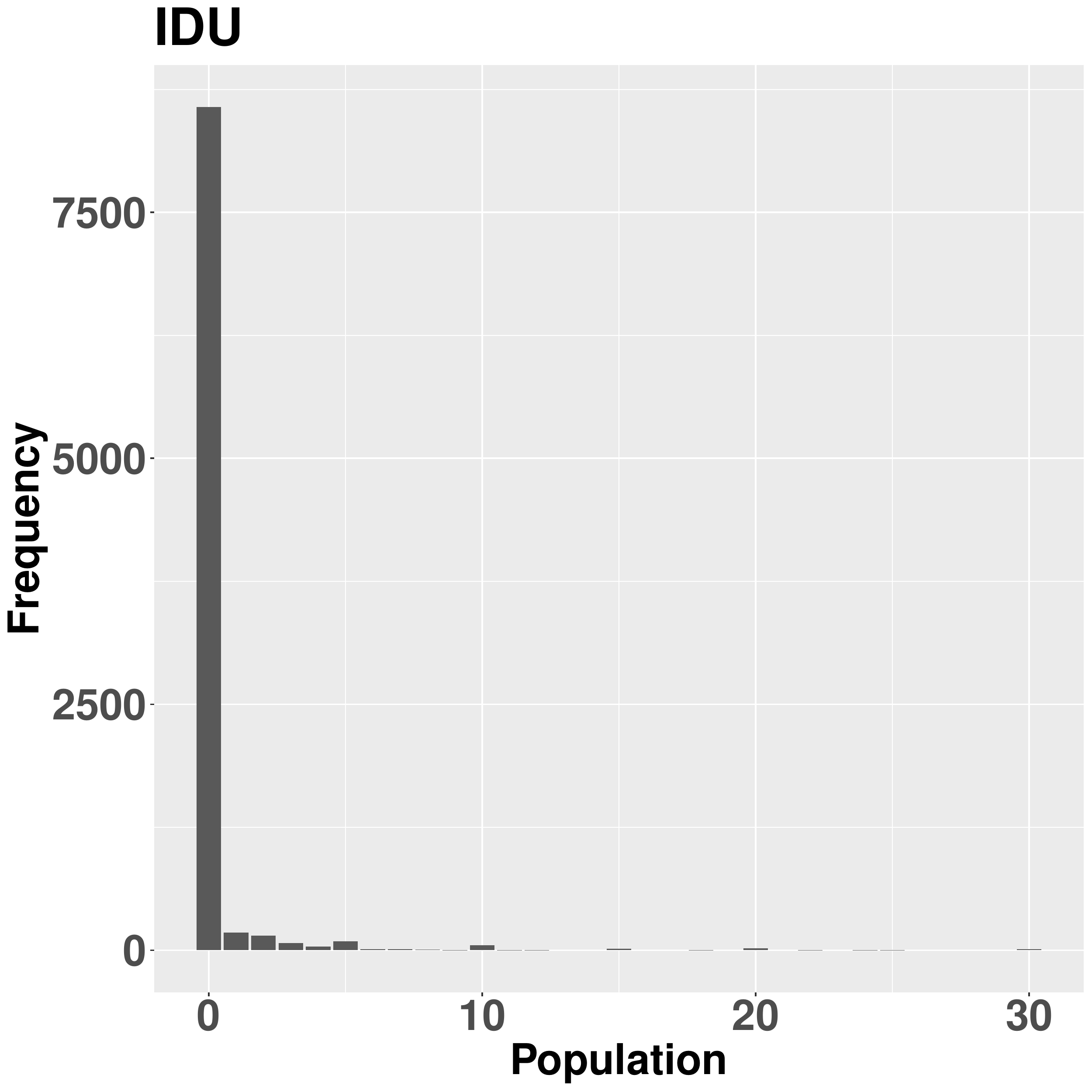}
        \caption{}
    \end{subfigure}
    \caption{Barplots of the observed Ukraine ARD for men named Pavlo (a), people who died in 2007 (b), and injection drug users (c). For visualization purposes only, the barplot for 30 represents responses greater than and equal to 30.}
    \label{fig:data_barplots}
\end{figure}

There are significant differences between the distributions of responses across subpopulations. Figure \ref{fig:data_barplots} shows frequency barplots for three subpopulations, men named Pavlo, people who died in 2007, and injection drug users (IDUs). For men named Pavlo and people who died in 2007, the average responses are relatively similar (2.11 and 2.98, respectively). However, compared to men named Pavlo, respondents tend to know either 0 people who died in 2007, or several people who died in 2007 -- there were roughly 1.4 times as many respondents who knew 0 people who died in 2007 than respondents who knew 0 men named Pavlo, despite Pavlo having a smaller average response. In this situation, the distribution of people who died in 2007 known by the respondents is more overdispersed than the distribution of men named Pavlo known by the respondents. For the hard-to-reach populations, the overdispersion is even more significant. Most respondents know 0 IDUs (92.6\%), while some respondents report knowing 40, 50, 60, and even 130 IDUs. The distribution of responses indicates that there are likely significant barrier effects for certain populations like IDUs that violate the random mixing assumption.

\section{Models}
\label{sec:model}

In this section, we introduce our correlated NSUM model, discuss its properties, and describe the estimation procedure. We first introduce the ARD notation used in the remainder of the manuscript. Given an ARD survey with $n$ respondents about $K$ groups, the number of people that respondent $i$ reports knowing in group $k$ is $y_{ik}$. Thus, the ARD matrix of responses $\bm{Y}$ is an $n \times K$ matrix. The demographics (e.g. age, gender, occupation, etc.) are denoted by $\bm{Z}$, where $z_{ij}$ is the information about respondent $i$ for variable $j$, $j \in \{1, \ldots, p\}$. In some surveys, respondents are also asked how they feel about members in group $k$ using questions of the form ``what is your level of respect towards group $k$?'' The exact phrasing of the question can vary, but the answers are denoted $\bm{X}$ with entries $x_{ik}$ for respondent $i$ and group $k$. The key distinction between $\bm{Z}$ and $\bm{X}$ is that $\bm{Z}$ is a respondent-level feature while $\bm{X}$ contains information about the interaction between the respondents and the groups. All columns of $\bm{Z}$ and $\bm{X}$ are centered to have mean zero.

\subsection{The Correlated NSUM Model}

Our correlated NSUM model is written as, for $i$ in $1, ..., n$, and $k$ in $1,..., K$,
\begin{eqnarray}
    y_{ik} &\sim& Poisson\left(exp\left\{\delta_i + \rho_k + \bm{\beta} \bm{z}_{i} + \alpha_k x_{ik} + b_{ik} \right\} \right),
\label{eq:correlated}\\
\delta_i &\sim& \mathcal{N}(0, \sigma_\delta^2), \hspace{2cm}  \rho_k \sim \mathcal{N}(\mu_\rho, \sigma^2_\rho), \nonumber \\
\bm{b}_i &\sim& \mathcal{N}_{K}(\bm{\mu},\Sigma_{K\times K} ), \nonumber 
\end{eqnarray}
where $\Sigma = \text{diag}({\bm{\tau}}) \Omega \text{diag}({\bm{\tau}})$. One key feature of the model is that after accounting for the covariate effects, we allow the biases ($b_{ik}$) to have group-specific variance ($\tau_k$) and within-person and between-group correlations ($\Omega$). Driven by network features such as homophily, the correlated bias indicates if someone knows more or less of a certain group of people, he/she tends to know more or less of a similar group. The estimated correlation reveals which groups have similarities. We can also separate the biases into the different terms (barrier, transmission, and recall) after all parameters have been estimated, and these details are shown in Supplementary Section 1. After scaling, $d_i = exp(\delta_i)$ represents the degree of respondent $i$, and $p_k = N_k / N = exp(\rho_k)$ represents the prevalence of group $k$. Depending on the application, the regression coefficients, $\bm{\beta}$, could potentially be divided into global ones ($\beta^{global}$, those that are constant across groups) and group-specific ones ($\beta_k^{group}$, those that vary with groups). This modeling choice allows researchers the flexibility of determining whether each covariate affects the responses in the same way for each group.

Here we treat all the group sizes as unknown and estimate them. As discussed in \cite{feehan_fb}, the so-called ``known'' groups need to satisfy several conditions for them to be reliably treated as ``known," including the sizes should be accurately known from census or administrative data, correct identification of memberships, representativeness of the known groups altogether, and several size requirements for each of them. In reality, most known groups do not meet those conditions. As a result, as shown in \cite{feehan_fb}, those ``known population method'' lead to various biases by treating those population sizes as known. Therefore, we choose to treat all group sizes as unknown. The known group sizes are used to help scale the estimates as detailed in the next section.

We complete the formulation with the following priors:
\begin{align*}
    \alpha_k &\sim \mathcal{N}(0, 100), &\beta_{kj} &\sim \mathcal{N}(0, 100),\\
    \sigma_\delta &\sim Cauchy(0, 2.5)I(\sigma_\delta > 0), & \mu_\rho &\sim \mathcal{N}(0, 100), \\
    \sigma_\rho &\sim Cauchy(0, 2.5)I(\sigma_\rho > 0), & & \\
    \Omega^{1/2} &\sim LKJ Cholesky(2), & \tau_{N,k} &\sim Cauchy(0, 2.5)I(\tau_{N,k} > 0),\\
    \bm{\mu} &= \log\left(1 / \sqrt{1 + \bm{\tau}_N^2}\right), & \bm{\tau} &= \sqrt{1 + \bm{\tau}_N^2}.
\end{align*}
Note that $\bm{\mu}$ and $\bm{\tau}$ are not sampled, and are only transformations of the sampled parameters $\bm{\tau}_N = (\tau_{N,1}, \ldots, \tau_{N,K})$. This parameterization is such that $E(b_{ik}) = 1$, a property shared by the Gamma prior in the \cite{zheng2006many} overdispersed model. The half-Cauchy priors on $\sigma_\delta$ and $\tau_{N,k}$ are recommended by \cite{gelman2006prior} to restrict the parameters away from very large values.

\subsection{Computation}

We provide the \code{networkscaleup} R package for readers to implement our proposed model \citep{networkscaleup}. Parameters are estimated using Markov Chain Monte Carlo (MCMC) via Stan. It is important to note that as presented above, the MCMC implementations would have trouble producing unbiased results without prohibitively long sampling chains. The hierarchical form of the model for the bias terms suffers from inefficient MCMC sampling. Specifically, when the $\mathbf{b}_i$ are all close to one another, the diagonal elements of $\Sigma = \text{diag}({\bm{\tau}}) \Omega \text{diag}({\bm{\tau}})$ will shrink towards 0. Then, since the diagonal elements of $\Sigma$ are small, $\mathbf{b}_i$ can only take very small MCMC steps, keeping both $\mathbf{b}_i$ close to one another and the diagonal elements of $\Sigma$ close to 0. To break this dependence between $\mathbf{b}_i$ and $\Sigma$, the model can be reparameterized through \textit{parameter expansion}, which allows the size of the residuals to move independently of the variance parameters \citep{van2001art, liu2003alternating, gelman2004parameterization, gelman2008using}.  Expanding our model, we reparameterize the bias as
\begin{equation}
    \mathbf{b}_i = \bm{\mu} + \text{diag}({\bm{\tau}}) \Omega^{1/2} \boldsymbol\epsilon_i,
\end{equation}
where $\Omega^{1/2}$ represents the lower-triangular Cholesky decomposition of the correlation matrix $\Omega$ and $\boldsymbol\epsilon_i \sim N(\boldsymbol{0}, I)$. By sampling $\boldsymbol\epsilon_i$ instead of the $\mathbf{b}_i$ directly, the values of $\mathbf{b}_i$ can comfortably jump around even when $\Omega^{1/2}$ is small because of the reparameterized error terms $\boldsymbol\epsilon_i$.

\subsection{Scaling}
\label{sec:scaling}

Here we introduce an equally important component of our proposed model, namely the scaling procedure. In order to convert the $\rho_k$ estimates to interpretable group size estimates, \cite{zheng2006many} proposed a scaling strategy that relies on groups of rare names, those believed to have the least biased ARD responses. However, this approach is dataset dependent and may lead to significantly biased results. In their modeling of the \cite{mccarty2001comparing} data, there were several male and female names to use for scaling. In the Ukraine data, there is only one group corresponding to a name, males named ``Pavlo.'' A scaling procedure that depends only on this group would bias the results significantly, as shown in Supplementary Figure 1. This is because the population size for men named ``Pavlo'' is significantly overestimated, so scaling by ``Pavlo'' leads to underestimating all other groups. On the other hand, in the \cite{mccarty2001comparing} data, the average bias of the rare female names is similar to the average bias of the remaining groups, which is a necessary condition for this scaling method to work.

\subsubsection{Correlated Scaling}
We propose a new scaling procedure that relies on scaling each group using correlated groups with known sizes. The idea behind this approach is simply that correlated groups have similar biases, so they should be scaled in a similar way. Specifically, we propose a weighted scaling procedure: the higher the correlation between populations A and B, the larger the weight A has on scaling B. Denoting the $m^{th}$ posterior sample for $\rho_k$ and $\Omega_{i,k}$ as $\rho^m_k$ and $\Omega^m_{i,k}$, respectively, and letting $n_{known}$ represent the number of groups with known size, our scaling procedure is outlined in Algorithm 1 below.

\begin{algorithm}[H]
\SetAlgoLined
\KwResult{Scaled $\rho'_k$ estimates}
    Set $N_{mc}$ equal to the number of posterior samples\;
 \For{each k in 1:$K$}{
    \For{each m in 1:$N_{mc}$}{
        Set $\bm{\omega} = (\Omega^m_{k,1}, \ldots, \Omega^m_{k,n_{known}})$\;
        Set negative elements of $\bm{\omega} = 0$\;
        Set $\omega_k = 0$\;
        Scale $\bm{\omega}$ such that the elements sum to $n_{known}$\;
        $C^m = \log\left(\frac{1}{n_{known}} \sum_{k \in known} \frac{e^{\rho^m_k} \omega_k}{N_k / N} \right)$\;
        $\rho'^m_k = \rho^m_k - C^m$\;
    }
 }
\caption{Correlated Scaling}
\end{algorithm}
This scaling approach inherently corrects biases such as transmission effects without requiring additional surveys like the game of contacts \citep{salganik2011game}. Collecting additional data from each hard-to-reach groups is still the most promising approach, however it is often infeasible, especially when estimating the size of several hard-to-reach groups simultaneously (like in the Ukraine dataset). While ambitious, scaling sizes using correlated groups has the potential to correct for large biases that would otherwise be impossible to account for without these additional datasets.

\section{Simulation Study}
\label{sec:simulations}

In this section, we implement a variety of simulation studies to better understand the properties of our model and scaling procedures. In Section \ref{sec:misspec}, we study how ignoring the  correlation structure affects other model parameters. In Section \ref{sec:sim_scaling}, we demonstrate the utility of the correlated scaling procedure by simulating data from two realistic scenarios.

\subsection{Missing Correlations}
\label{sec:misspec}

We simulate data from the correlated model in Equation (\ref{eq:correlated}) excluding covariates, where $K = 5$, $\sigma_\delta = 0.7$, $\rho_k = \log(2.5)$ for all $k$, $\bm{\tau}=(2, 1.05, 0.7, 1, 1.2)$, and 
\begin{equation*}
    \Omega = \begin{pmatrix}
1 & 0.9 & 0.8& -0.05& 0\\
0.9 & 1 & 0.75& 0& -0.1\\
0.8 & 0.75 & 1 & 0& 0\\
-0.05 & 0 & 0& 1& 0.85\\
0 & -0.1 & 0& 0.85& 1\\
\end{pmatrix}
\end{equation*}
This correlation matrix is designed to replicate the situation in which there are two clusters of groups, for example for stigmatized groups and for unstigmatized groups. In existing ARD surveys, the number of respondents varies widely from around 200 to over 10,000, so several different sample sizes are used. We perform the simulations at five different respondent sample sizes, $n =$ 100, 300, 1000, 3000, and 10,000. We fit the datasets using an uncorrelated version and the correlated parameterization. 

When fitting models that do not estimate a group correlation matrix, the proposed correlated scaling procedure is not possible, so we scale using all known group sizes after standardizing by group size (i.e. so larger groups do not have more weight than smaller groups). Specifically, we define a constant for each posterior sample $m$
\begin{equation}
    C^m = \log\left(\frac{1}{n_{known}} \sum_{k \in known} \frac{e^{\rho^m_k}}{N_k / N} \right),
\end{equation}
where $known$ represents the set of known groups. For each posterior sample, we scale $\rho'_k$ by $\rho'^m_k = \rho^m_k - C^m$. This scaling procedure yields group size estimates that have an average relative error of zero across all groups for each posterior sample.

\begin{figure}[!t]
    \centerline{\includegraphics[width=1\textwidth]{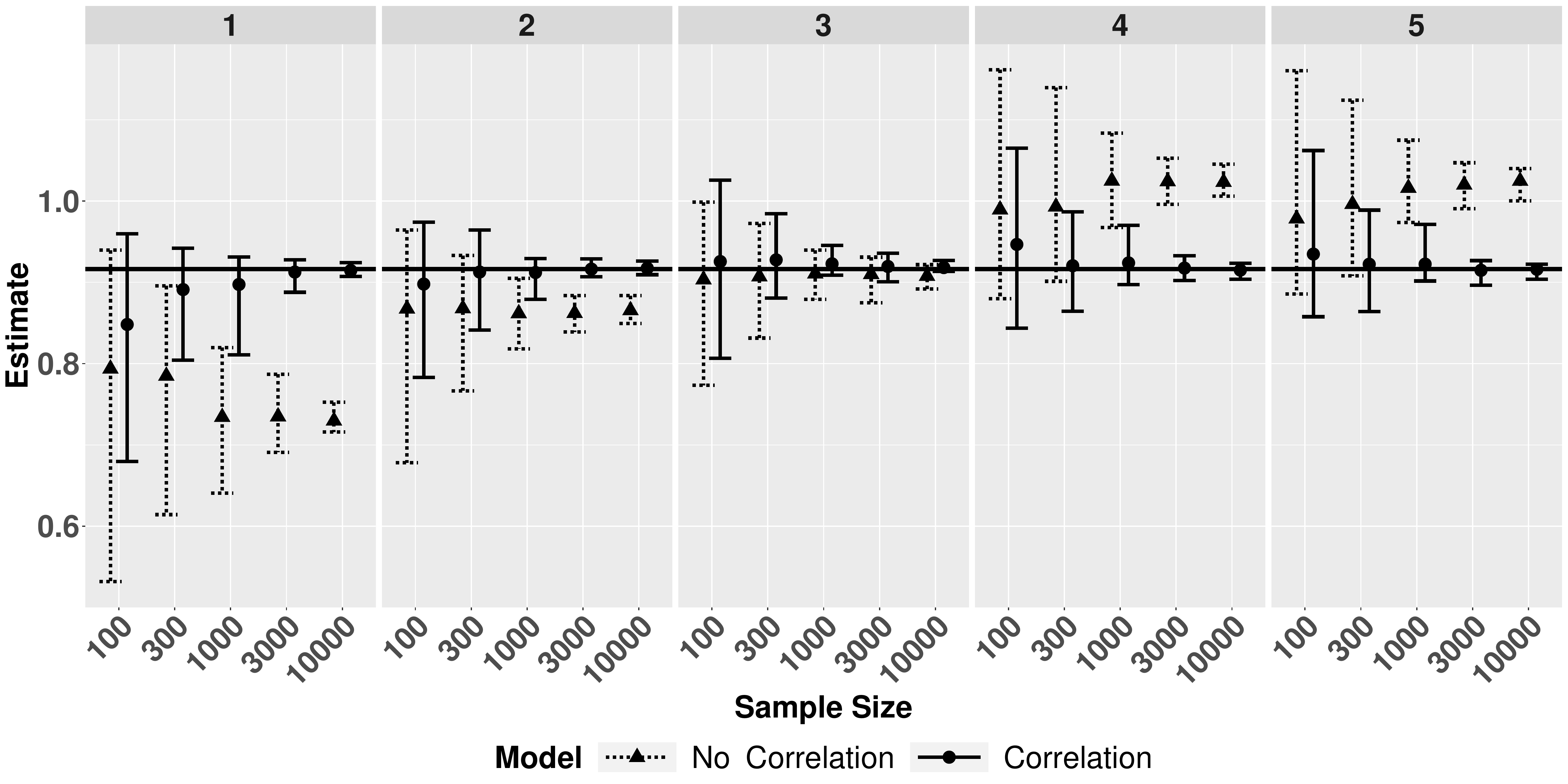}}
    \caption{95\% interval of posterior means of $\bm{\rho}$ across 100 simulations for the missing correlation simulations. The true size is represented by the horizontal black line.}
    \label{fig:sim_corr_mean}
\end{figure}

We study the distribution of the point estimates from 100 simulations and the results are shown in Figure  \ref{fig:sim_corr_mean}. The points represent the mean of the estimates across the 100 simulations while the 95\% intervals represent the upper and lower $2.5\%$ and $97.5\%$ quantiles of the estimates. Across all sample sizes, $\hat{\rho}_k$ estimates are biased when an uncorrelated model is assumed, but the data come from a correlated model. The effects are larger when the group variance $\tau_k$ is larger. 

To further study the importance of accounting for the correlation structure, we also perform simulation studies for two versions of the \cite{zheng2006many} models and two versions of the \cite{maltiel2015estimating} models. Details are shown in Supplementary Materials Section 5. We find that for NSUM models that sample random effects directly (\cite{zheng2006many} Poisson model and \cite{maltiel2015estimating} barrier effects model (sampled version)), the size estimates are biased when the ARD is correlated. In some cases, integrating out the random effects can produce unbiased point estimates (the \cite{zheng2006many} negative binomial model), while in other cases the integration does not improve estimates (the \cite{maltiel2015estimating} barrier effects model (integrated version)). We conjecture that models which have separate parameters to estimate the mean of the data and the overdispersion can produce unbiased estimates when data are correlated, while models that have parameters that influence both the mean and the variance simultaneously may lead to biased size estimates. In general, it is important to model the correlation directly, both for obtaining reliable inference results and for understanding the network structure.

\subsection{Correlated Scaling}
\label{sec:sim_scaling}

We demonstrate the utility of the correlated scaling through two simulation studies. In the first, we include systematic transmission error through correlated covariates. Specifically, we simulate ARD with $n = 1000$ from the same parameter and hyperparameter setup as before with $\Omega$ as the identity matrix. Now, we simulate each row of $\bm{X}$ as independent multivariate normal random variables, with mean $\bm{\mu} = (0, 0, 0, -2, -2)$ and $\Sigma$ equal to the correlation matrix used in the missing correlation simulation study. Then, we fit a model that does not include $\bm{X}$. This setup simulates the situation where an unobserved respondent-group level covariate explains both the group correlation and a systematic bias like transmission error (i.e. the two columns with mean -2 correspond to groups where members reveal their status to only a small percent of their social network).

\begin{figure}[!ht]
    \centerline{\includegraphics[width=0.75\textwidth]{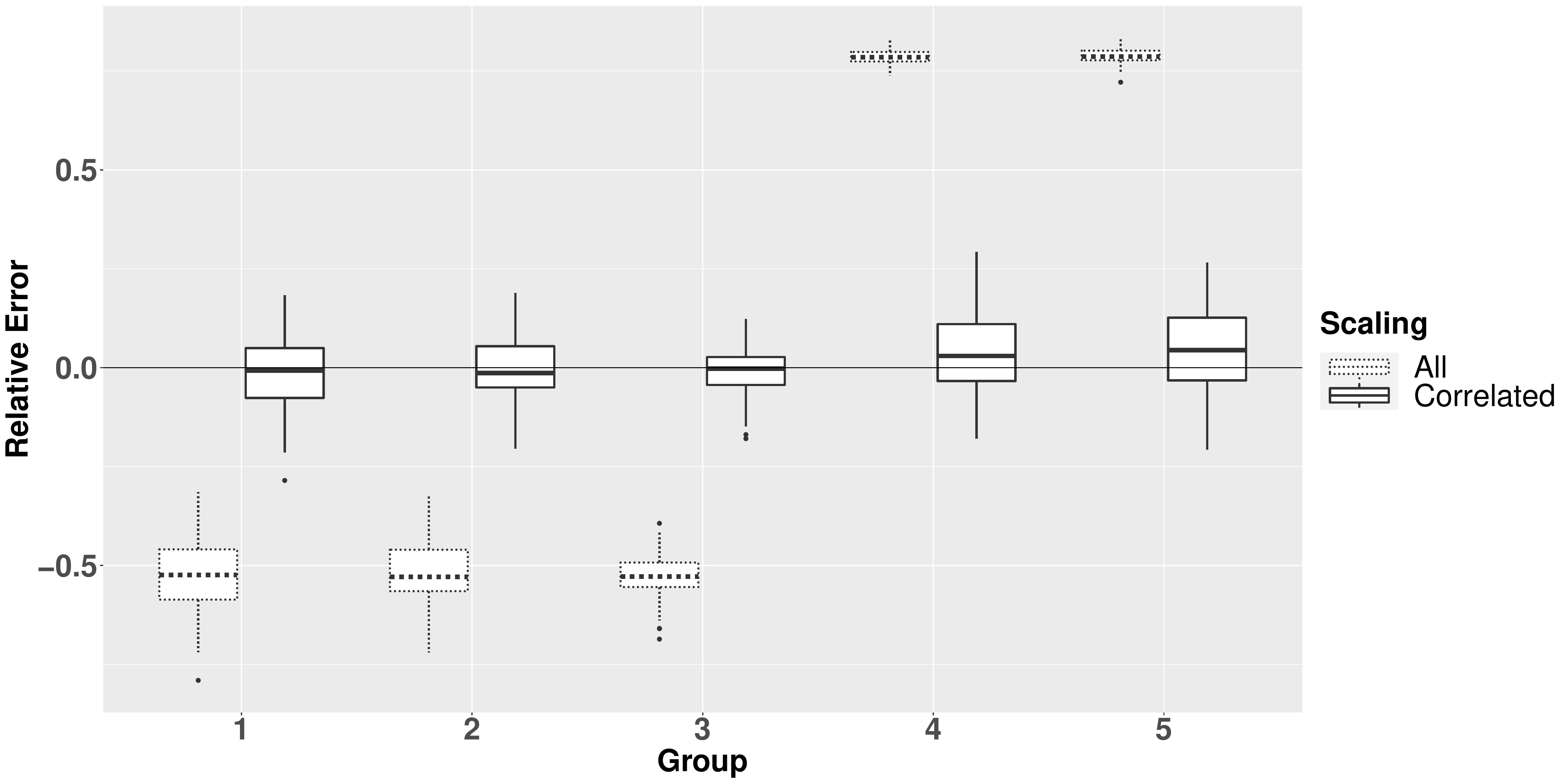}}
    \caption{Boxplot of relative errors of scaled $\bm{\rho}$ estimates across 100 simulations for missing covariate with transmission error simulations.}
    \label{fig:sim_miss_cov_corr}
\end{figure}

\begin{figure}[!ht]
    \centerline{\includegraphics[width=0.75\textwidth]{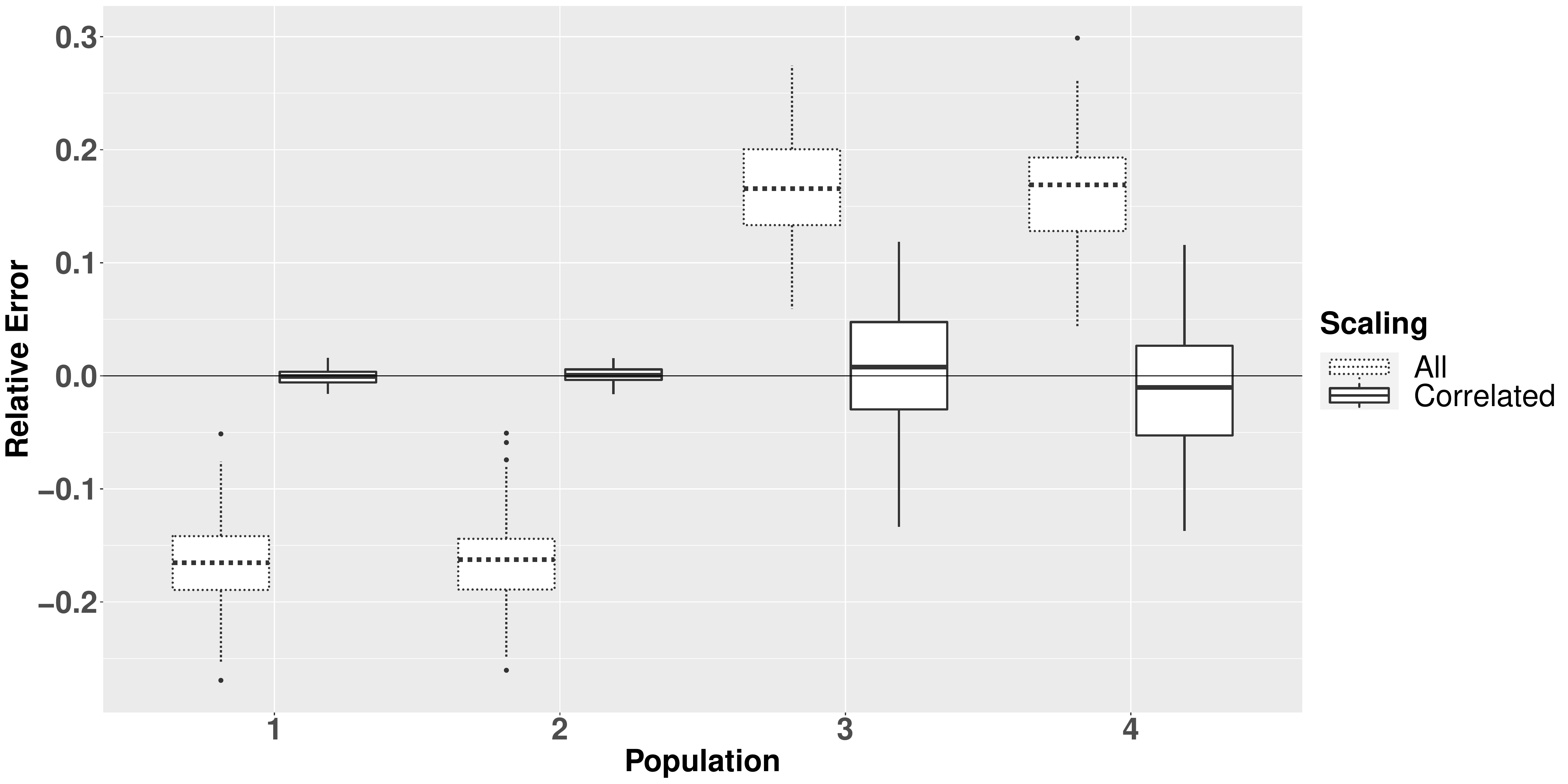}}
    \caption{Boxplot of relative errors of scaled $\bm{\rho}$ estimates across 100 simulations for SBM with transmission error simulations.}
    \label{fig:sim_sbm}
\end{figure}

Second, we simulate data from a full network model (a stochastic block model) and introduce transmission error, again where $n = 1000$. The simulation design is intentionally complex in order to best resemble a realistic network. For each simulation, first, a network is simulated from a stochastic block model with group proportions $(0.5, 0.5, 0.25, 0.25)$ and connectivity matrix $P$ provided in Equation (\ref{eq:sbm}). Then, for each true link, there is a probability that a respondent does not report the link. The probability of this missing link between respondents is given by matrix $T$, where $T_{i,j}$ denotes the probability that a respondent in group $i$ correctly reports each link they have to a member of group $j$, and $inv-logit(\infty) = 1$ for convenience. This design replicates the situation where respondents are likely to accurately recall links from certain groups (e.g. men named Pavlo), while they are likely to underestimate the number of people they know from other groups (e.g. female sex workers). Furthermore, members in these groups or adjacent groups will provide more accurate answers (female sex workers will more likely know the status of other female sex workers \textit{and} drug users).
\begin{align}
\label{eq:sbm}
   P =  \begin{pmatrix}
0.2 & 0.2 & 0.05 & 0.05 \\
0.2 & 0.2 & 0.05 & 0.05 \\
0.05 & 0.05 & 0.5 & 0.3 \\
0.05 & 0.05 & 0.3 & 0.5 \\
\end{pmatrix} & \qquad T = inv-logit\begin{pmatrix}
\infty & \infty & -1 & -1 \\
\infty & \infty & -1 & -1 \\
\infty & \infty & 2 & 2 \\
\infty & \infty & 2 & 2 \\
\end{pmatrix}
\end{align}
The results from the missing covariate and the stochastic block model are shown in Figures \ref{fig:sim_miss_cov_corr} and \ref{fig:sim_sbm}, respectively. In both cases, we plot boxplots of the relative error of each estimated and scaled $\rho_k$, scaling by either all groups or by our proposed correlated scaling. In both simulations, the correlated scaling clearly accounts for the transmission error and produces unbiased results. Thus, including weights in the scaling procedure can account for unobserved errors when the overall bias is similar for correlated populations.

\section{Ukraine Analysis}
\label{sec:application}

In this section, we fit our correlated NSUM model to the Ukraine data in order to better understand the behavior of the key populations. We used RStan to fit both models \citep{stan}. The code used is available in our \code{networkscaleup} R package \citep{networkscaleup}. We run three parallel chains for 10,000 iterations each, remove 2,000 iterations for burn-in, and thin each chain by keeping every fifth sample. In all cases the $\hat{R}$ measure of convergence is well below 1.05, indicating convergence. Additional MCMC diagnostics are included in the Supplementary Material. We include only the main results in the main text. Additionally, in Supplementary Material Section 6, we adapt surrogate residuals, first proposed by \cite{liu2018residuals}, to the Bayesian setting and use these residuals to further evaluate model fits. We observe no significant lack-of-fit in all of our diagnostic checks, showing the reliability of our model estimates reported below.

\subsection{Parameter Estimates}

We first show the results of the size estimates for the hidden populations and corresponding 95\% uncertainty intervals in Table \ref{tab:ukr_sizes}. We also include the original estimates (raw and adjusted) from the Ukraine study. The adjusted size estimates are obtained by multiplying the average estimates in each subpopulation by a weight calculated based on the level of the respect answers, and are believed to be closer to the true sizes than the raw estimates. Our model produced correlated size estimates are very similar to the Ukraine adjusted size estimates with much wider uncertainty intervals. This is a desired outcome, since simpler models often lead organizations to put too much confidence in size estimates, while the uncertainty around hard-to-reach population sizes is typically very large. While a larger uncertainty interval does not imply a more trustworthy size estimate, it is important to accurately propagate the uncertainty in the data collection method and the modeling.
\begin{table}[!ht]
\centering
\caption{Estimated unknown subpopulation sizes and 95\% credible intervals. Values are rounded to the nearest 100. ``Correlated'' indicates the results from our correlated model with the correlated scaling, ``\cite{paniotto2009estimating} Raw`` indicates the estimates reported in \cite{paniotto2009estimating} using the NSUM MLE, and ``\cite{paniotto2009estimating} Adjusted'' indicates estimates reported using their method which adjusts for the level of respect.}
\begin{tabular}{c|c|c|c}
\hline
\textbf{Subpopulation}    & \textbf{Correlated} & \begin{tabular}{@{}c@{}} \textbf{\cite{paniotto2009estimating}}\\ \textbf{Raw} \end{tabular}  & \begin{tabular}{@{}c@{}} \textbf{\cite{paniotto2009estimating}} \\ \textbf{Adjusted} \end{tabular}  \\ \hline \hline
FSW          & \begin{tabular}{@{}c@{}} 85,200 \\ (48,100\,-\,150,000)  \end{tabular}           & \begin{tabular}{@{}c@{}} 34,000  \\ (27,000\,-\,39,000)  \end{tabular} &  \begin{tabular}{@{}c@{}} 81,000 \\  (65,000\,-\,93,000) \end{tabular}     \\ \hline
MSW          &  \begin{tabular}{@{}c@{}} 6,190 \\  (1,950\,-\,20,700) \end{tabular}               &  \begin{tabular}{@{}c@{}} 2,400  \\ (1,800\,-\,3,400)  \end{tabular} & \begin{tabular}{@{}c@{}} 3,700  \\  (2,800\,-\,5,200)  \end{tabular}     \\ \hline
MSM    &\begin{tabular}{@{}c@{}}  12,300 \\   (5,160\,-\,28,800)   \end{tabular}             & \begin{tabular}{@{}c@{}} 7,200  \\ (5,300\,-\,9,100)  \end{tabular}  &  \begin{tabular}{@{}c@{}} 14,000 \\   (10,000\,-\,17,000)    \end{tabular}      \\ \hline
IDU       &  \begin{tabular}{@{}c@{}} 401,000   \\ (242,000\,-\,643,000)   \end{tabular}            &  \begin{tabular}{@{}c@{}} 103,000  \\ (85,000\,-\,112,000)  \end{tabular} &  \begin{tabular}{@{}c@{}} 358,000  \\ (285,000\,-\,389,000)  \end{tabular}    \\ \hline
\end{tabular}
\label{tab:ukr_sizes}
\end{table}

Next, we consider the covariate parameters estimates, $\bm{\alpha}$ and $\bm{\beta}$. Table \ref{tab:group_par} includes the group-specific regression coefficients corresponding to age, age$^2$, and level of respect. Age is standardized with the mean age of 43.7 and standard deviation of 19.0. The standardized age is then squared and centered to create Age$^2$. Level of respect is centered for each group. Overall, the parameter estimates are consistent with the expected results. For example, younger people are more likely to know kids. The chance that someone knows a prisoner peaks at about 34 and the chance that someone knows divorced men peaks at about 37. For the hard-to-reach populations, younger respondents are more likely to know people in all unknown groups, which could potentially provide some guidance for future sampling.

\begin{table}[!t]
\centering
\caption{Table of selected group parameter estimates for the correlated NSUM model. Age is standardized with mean 43.7 and standard deviation 19.0. Significance at $\alpha=0.05$ is denoted by *. The level of respect question was not asked for ``people who died in 2007.''}
\label{tab:group_par}
\begin{tabular}{l|l|l|l}
\hline
\textbf{Subpopulation}     & \textbf{Age}   & \textbf{Age$^2$} & \textbf{Level of Respect} \\ \hline \hline
Men 20-30         & -0.38* & -0.17*                  & 0.04*             \\ \hline
Female 20-30      & -0.37* & -0.14*                  & 0.07*             \\ \hline
Kids              & -0.31* & 0.00                   & 0.12*             \\ \hline
Prisoners 2007    & -0.24* & -0.24*                  & 0.21*             \\ \hline
Divorced Men 2007 & -0.22* & -0.32*                  & -0.01            \\ \hline
Birth 2007        & -0.15* & -0.12*                  & 0.16*             \\ \hline
FSW               & -0.69* & -0.10                  & 0.22*             \\ \hline
MSW               & -0.72* & 0.20                   & 0.02             \\ \hline
MSM               & -1.05* & -0.20                  & 0.58*             \\ \hline
IDU               & -0.57* & -0.31*                  & 0.04             \\ \hline
\end{tabular}%
\end{table}

Regarding level of respect, all significant parameters are positive, consistent with the belief that respondents with a more positive perception of a subpopulation will tend to know more people from the subpopulation. The two largest significant parameters for known groups are for prisoners and kids, which are perhaps more likely to have significant barrier effects. The parameters for FSW and MSM are the largest across all groups, which is also consistent with our intuitions.

We also report the global level regression coefficients (gender, education, nationality, profession, and access to internet). The parameter estimates corresponding to male was 0.01, 0.18 for Ukraine, 0.17 for employed, 0.12 for access to internet, 0.18 for secondary education, and 0.21 for vocational education, with a baseline of candidate of sciences or doctor of sciences. Only the 95\% credible interval for male included 0. While previous studies have found that men have larger network sizes than women, for the populations in this study, there is not a significant difference between the number of people reported between the gender of the respondents.  The network literature has shown that employed and educated individuals typically have larger network sizes. While we are not aware of literature that studies how access to internet affects network sizes, our finding is intuitive since access to internet typically means the individual can reach a broader range of contacts, for example through email.

\begin{figure}[!ht]
    \centerline{\includegraphics[height=0.8\textwidth]{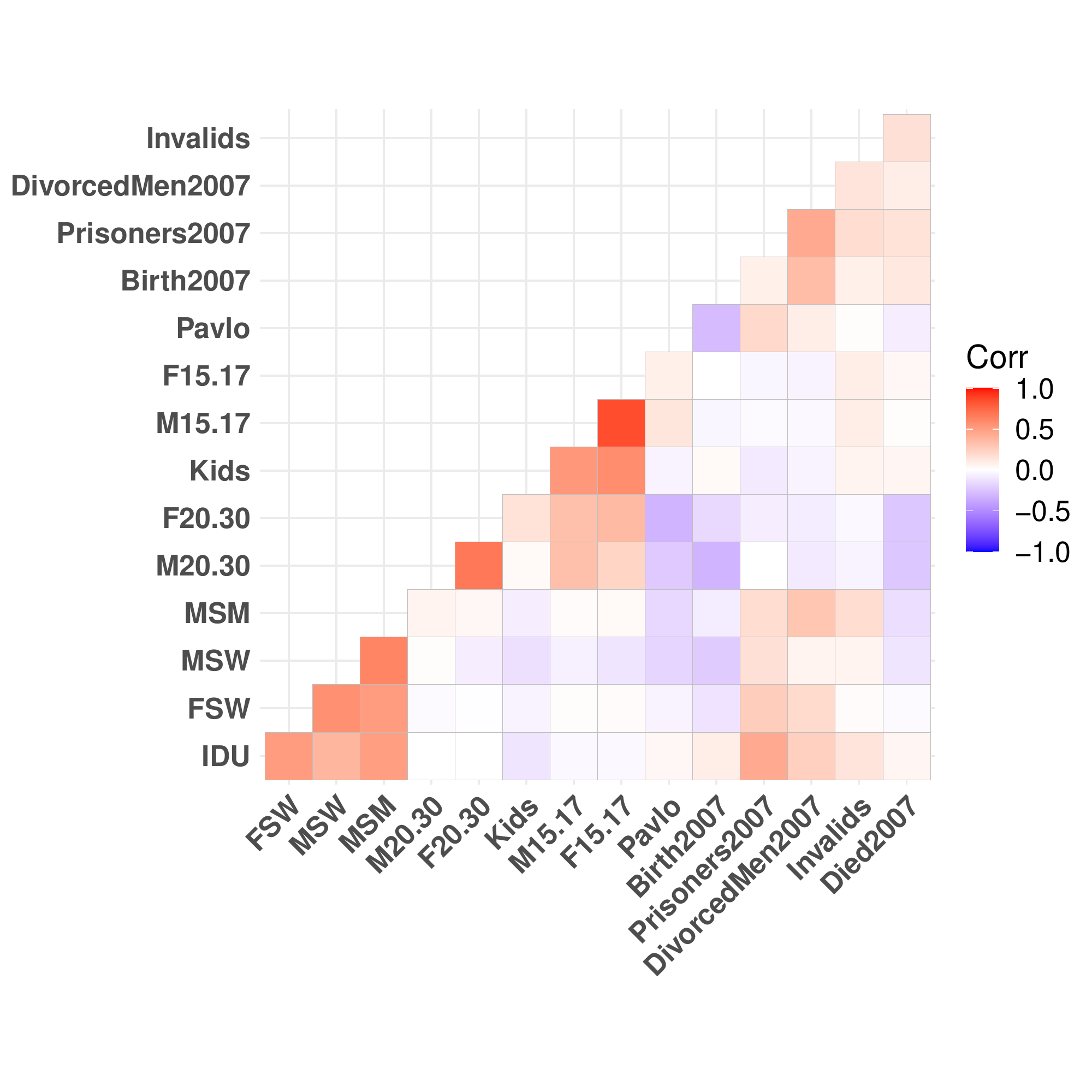}}
    \caption{Estimated correlation matrix for the Ukraine data, arranged by a hierarchical clustering algorithm.}
    \label{fig:Ukr_corr}
\end{figure}

Finally, we look at the estimated correlations, the key feature of our model. The correlation matrix is shown in Figure \ref{fig:Ukr_corr} and is sorted using a hierarchical clustering algorithm. Our model produces many expected correlations, e.g. respondents who know more men aged 15-17 also know more women 15-17; respondents who know more people aged 20 to 30 are less likely to know people who died in 2007. The results also highlight some interesting relationships that are less obvious. First, we find that respondents who know more men named Pavlo are less likely to know young men, and even less likely to know young women. Without census information about the birth records of men named Pavlo in Ukraine, we can guess that Pavlo is more common among older men in Ukraine. Second, respondents who know more/fewer men who got divorced in 2007 also know more/fewer women who gave birth in 2007. This correlation reflects that divorce and birth are both family issues that have similar barrier effects.

The key populations of the Ukraine survey are all highly correlated with one another, prisoners, and, to a lesser extent, divorced men. 

It is important to note that our correlation estimates correspond to the correlation of the \textit{reported} number of connections. Therefore, some of the correlation between the hard-to-reach populations may be an indication of a respondent's willingness to answer questions truthfully. That is, if respondents are unwilling to truthfully divulge how many FSW they know, then they may also be unwilling to answer honestly about IDU, MSW, and MSM, leading to two groups of people: those who are willing to report knowing members of hard-to-reach populations and those not willing, potentially increasing the observed correlation.

\section{Practical Advice}
\label{sec:advice}

In this section, we offer guidance on how to better collect and analyze ARD. 

\textit{Matching target groups with objectives:} The first suggestion is to align the questions about the hard-to-reach groups with how the size estimates will be used in practice. In the \cite{paniotto2009estimating} study, the question involving drug users, for example, is ``Do you know people that used drugs by injection for the last 12 months? How many of them?'' This question is phrased properly, because government organizations can use the size estimate based on this question to efficiently implement services for current or recent people who used drugs by injection. On the other hand, consider phrasing ``Do you know people that \textit{ever} used drugs by injection? How many of them?'' In this case, the size estimates cannot be used directly to allocate resources. Therefore, it is vital to phrase ARD questions to correspond to the public health objective.

\textit{Level of respect:} Second, we suggest phrasing ``level of respect'' questions to maximize the correlation between the level of respect and the ARD. In Ukraine, the level of respect question is phrased as, ``What level of respect do following groups have in Ukraine...''. The phrasing focuses on how people in Ukraine feel about the groups rather than how the respondent feels, leading to a weaker relationship between the level of respect and ARD responses. However, in \cite{teo2019estimating}, their measure of respondent social acceptability rating of the hard-to-reach populations results in a closer connection between the ARD and the level of respect. For the three shared hard-to-reach groups (FSW, MSM, and IDU), the pairwise correlations in Ukraine between the number of people the respondent reports knowing and their level of respect for the three groups are 0.038, 0.031, and 0.020, respectively. On the other hand, the pairwise correlations in Singapore are 0.040, 0.129, and 0.131, respectively. Thus, the level of respect does seem to be a relatively strong predictor for MSM and IDU in the Singapore data. It is important to ensure that questions are phrased in order to detect as much correlation as possible.

We are not aware of any study that shows how the phrasing of the level of respect question affects the correlation between the ARD responses and the level of respect responses. The phrasing used in the Ukraine study may actually be preferable, and the low correlation is simply a property of the population. This may be an interesting direction of future research.

Similarly, it is important to collect the level of respect questions for all groups. While the \cite{teo2019estimating} level of respect responses are more correlated with the ARD than in the Ukraine dataset, the authors only include information about the hard-to-reach populations. This results in a loss of information. In our analysis, we are able to account for the level of respect for all groups, further improving the results.

\textit{Inclusion of similar groups:} Finally, we recommend including more known groups which face similar stigma to hard-to-reach populations. \cite{zheng2006many} find that other populations associated with negative experiences (e.g. prisoners, homicide victims, rape victims, people who have committed suicide, and people who were in auto accidents) are highly correlated. While including correlated groups improves the size estimates in our correlated model, understanding how connected different groups are is also important. If groups are identified as being highly correlated with hard-to-reach populations, future researchers can better understand the social networks of members of hard-to-reach populations, making future survey sampling easier and more efficient.

\section{Discussion}
\label{sec:discussion}

Aggregated relational data (ARD) is an extremely useful tool not only to estimate population sizes, but also to learn about properties of social networks. Many models have been developed to better capture the behavior of the data and account for the different sources of bias. One major limitation of the models is that uncertainty estimates are far too small, so researchers are too confident in their estimates. We improve upon these models by incorporating covariates and addressing the empirical correlation between groups, and advocating the idea of correlated scaling. Another benefit of our model is that we make very few assumptions about the biases in the model, allowing the data to drive the parameter estimates. The proposed various model diagnostics are also useful in other general settings when there is not a ground-truth to compare with. Satisfactory diagnostic results will increase the credibility of the new NSUM estimates and make them more acceptable by decision makers.

From this Ukraine study, our results can be used to inform government HIV prevention policy. Of the four key populations we considered, we estimate that there are nearly four times as many injection drug users as there are FSW, MSW, and MSM. This is consistent with other studies that estimate IDUs and their sexual partners make up 64\% of people living with HIV in Ukraine \citep{des2009hiv}. Combined with estimates of HIV prevalence among key populations, our size estimates illustrate the severity of the HIV epidemic in Ukraine.

Our analysis hints at, but does not explicitly model, the increased risk in individuals that belong to more than one key population. Our correlated model has shown that respondents who report knowing more people in one hidden population are more likely to know people in the other hidden populations. \cite{who2011global} estimated that the HIV prevalence among female sex workers who inject drugs is around 43\%, while only around 8.5\% in female sex workers who do not inject drugs. Based on these relationships, it is clear that is it not sufficient to understand only the behavior of these key populations as a whole, but it is also necessary better understand the relationship between populations in order to effectively lower new HIV infections.

We believe that future ARD models should better exploit the relationship between populations, as illustrated by both our estimated correlation matrix and the covariate effects. It is clear that some individuals are closer to the key populations than others, either because of their age, gender, and similar characteristics, or because of their existing social networks. We would be able to better understand the properties and behavior of the key populations if we were able to survey respondents who were more familiar with the populations of interest.



\if1\blind
{
    \section{Acknowledgements}
    
    The work is supported by the National Institute of Allergy and Infectious Diseases of the National Institutes of Health under award number R01AI136664. Computational efforts were performed on the Hyalite High Performance Computing System, operated and supported by University Information Technology Research Cyberinfrastructure at Montana State University.
    
} \fi

\newpage

\if1\blind
{
  \begin{center}
    {\LARGE\bf Supplementary Material to: ``A Correlated Network Scale-up Model: Finding the Connection Between Subpopulations''}\\
      {\Large \author{Ian Laga, Le Bao, and Xiaoyue Niu }}
\end{center}

  \maketitle
} \fi

\if0\blind
{
  \bigskip
  \bigskip
  \bigskip
  \begin{center}
    {\LARGE\bf Supplementary Material to: ``A Correlated Network Scale-up Model: Finding the Connection Between Subpopulations''}
\end{center}
  \medskip
} \fi

\tableofcontents

\newpage
\spacingset{1.5} 
\section{Ukraine Data}

\begin{table}[!ht]
\caption{Ukraine subpopulations and sizes}
\centering
\resizebox{!}{0.43\textheight}{
\begin{tabular}{|p{8cm}|p{4cm}|} \hline
\textbf{Subpopulation} & \textbf{Known Size}    \\ \hline
Men aged 20-30&  4,088,438 \\\hline
Men aged 15-17&  935,153 \\\hline
Men above 70&  1,328,606 \\\hline
Women aged 20-30&  3,966,956 \\\hline
Women aged 15-17&  899,443 \\\hline
Women above 70&  2,993,846 \\\hline
Children (boys and girls) aged 10-13&  1,869,098 \\\hline
Moldavians &  258,619 \\\hline
Romanians& 150,989  \\\hline
Poles&  144,130 \\\hline
Jews&  104,186 \\\hline
Romany&  47,587 \\\hline
1st group invalids&  278,195 \\\hline
Doctors of any speciality&  222,884 \\\hline
People who died in 2007& 762,877  \\\hline
Men named Pavlo&  323,863 \\\hline
Men who served sentences in places of imprisonment in  2007&  108,511 \\\hline
Men who officially divorced in 2007&  178,364 \\\hline
Women who gave birth to a child in 2007&  472,657 \\\hline
Doctors and Candidates of Science who received a scientific degree in Ukraine over the last 15 years&  69,471 \\\hline
Nurse women, nurse men, aid-men and aid-women&  487,148 \\\hline
Militiamen& 273,200 \\  \hline
\end{tabular}
\label{tab:ukraine_dat}
}
\end{table}
\FloatBarrier

\section{Bias Decomposition}
\label{sec:bias_decomp}

In this section, we show how to decompose our bias estimates into the transmission, barrier, and recall effects. While our main focus is on subpopulation size estimates, the effect of each bias is also an important question. We have found that without overly strict assumptions on the structure of the biases, estimating each individual bias separately is infeasible, but imposing incorrect assumptions can negatively affect the other parameter estimates. Therefore, we estimate only an overall bias term that accounts for all sources of bias and decompose the bias into the individual biases after fitting the model, leaving the other model parameters unaffected by the bias assumptions. The form of our bias gives the model more flexibility in capturing the heterogeneity present in the data.

Our decomposition approach assumes the probability component of the binomial likelihood of the \cite{maltiel2015estimating} recall error model is equivalent to the mean parameter of our Poisson likelihood. The recall error model is given by:
\begin{align}
\begin{split}
    y_{ik} &\sim Binom(d_i, e^{r_k}\tau_k q_{ik}) \\
    d_i &\sim Lognormal(\mu, \sigma^2) \\
    r_k &\sim N(a + b \cdot \log(N_k), \sigma^2_r) \\
    \tau_k &\sim Beta(mean = \nu_k, dispersion = \eta_k) \\
    q_{ik} &\sim Beta(mean = N_k / N, dispersion = \rho_{ik}),
\end{split}
\end{align}
Then, we fit a Bayesian regression model with the posterior means of our probabilities as the response. Specifically, we first calculate the posterior means as
\begin{equation}
    \bar{b}_{ik} = \sum_{m = 1}^{N_{mc}} \frac{N_k^m}{N} b_{ik}^m,
\end{equation}
where $N_{mc}$ is the number of posterior samples. Then, using the prior assumptions from \cite{maltiel2015estimating}, we assume the model
\begin{align}
\begin{split}
    \bar{b}_{ik} &\sim N(e^{r_k}\tau_k q_{ik}, \sigma^2) \\
    r_k &\sim N(a + b \cdot \log(N_k), \sigma^2_r) \\
    a &\sim 1, b \sim 1 \\
    \sigma^2_r &\sim Cauchy(0, 2.5)I(\sigma^2_r > 0) \\
    \tau_k &\sim Beta(mean = \nu_k, dispersion = \eta_k) \\
    q_{ik} &\sim Beta(mean = N_k / N, dispersion = \psi_{ik})\\
    \psi_{ik} &\sim U(0, 1),
\end{split}
\end{align}
where $\nu_k$ and $\eta_k$ are fixed from prior information and flat non-informative priors are placed on the remaining hyperparameters. The above priors can easily be adapted to suit new prior information and different assumptions about the structure of the biases.

\section{Scaling}
\label{sec:supp_scaling}

Here we offer a more detailed examination of how problematic the original scaling in \cite{zheng2006many} may be. We fit our proposed Correlated NSUM model to the Ukraine dataset. We scale the $\hat{\rho}_k$ estimates via four different approaches:
\begin{itemize}
    \item Scale using $G_1$ with the only ``name'' subpopulation in the Ukraine dataset, ``Men named Pavlo.''
    \begin{itemize}
        \item[$\diamond$] $C = \log \left(\sum_{k \in G_1} \left( \frac{e^{\rho_k}}{\sum_{k \in G_1} N_k / N }  \right) \right)$
    \end{itemize}
    \item Scale using $G_1$ with ``people who died in 2007'' and ``women who gave birth to a child in 2007'', $G_2$ with ``Women aged 20-30'' and $B_2$ with ``Men aged 20-30.''
    \begin{itemize}
        \item[$\diamond$] $C_1 = \log \left(\sum_{k \in G_1} \left( \frac{e^{\rho_k}}{\sum_{k \in G_1} N_k / N }  \right) \right)$
        \item[$\diamond$] $C_2 = \log \left(\sum_{k \in B_2} \left( \frac{e^{\rho_k}}{\sum_{k \in B_2} N_k / N }  \right) \right) - \log \left(\sum_{k \in G_2} \left( \frac{e^{\rho_k}}{\sum_{k \in G_2} N_k / N }  \right) \right)$
        \item[$\diamond$] $C = C_1 + \frac{1}{2}C_2$
    \end{itemize}
    \item Scale using $G_1$ with all subpopulations
    \begin{itemize}
        \item[$\diamond$] $C = \log \left(\sum_{k \in G_1} \left( \frac{e^{\rho_k}}{\sum_{k \in G_1} N_k / N }  \right) \right)$
    \end{itemize}
    \item Scale using $G_1$ with all subpopulations, but standardizing by size
    \begin{itemize}
        \item[$\diamond$] $C = \log \left( \frac{1}{n_{known}} \sum_{k \in G_1} \left( \frac{e^{\rho_k}}{N_k / N }  \right) \right)$
    \end{itemize}
\end{itemize}
where $N_k$ is the known subpopulation size for group $k$ and $n_{known}$ denotes the number of known subpopulations.

The results are shown in Supplementary Figure \ref{fig:overdisp_scaling}. For this dataset, scaling by the only ``name'' subpopulation yields reasonable results. However, out of the remaining 9 known subpopulations, 7 are underestimated (positive relative error). Now suppose we wanted to use subpopulations we believed had relatively even mixing and low transmission error. This is the motivation between the second approach, where using the two populations of ``people who died in 2007'' and ``women who gave birth to child in 2007'' seems like they should have low bias and represents both the older and younger respondents. We realize that there might be a gender discrepancy because one of the subpopulations corresponds to women, so we also should secondary groups to scale with. This seems like a reasonable approach, but it turns out that both people who died in 2007 and women who gave birth to child in 2007 are produce significantly smaller estimates than they should compared to other subpopulations, so scaling by these subpopulations leads to overestimating every other subpopulations. While this example was chosen to demonstrate a point, it is clear that if researchers chose the wrong subpopulations to scale with, it can significantly affect the size estimates of the remaining subpopulations.

Therefore, we suggest a more robust method by using all available known $N_k$ in the scaling procedure. Using all subpopulations on the original scale results in underestimates only 6 out of 9 known subpopulations, with one of those 6 being very close to having zero relative error (prisoners 2007). This is an improvement on previous scaling methods, but now the larger subpopulations have more influence in the scaling procedure, while not necessarily corresponding to more accurate estimates. Therefore, we suggest standardizing by size so that all subpopulations have equal weight. In other words, the average relative error across all subpopulations should be zero. In this dataset, using the last scaling procedure with all subpopulations on the relative error scale also results in overestimating 6 out of 9 known subpopulations. However, if the largest known subpopulations were consistently overestimated or underestimated, this would significantly effect the previous scaling method, but not the scaling method that standardizes by scale.

\begin{figure}[!t]
    \centerline{\includegraphics[width=\textwidth]{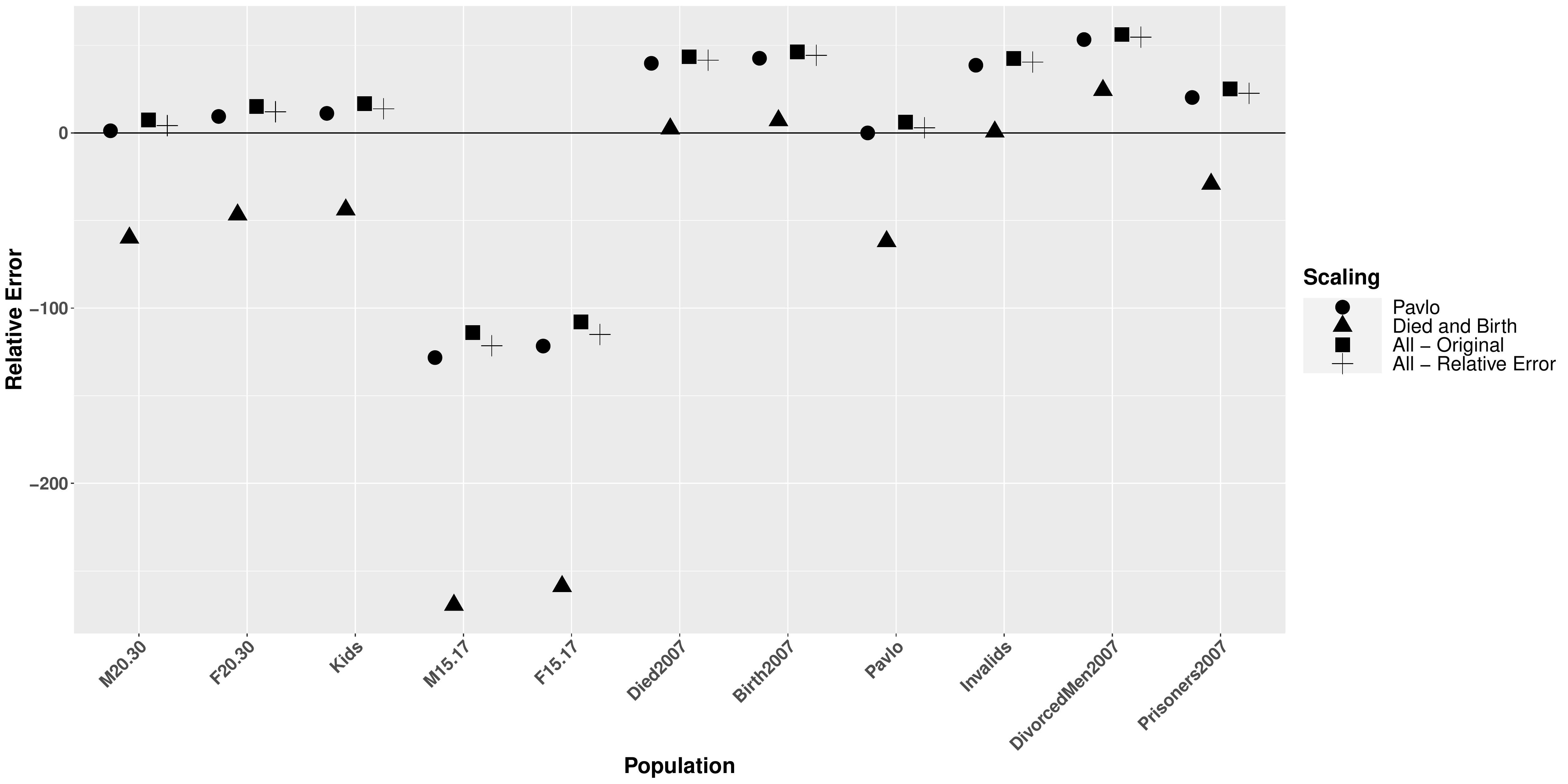}}
    \caption{Relative error point estimates in the Ukraine dataset from four different scaling methods on the $\hat{\rho}_k$ estimates. Relative error is calculated by $100*(Truth - estimates) / Truth$. Subpopulations are ordered from largest to smallest.}
    \label{fig:overdisp_scaling}
\end{figure}

\FloatBarrier
\newpage
\section{Missing Data}
\label{sec:missing_data}

While not exhaustive, here we provide additional information about the missing responses. Across all 10,866 rows and columns of $\bm{Y}$ (26), $\bm{X}$ (26), and $\bm{Z}$ (7), only $0.72\%$ of the responses are missing. However, $17.52\%$ of rows have at least one missing value. After subsetting data to the columns used for analysis (removed unused ARD groups, level-of-respect groups, and demographic questions), only $14.95\%$ of rows have at least one missing value. A barplot showing the percent of responses missing for each subpopulation is shown in Supplementary Figure \ref{fig:ukraine_miss_ard_barplot}. Perhaps surprisingly, responses for the hard-to-reach populations have roughly the same percent missing as other less stigmatized populations. Supplementary Figures \ref{fig:ukraine_miss_lor_barplot} and \ref{fig:ukraine_miss_dem_barplot} show the percent of responses missing for the level-of-respect and demographic questions, respectively.

We examine the interaction between missing responses for the 20 most frequently occurring intersections in Supplementary Figure \ref{fig:ukraine_miss_intersect}. Here we see that the majority of the missing responses are from respondents ignoring a single missing variable. For example, 198 respondents did not answer whether they have access to internet, but did provide an answer to all other questions we included in our final analysis. The most worrying intersection is that some respondents did not answer questions for how many MSW, FSW, or MSM they knew. However, there were only 18 respondents in this intersection, so any bias introduced by these respondents is likely trivial. There also appears to be very little intersections between the level-of-respect questions and the ARD. In these 20 most frequently occurring intersections, there is no overlap between missing ARD and missing LoR responses. Based on this figure, we do not worry about the missingness between multiple variables.

In Supplementary Figure \ref{fig:Missing_matrix_men_ages}, we plot the data for four gender-age groups (men and women aged 15-17 or 20-30) against the age of the respondents. Here we clearly see that a respondents age does play a role in whether a response is missing for each group. In particular, responses corresponding to men and women aged 15-17 are more likely to be missing for older respondents, while responses corresponding to men and women aged 70+ are more likely to be missing for younger respondents.

Next, we consider we extend this investigation to the hard-to-reach subpopulations, FSW, MSW, MSM, and IDUs. Supplementary Figure \ref{fig:Missing_matrix_htr_ages} presents the analogous results. Here, there is no clear trend that age has any affect on whether a respondent is more or less likely to have missing responses for the hard-to-reach subpopulations.

Next, we consider how the level-of-respect questions ($\bm{Z}$) correspond to the missing responses.  In Supplementary Figure \ref{fig:Missing_matrix_men_ages}, we see that the only clear relationship between level-of-respect and the probability of missing responses is that respondents who had a higher level of respect for men aged 15-17 were more likely to have missing responses, which is unexpected. For the hard-to-reach populations, Supplementary Figure \ref{fig:missing_htr_lor} shows that for FSW, MSW, and MSM, again, respondents who had a higher level of respect were more likely to have missing responses. It is not clear why this relationship exists, since it is expected that respondents who view a subpopulation more favorably would be more willing to reveal how many people they know in the hard-to-reach subpopulation.

We also examine the distribution of the estimated degrees and the ARD responses between the data kept for analysis and the data removed because at least one column contained a missing response. As in the main manuscript, we set all $y_{ik} > 150$ equal to 150. For the degrees, we estimate the degrees using the \cite{killworth1998estimation} MLE estimator in two ways: (1) using only data kept for analysis, or (2) using all available data. For this second approach, we estimate the degree using
\begin{equation*}
    \hat{d}_i = N \frac{\sum_{k \in A_i} y_{ik}}{\sum_{k \in A_i} N_k},
\end{equation*}
where $A_i$ denotes the set of subpopulations that both have known $N_k$ and have available responses for respondent $i$. The boxplots of the log of these degree estimates are shown in Supplementary Figure \ref{fig:ukraine_miss_degree}. The boxplots indicate that the distributions of the log-transformed degree estimates are almost identical for the respondents kept for final analysis and the respondents removed due to one or more missing responses.

Similarly, we also compare the \cite{killworth1998estimation} size estimates from either all respondents using the degrees estimated using the approach above, or from the respondents with no missing data, i.e. the respondents we used for our final analysis. The size estimates are plotted against each other on the log-scale in Supplementary Figure \ref{fig:ukraine_miss_size}. We can see that removing the respondents with missing responses does not change the size estimates from the \cite{killworth1998estimation} model.

Finally, we compare the distributions of $y_{ik} / \hat{d}_i$ for the complete data and for the removed data. We plot histograms of these two groups in Supplementary Figure \ref{fig:ukraine_miss_yik}. Note that for these plots, we remove the responses that are equal to 0 and standardize the histograms so that the maximum density is 1. We find that almost all groups, the distributions of the responses are almost identical. The notable exceptions are for the hard-to-reach populations, but this is mostly do to the small number of non-zero values. In particular, we find that the complete data has several larger responses for FSW and MSM, while the MSW and IDU responses are fairly similar for the complete and removed data. These differences are reasonable given the small sample size, and we believe the bias introduced by the missing data is minimal. We also compare the percent of responses that are 0 in Supplementary Table \ref{tab:ukraine_miss_yik_0}. For all subpopulations, the percents are very similar. The largest difference is for people who died in 2007, which is surprising given the almost identical distribution of positive values.

To summarize the results, we do not believe that removing the respondents with missing responses will significantly alter the results. From what we can observe, the removed data behaves very similarly to the data used for analysis. The missing data appears to be mostly from respondents simply not answering one question in the survey, although we are unable to determine why the respondents would behave this way. There does appear to be a weak relationship between age and the level of missingness for age-related groups like men aged 15-17, but this behavior does not extend to the hard-to-reach subpopulations. We recognize the importance of appropriately handling missing data, and we recommend this as a future direction of research for the Network Scale-up Method.

\begin{figure}[!t]
    \centerline{\includegraphics[width=0.8\textwidth]{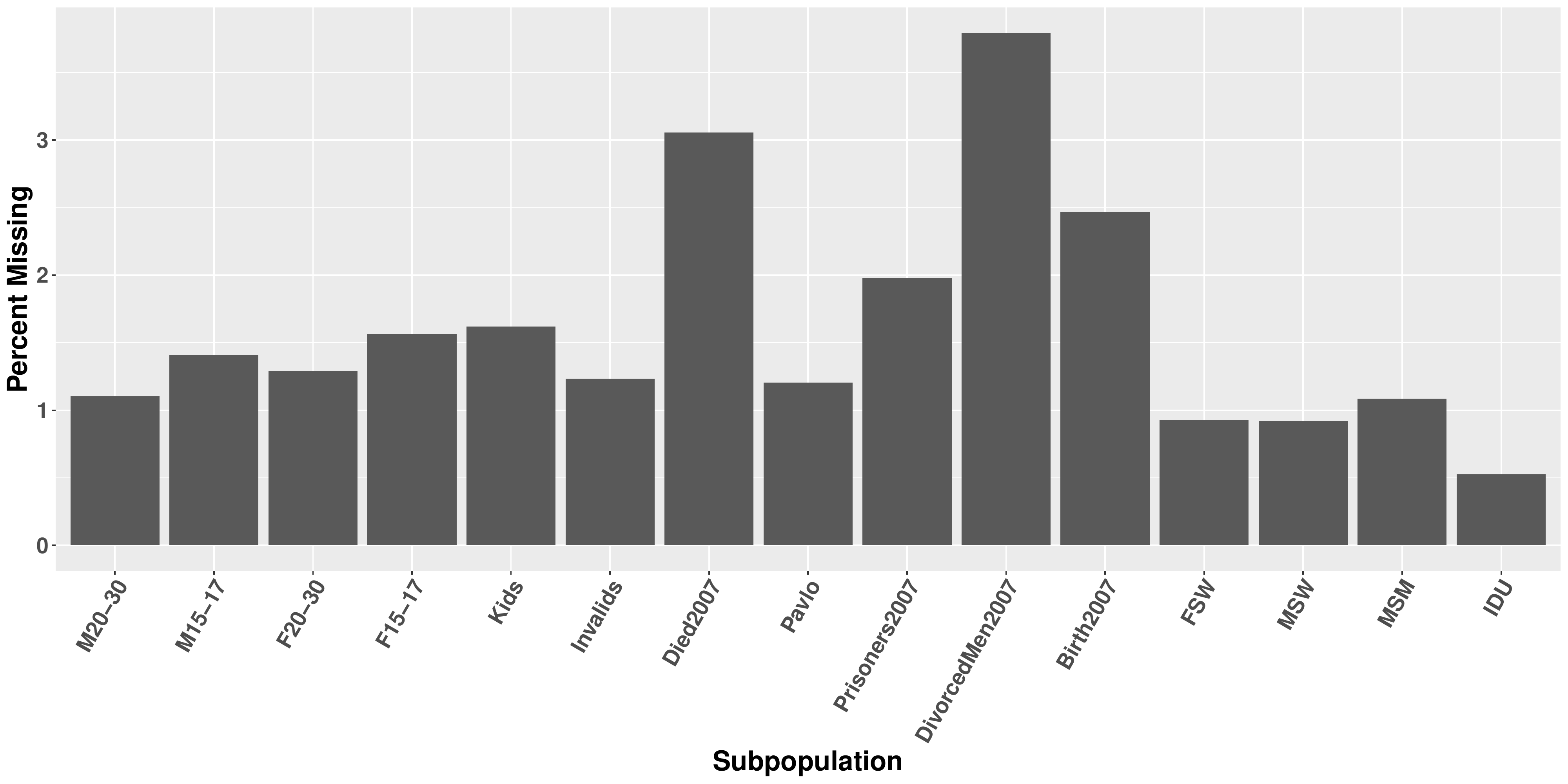}}
    \caption{Barplot showing the percent of missing data for ARD responses in the Ukraine dataset.}
    \label{fig:ukraine_miss_ard_barplot}
\end{figure}

\begin{figure}[!t]
    \centerline{\includegraphics[width=0.8\textwidth]{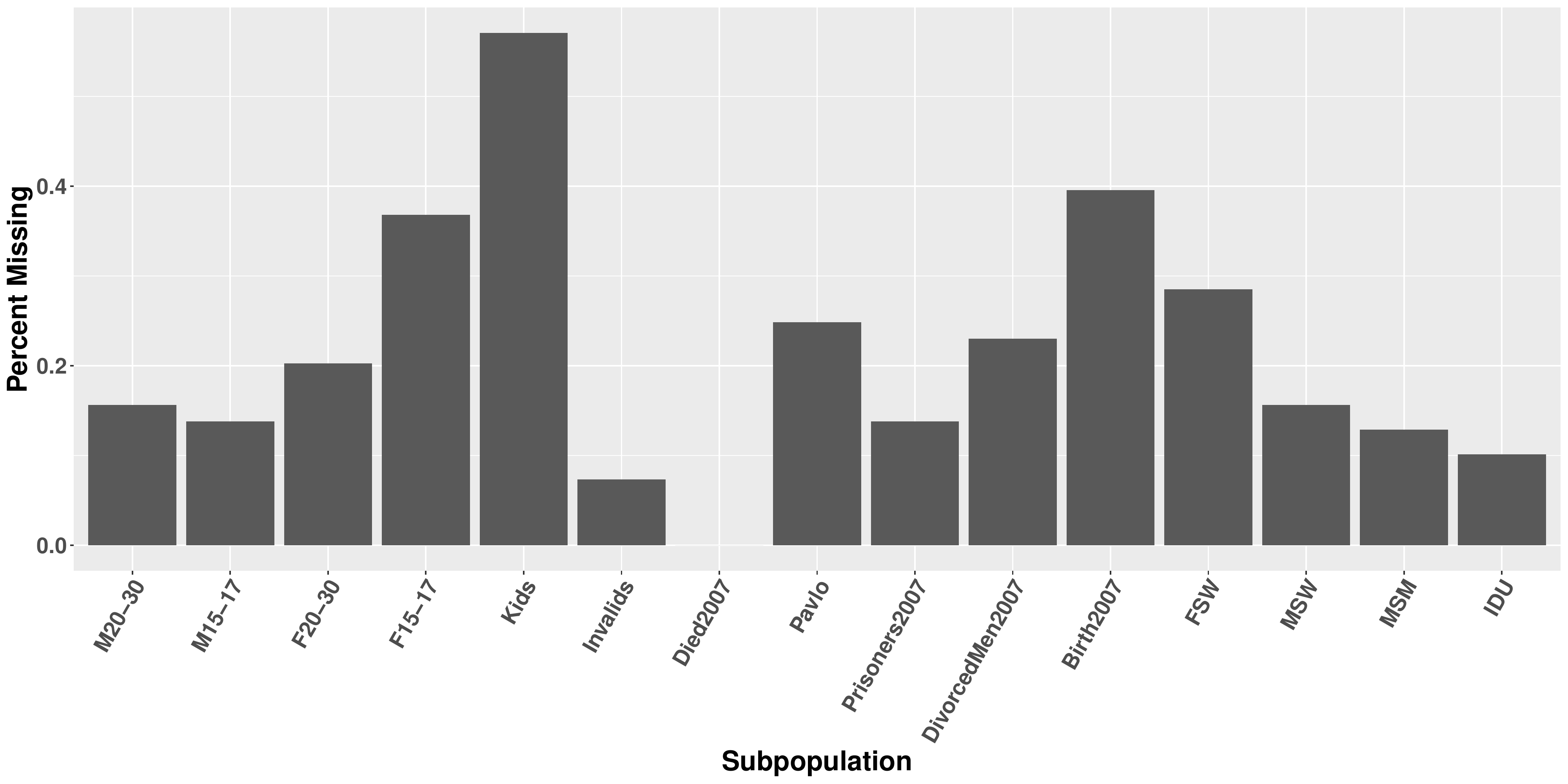}}
    \caption{Barplot showing the percent of missing data for level-of-respect responses in the Ukraine dataset.}
    \label{fig:ukraine_miss_lor_barplot}
\end{figure}

\begin{figure}[!t]
    \centerline{\includegraphics[width=0.8\textwidth]{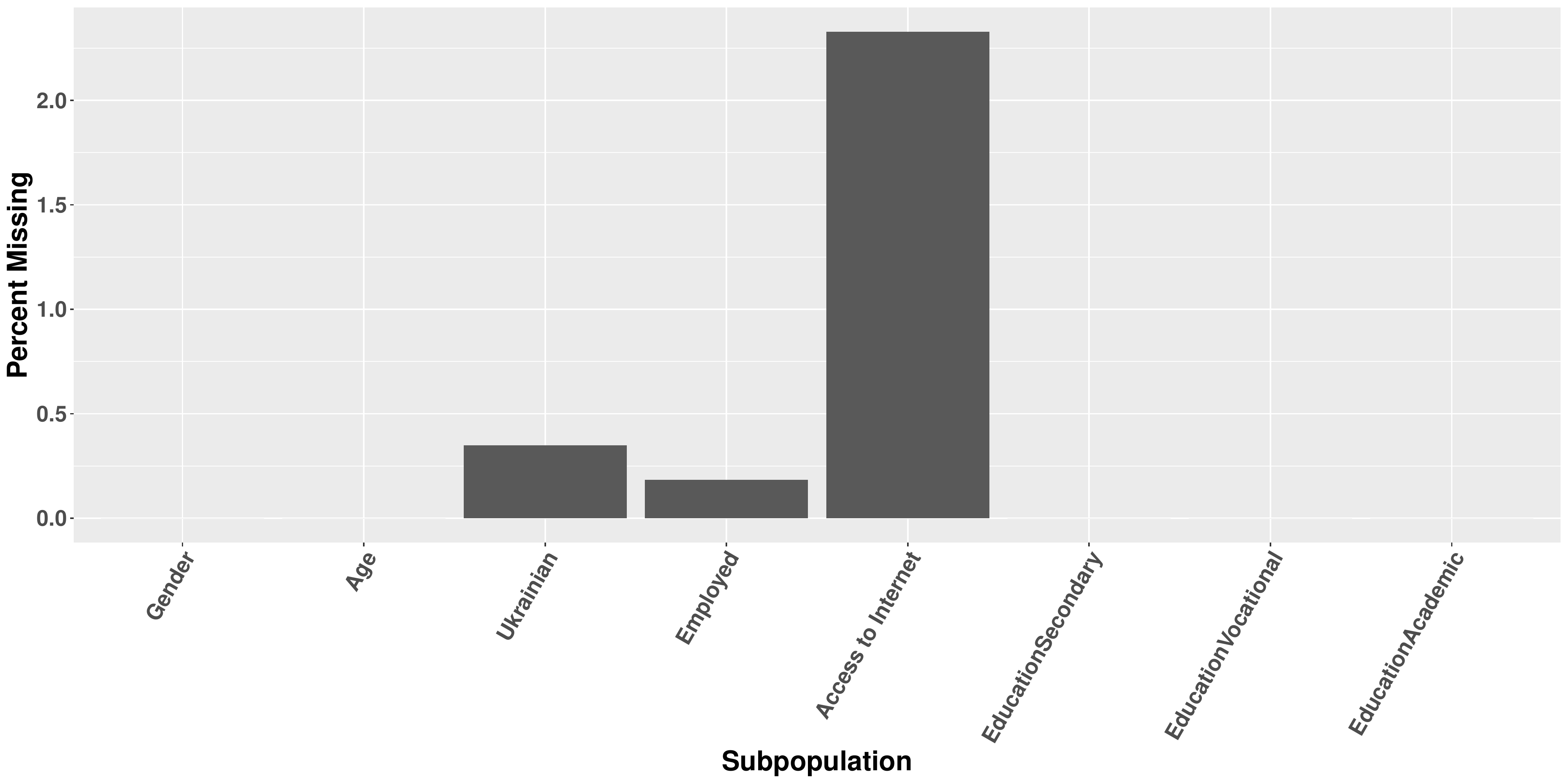}}
    \caption{Barplot showing the percent of missing data for demographic responses in the Ukraine dataset.}
    \label{fig:ukraine_miss_dem_barplot}
\end{figure}

\begin{figure}[!t]
    \centerline{\includegraphics[height=0.8\textheight]{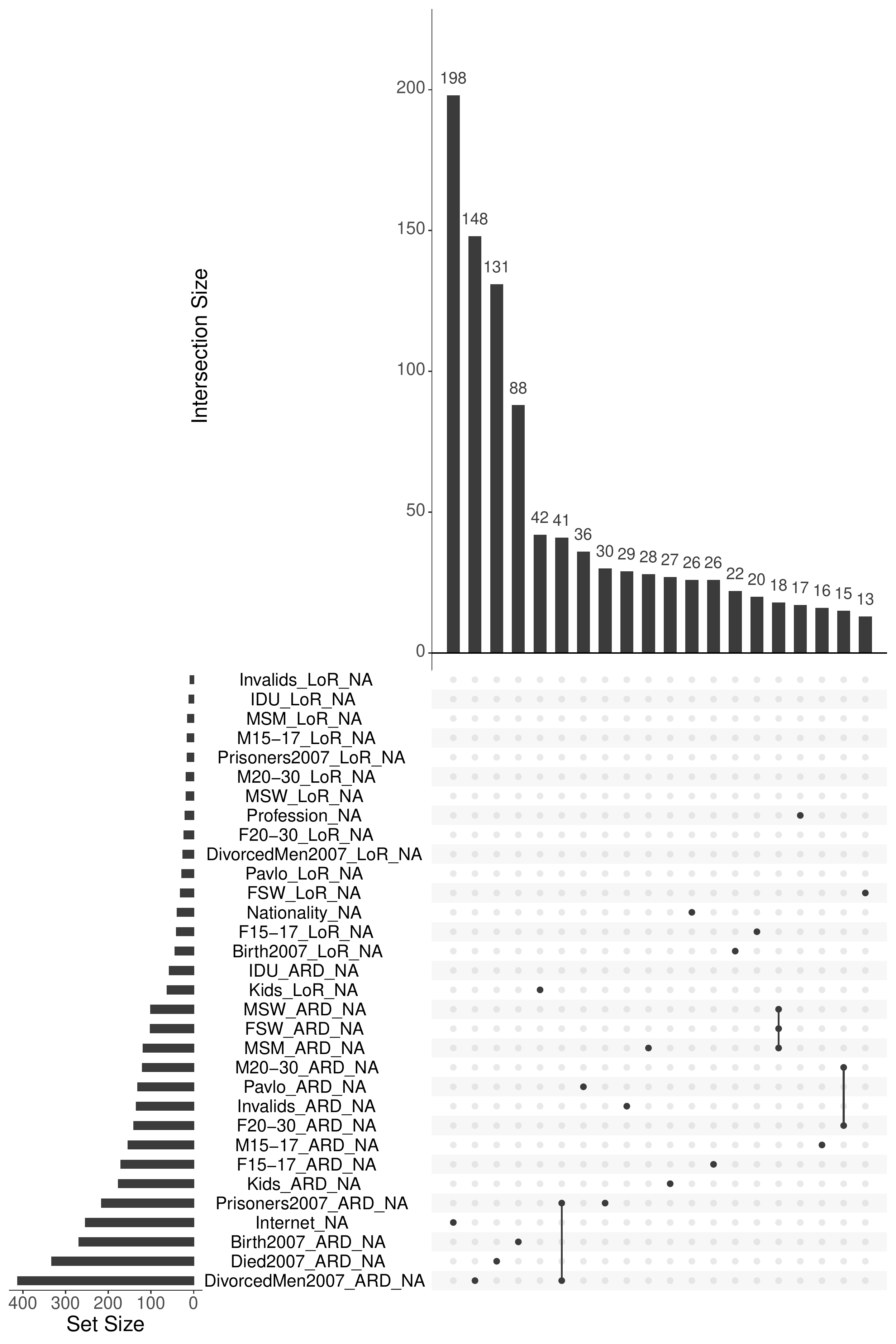}}
    \caption{Plots showing the number of missing responses for intersections between all variables with at least one missing value.}
    \label{fig:ukraine_miss_intersect}
\end{figure}

\begin{figure}[!t]
    \centerline{\includegraphics[width=0.8\textwidth]{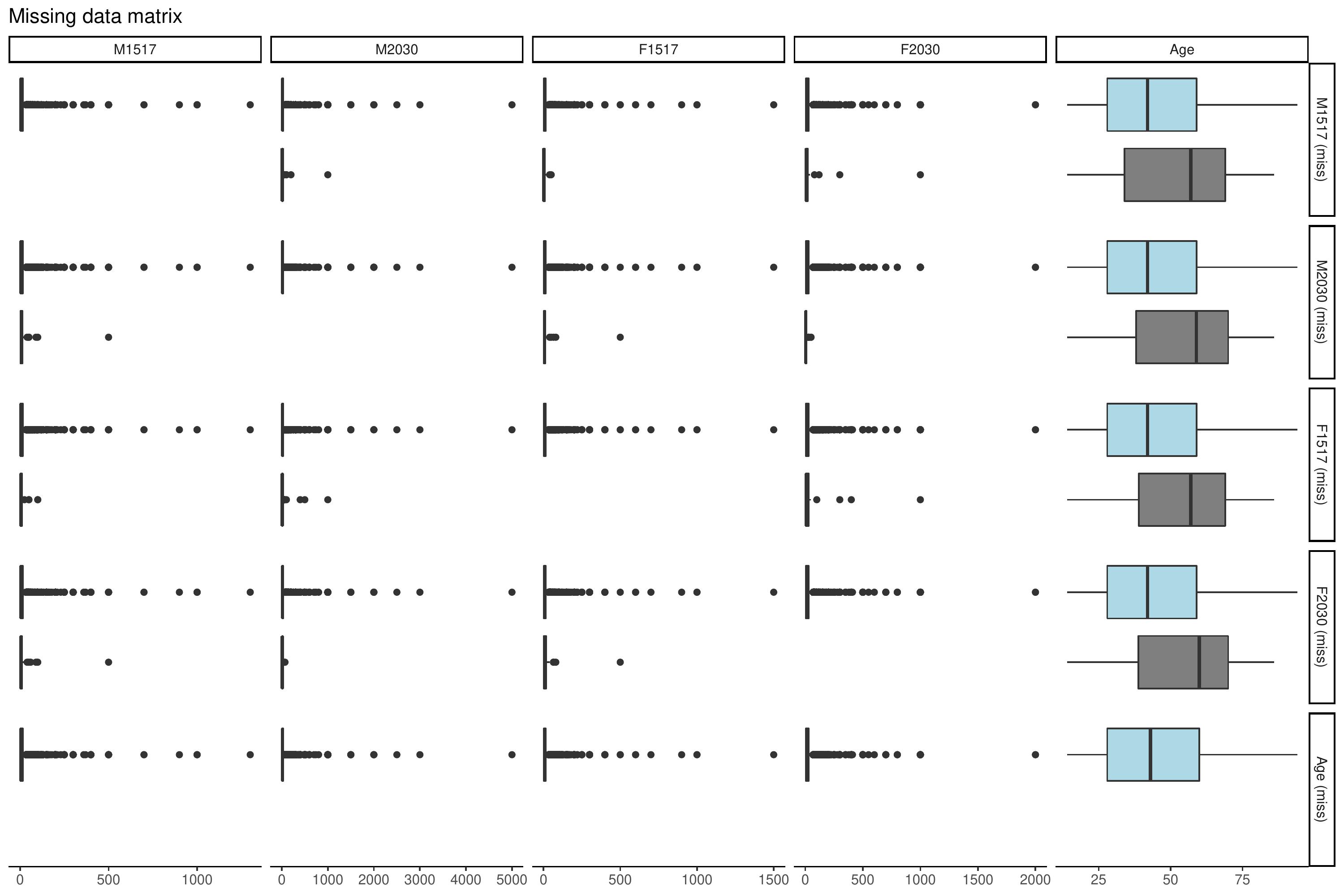}}
    \caption{Missing data matrices for gender-age groups with respect to age.}
    \label{fig:Missing_matrix_men_ages}
\end{figure}

\begin{figure}[!t]
    \centerline{\includegraphics[width=0.8\textwidth]{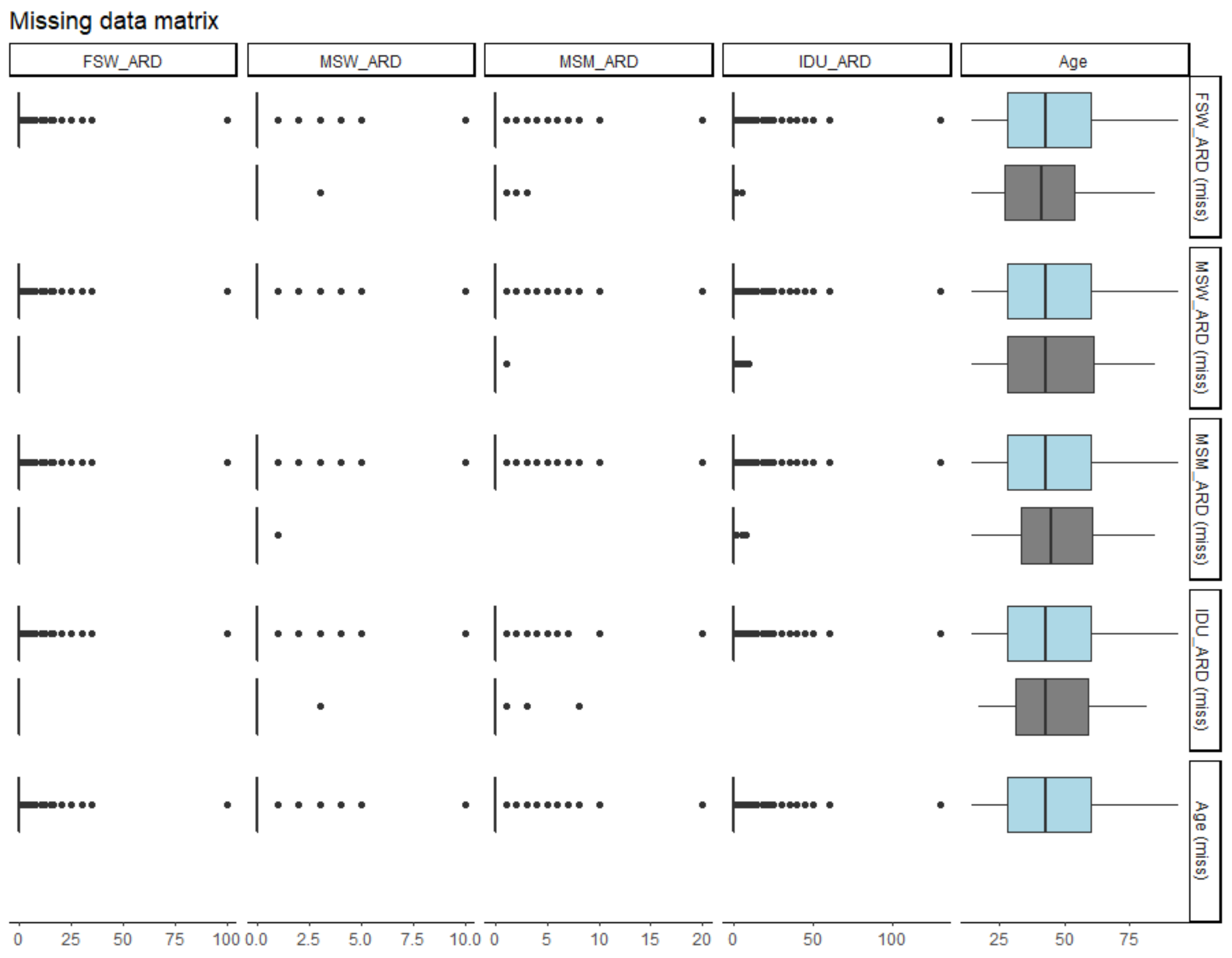}}
    \caption{Missing data matrices for the hard-to-reach groups with respect to age.}
    \label{fig:Missing_matrix_htr_ages}
\end{figure}

\begin{figure}[!tb]
\centering
\begin{subfigure}{.49\textwidth}
  \centering
  \includegraphics[width=1\linewidth]{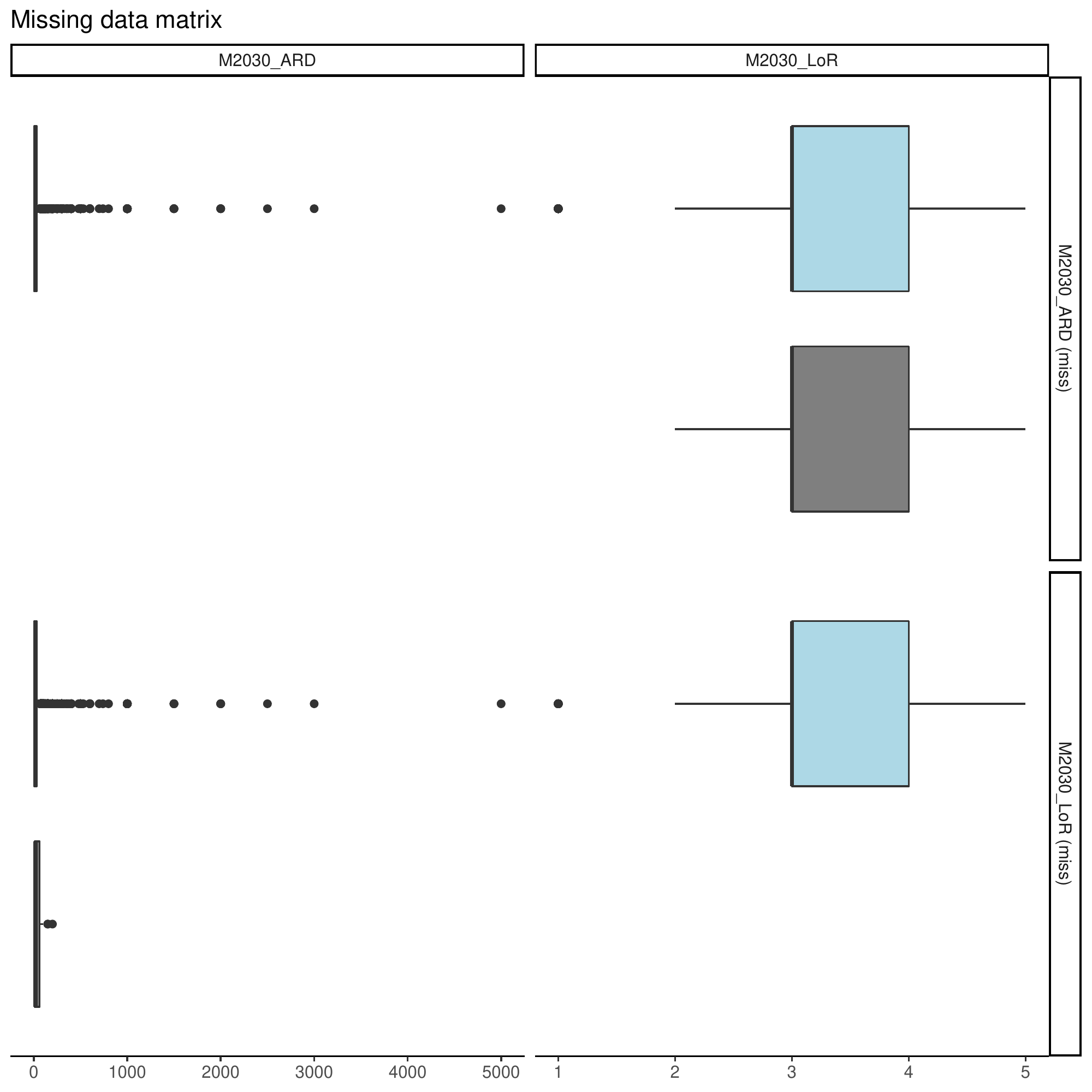}
  \caption{}
  \label{fig:diaga}
\end{subfigure}%
\begin{subfigure}{.49\textwidth}
  \centering
  \includegraphics[width=1\linewidth]{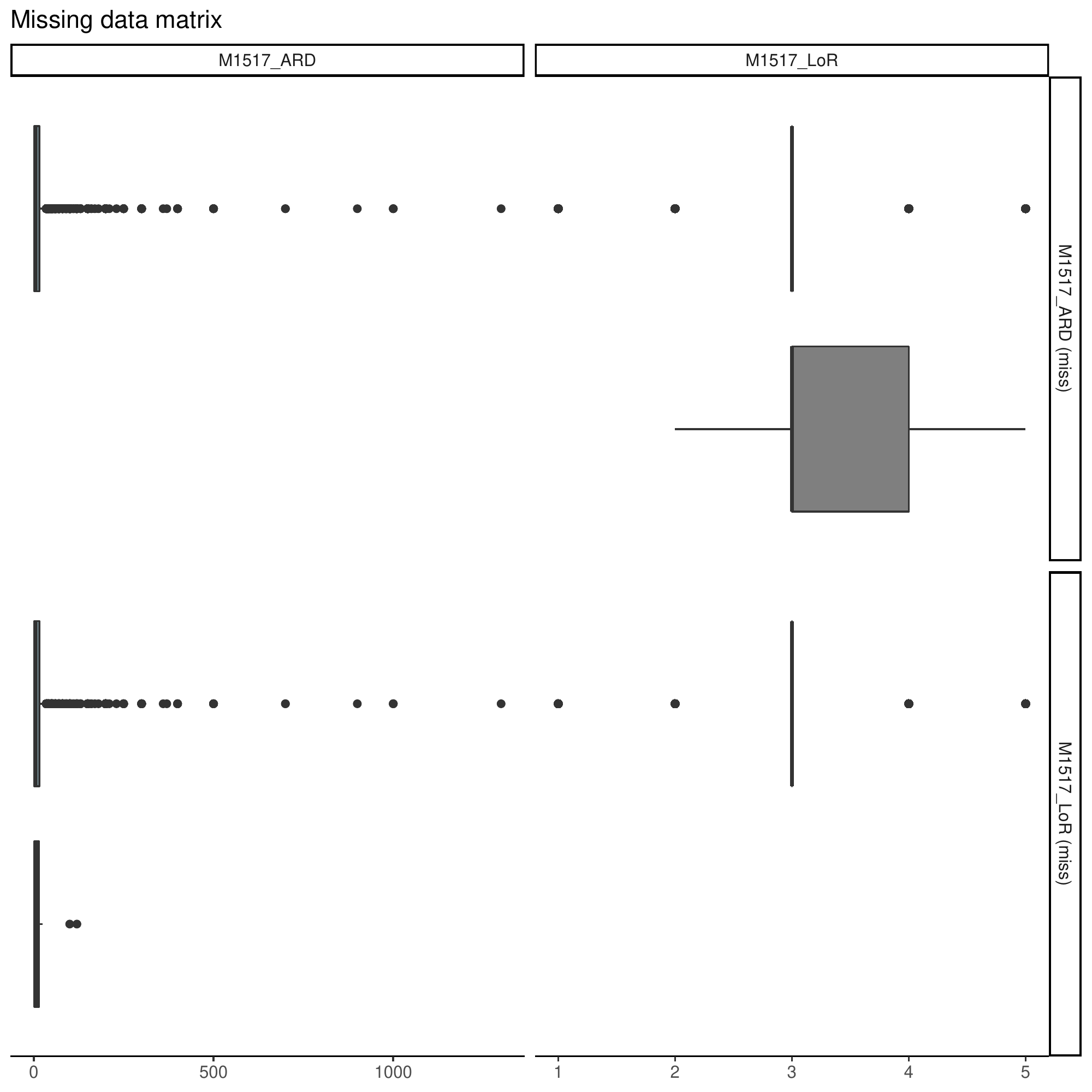}
  \caption{}
  \label{fig:diagb}
\end{subfigure}
\caption{Missing data matrices for the male subpopulations subpopulations, men aged 15-17 (a) and 20-30 (b), with respect to level-of-respect.}
\label{fig:missing_known_lor}
\end{figure}

\begin{figure}[!tb]
\centering
\begin{subfigure}{.4\textwidth}
  \centering
  \includegraphics[width=1\linewidth]{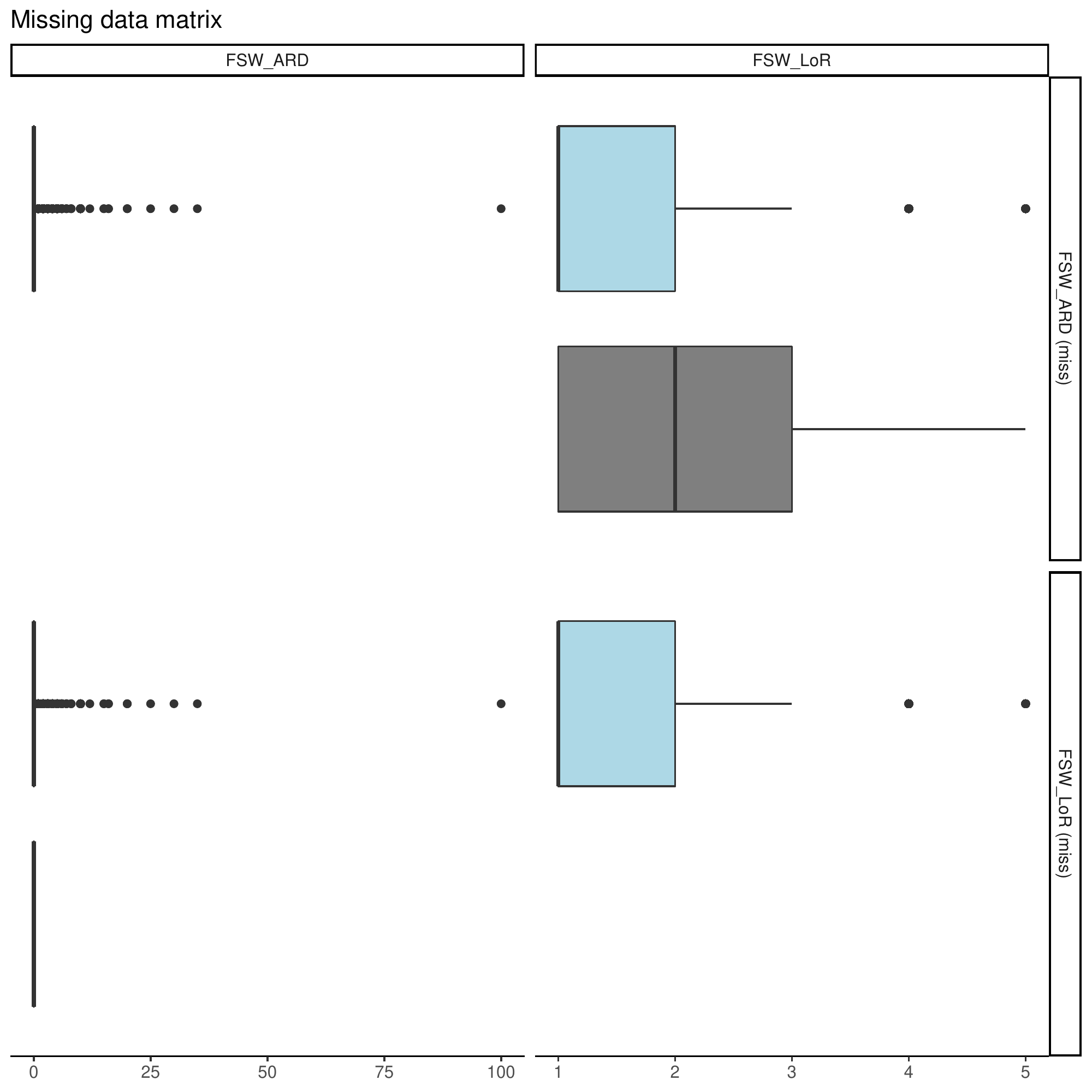}
  \caption{}
\end{subfigure}%
\begin{subfigure}{.4\textwidth}
  \centering
  \includegraphics[width=1\linewidth]{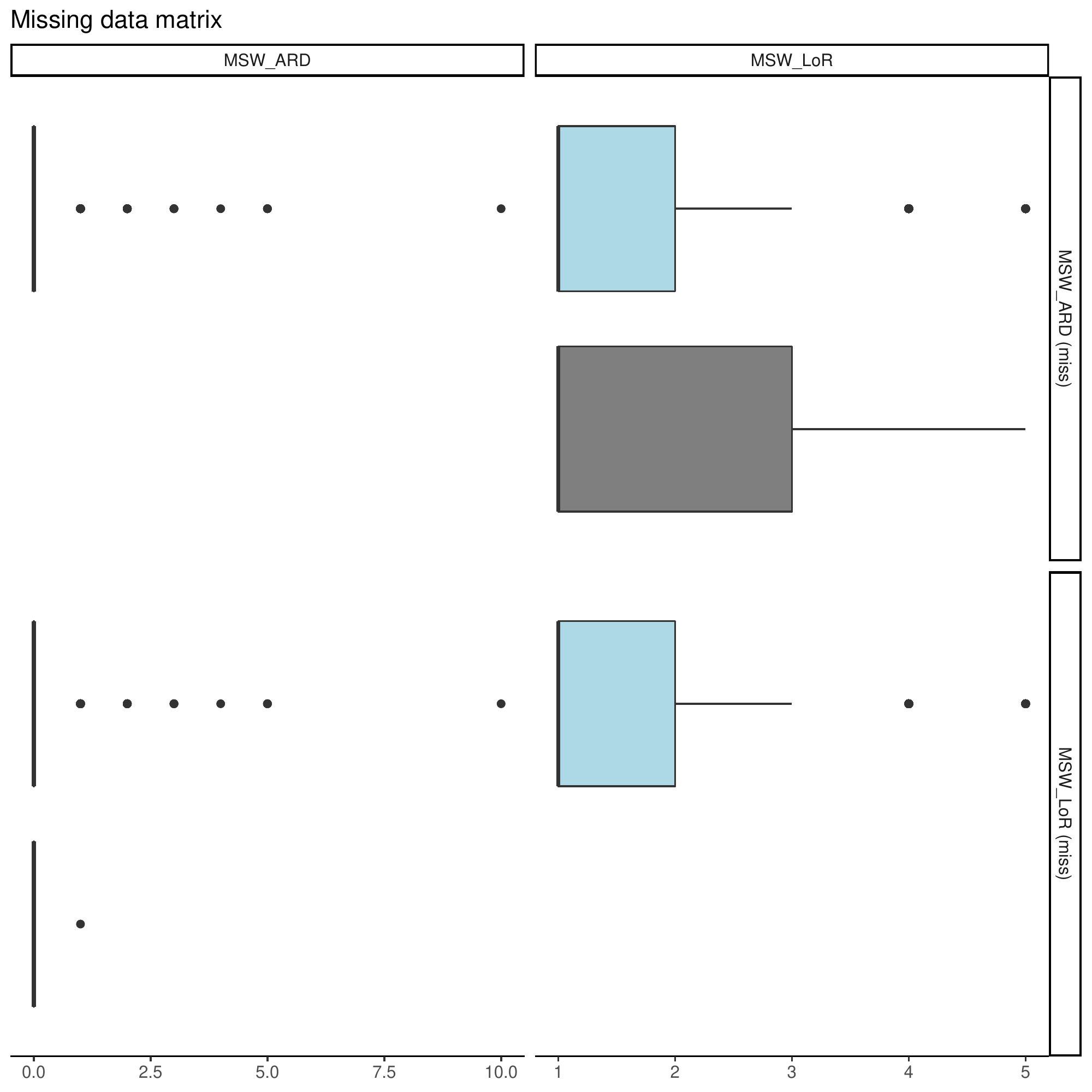}
  \caption{}
\end{subfigure}
\\
\begin{subfigure}{.4\textwidth}
  \centering
  \includegraphics[width=1\linewidth]{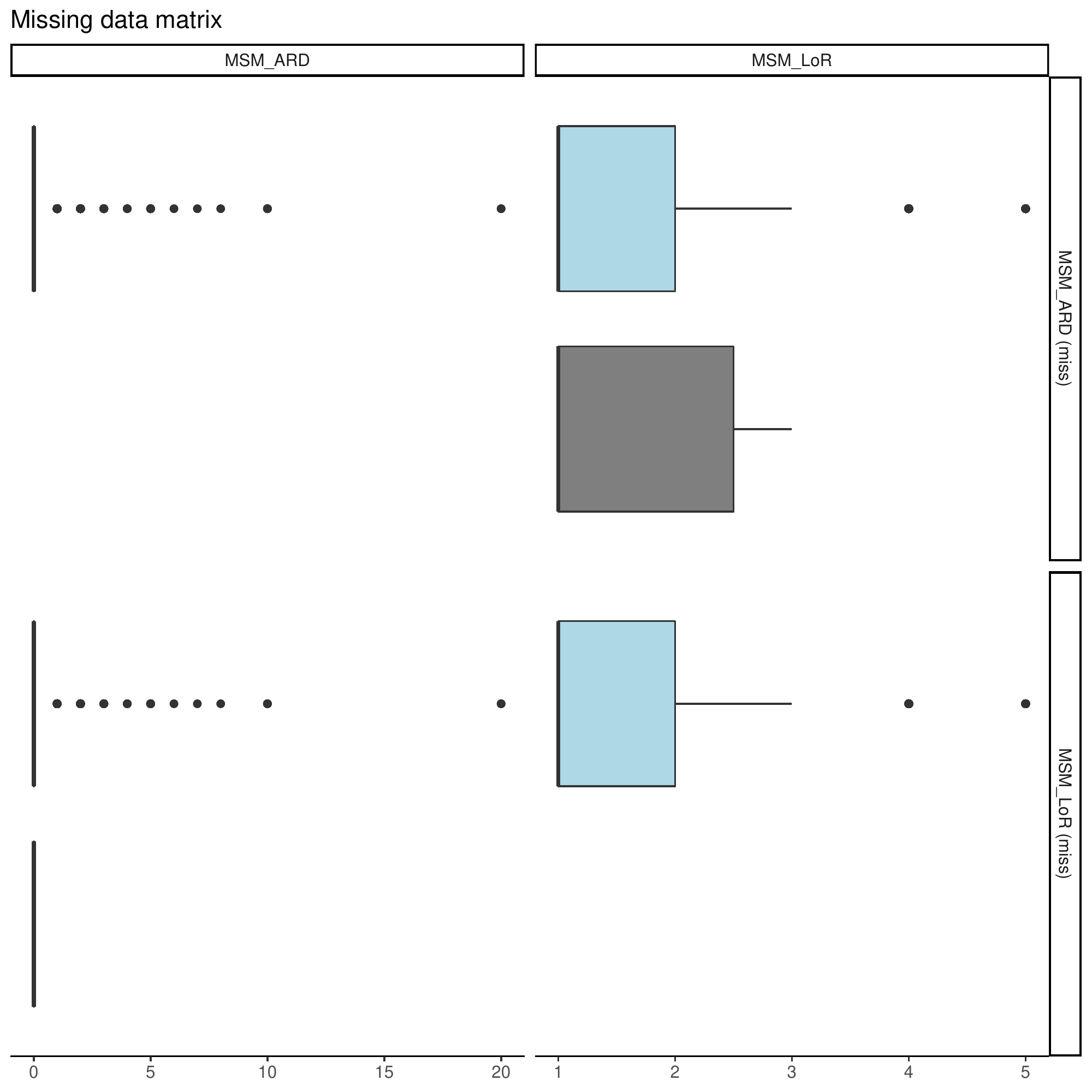}
  \caption{}
\end{subfigure}
\begin{subfigure}{.4\textwidth}
  \centering
  \includegraphics[width=1\linewidth]{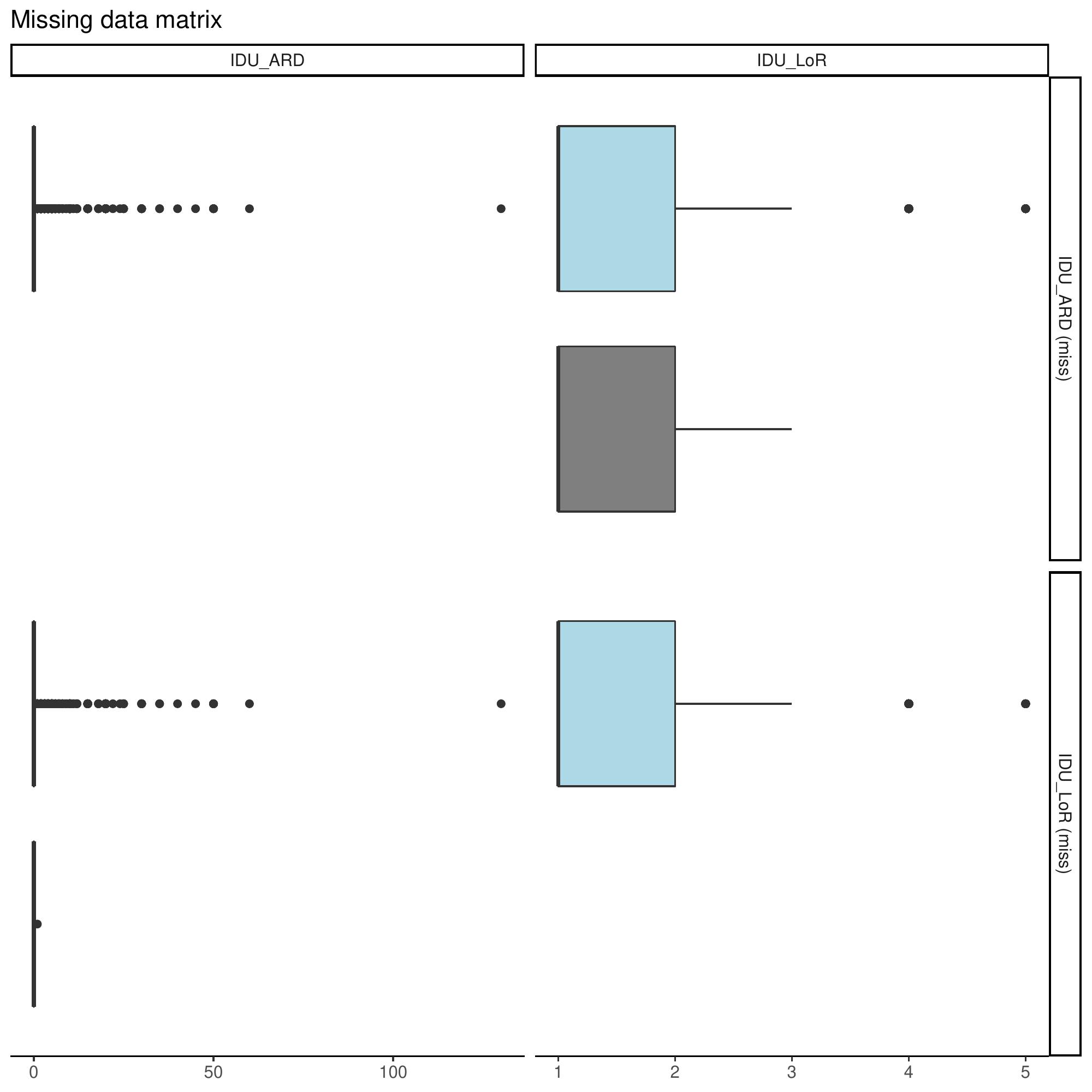}
  \caption{}
\end{subfigure}
\caption{Missing data matrices for the unknown subpopulations, female sex workers (a), male sex workers (b), men who have sex with men (c), and intravenous drug users (d), with respect to level-of-respect.}
\label{fig:missing_htr_lor}
\end{figure}

\begin{figure}[!t]
    \centerline{\includegraphics[width=0.8\textwidth]{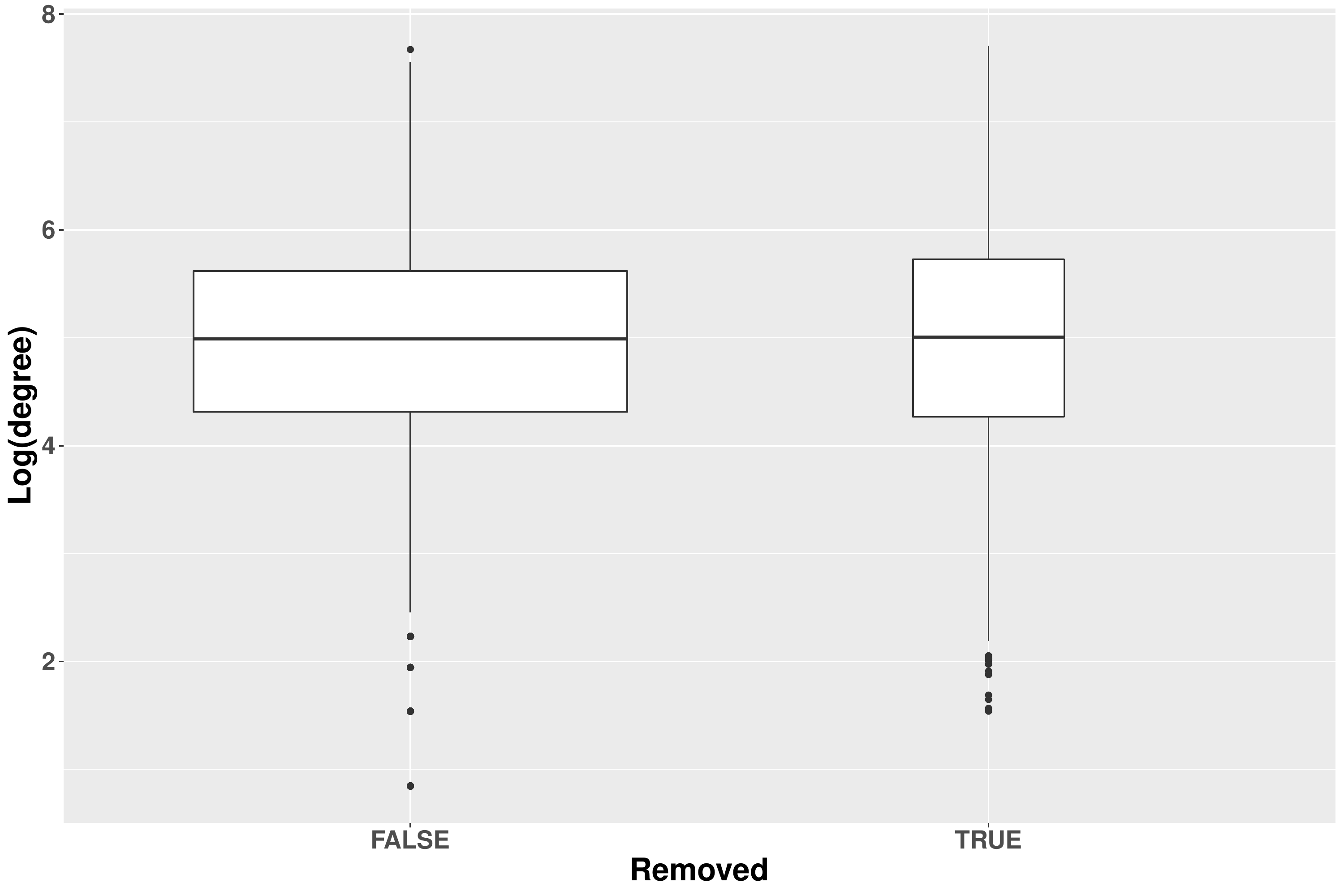}}
    \caption{Boxplots of the log-transformed degree estimates from either the data that was removed from the final analysis (TRUE) or the data that was used in the final analysis (FALSE).}
    \label{fig:ukraine_miss_degree}
\end{figure}

\begin{figure}[!t]
    \centerline{\includegraphics[width=0.8\textwidth]{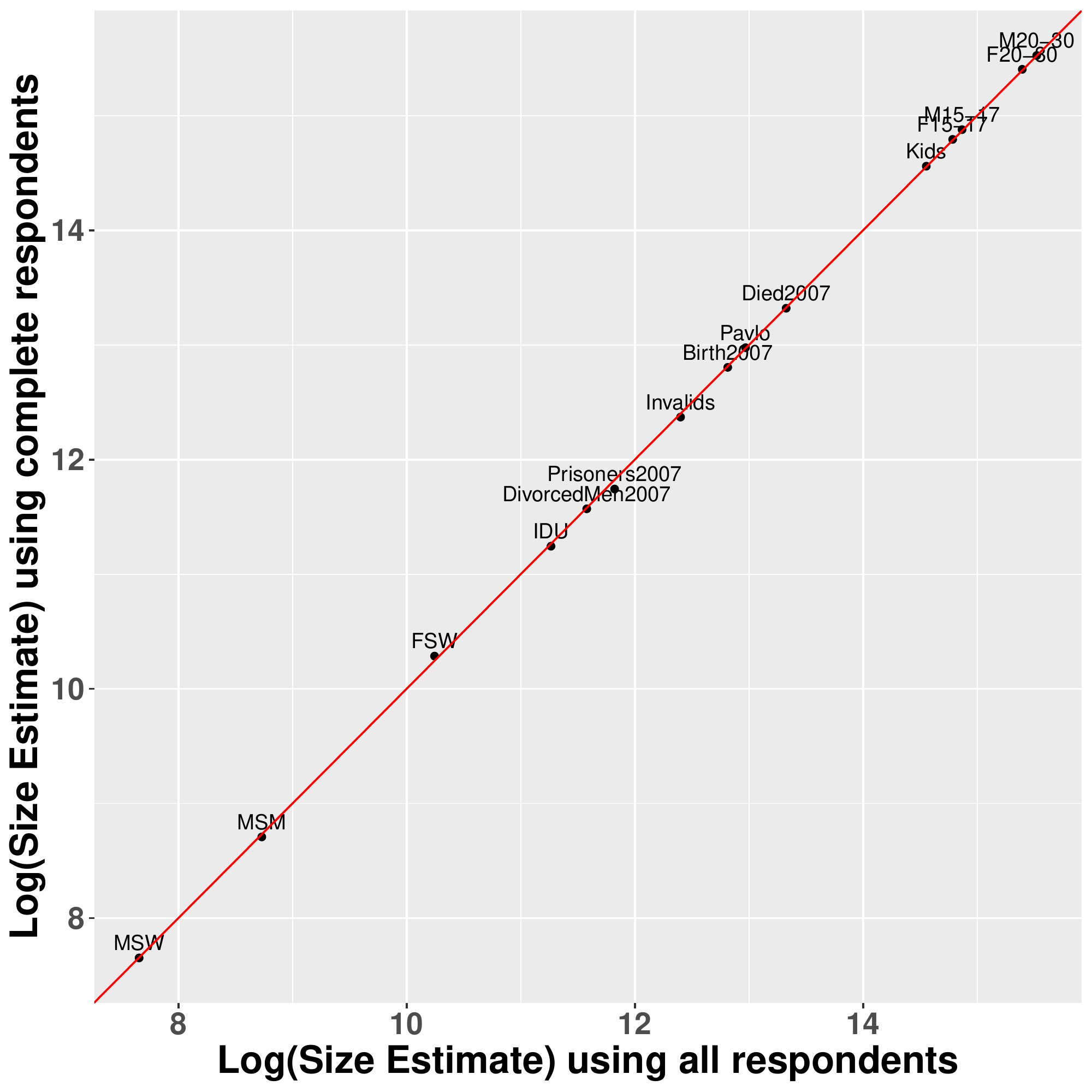}}
    \caption{Scatter plot of the log-transformed size estimates from using either all respondents (x-axis) or using only the complete respondents, i.e. the data that was used in the final analysis (y-axis). A line with intercept 0 and slope 1 is plotted for reference.}
    \label{fig:ukraine_miss_size}
\end{figure}

\begin{figure}[!t]
    \centerline{\includegraphics[width=1\textwidth]{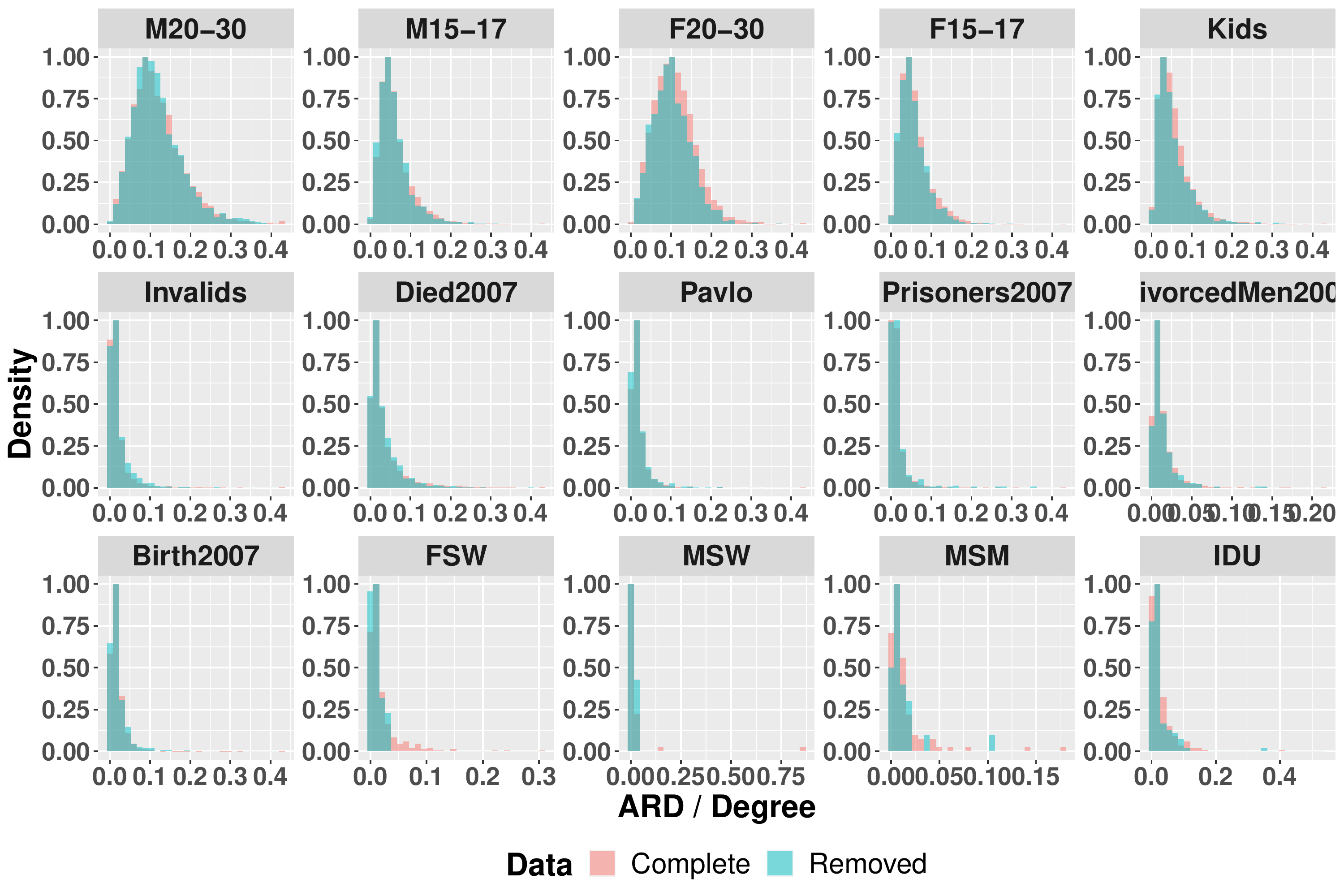}}
    \caption{Histograms of $y_ik / \hat{d}_i$ by subpopulation for both the data used for final analysis (``Complete'' in pink) and the data removed from final analysis (``Removed'' in blue).}
    \label{fig:ukraine_miss_yik}
\end{figure}

\begin{table}[!t]
\centering
\caption{Percent of $y_{ik} / \hat{d}_i$ values that are equal to 0 for the complete or missing data.}
\label{tab:ukraine_miss_yik_0}
\begin{tabular}{l|l|l}
\hline
\textbf{Subpopulation}     & \textbf{Complete}   & \textbf{Missing}  \\ \hline \hline
Men 20-30         & 0.042 &   0.034                    \\ \hline
Men 15-17         & 0.153 &     0.140                  \\ \hline
Female 20-30      & 0.050 & 0.043                           \\ \hline
Female 15-17      & 0.162 & 0.150                        \\ \hline
Kids              & 0.160 & 0.144                         \\ \hline
Invalids              & 0.584 &     0.526                      \\ \hline
Died 2007              & 0.345 &    0.273                    \\ \hline
Pavlo              & 0.294 &    0.258                      \\ \hline
Prisoners 2007    & 0.807 &     0.749                     \\ \hline
Divorced Men 2007 & 0.781 &     0.746                 \\ \hline
Birth 2007        & 0.386 &     0.339                     \\ \hline
FSW               & 0.957 &     0.964                       \\ \hline
MSW               & 0.994 &     0.993                       \\ \hline
MSM               & 0.987 &     0.984                        \\ \hline
IDU               & 0.927 &     0.919                  \\ \hline
\end{tabular}%
\end{table}

\FloatBarrier
\newpage
\section{Simulation Study}
\label{sec:sim_figs}

Here we present simulation studies of other existing NSUM models under correlated data. The \cite{zheng2006many} model is fit using our \code{networkscaleup} package for the negative binomial parameterization and using RStan for the Poisson parameterization, and the \cite{maltiel2015estimating} models are fit using hand-written MCMC code in C++. Note that given the differences between models, model parameters are not identical, but are chosen to have roughly the same mean and variance for each column of the responses.

For each model, we evaluate the performance using two metrics. First, we consider the 95\% interval of the posterior means for the size estimates across the 100 simulations to show whether models produce biased or unbiased size estimates when the true underlying data has correlated random effects. Second, we consider the correlation matrix estimated from the posterior residuals to see whether the underlying correlation matrix can be accurately estimated using only the residuals.

\subsection{\cite{zheng2006many} Model}

\textit{Poisson Model}: The Poisson model was parameterized as follows, adjusting notation to ours used in the main manuscript:
\begin{align*}
    y_{ik} &\sim Poisson(exp\{\delta_i + \rho_k \} \gamma_{ik}) \\
    \delta_i &\sim \mathcal{N}(0, \sigma_\delta^2) \\
    \rho_k &\sim \mathcal{N}(\mu_\rho, \sigma_\rho^2) \\
    \gamma_{ik} &\sim gamma(\lambda_k, \lambda_k) \\
    \lambda_k &\propto 1 I(\lambda_k > 0)
\end{align*}
In this model, the random effects $\gamma_{ik}$ and the parameters $\lambda_k$ are sampled directly.

\noindent \textit{Negative Binomial Model}: The negative binomial model was parameterized as follows, adjusting notation to ours used in the main manuscript:
\begin{align*}
    y_{ik} &\sim Negative-Binomial\left(\frac{exp\{\delta_i + \rho_k \}}{\omega_k - 1}, \frac{1}{\omega_k - 1}\right) \\
    \delta_i &\sim \mathcal{N}(0, \sigma_\delta^2) \\
    \rho_k &\sim \mathcal{N}(\mu_\rho, \sigma_\rho^2) \\
    \frac{1}{\omega_k} &\propto 1 I(0 < \omega_k < 1)
\end{align*}
In this model, the random effects $\gamma_{ik}$ are integrated out, meaning only the $\omega_k$ are sampled.

We simulate correlated and uncorrelated data from the Poisson parameterization of the Overdispersed model, with $n = 500$, $K = 5$, $\delta_i \sim N(5, 0.7^2)$, and $\rho_k = 0.015$ for all $k$. We simulate random effects for respondent $i$, $\bm{\gamma}_i$, from a gamma distribution with mean 1 using the inverse CDF method. First, we simulate a vector of random variables from a multivariate normal random variable with mean $\bm{0}$ and variance $\Omega$. Second, we transform these marginal normal mean 0, variance 1 random variables to a vector of correlated uniform random variables by taking the quantile of the random variables. Finally, we transform the vector of uniform random variables to a vector of correlated gamma random variables by taking the inverse CDF of the gamma distribution with shape and rate equal to $\bm{\lambda} = (0.5, 0.95, 1.5, 0.98, 0.82)$. We fit the datasets using both the Poisson and negative binomial parameterizations.

\subsection{\cite{maltiel2015estimating} Model}

\textit{Integrated Barrier Effects Model}: The integrated barrier effects model was parameterized as follows, adjusting notation to ours used in the main manuscript:
\begin{align*}
    y_{ik} &\sim Binom\left(d_i, q_{ik}\right) \\
    d_i &\sim \text{Log Normal}(\mu, \sigma^2) \\
    q_{ik} &\sim Beta(m_k, \rho_k) \\
    \pi(m_k) &\propto \frac{1}{m_k} \\
    \rho_k &\sim U(0, 1)
\end{align*}
In this model, we follow the proposed MCMC algorithm in \cite{maltiel2015estimating}, using the beta-binomial distribution to integrate out the $q_{ik}$, only sampling $m_k$ and $\rho_k$.

\noindent \textit{Sampled Barrier Effects Model}: The sampled barrier effects model is formulated as above, except the MCMC algorithm samples the $q_{ik}$ directly instead of using the beta-binomial distribution.

We simulate correlated and uncorrelated data from the \textit{barrier effects model} described in \cite{maltiel2015estimating}, with $n = 500$, $K = 5$, $d_i \sim N(5, 0.7^2)$ and then round to the nearest integer, and $m_k = 0.015$ for all $k$. We again use the inverse CDF method to simulate correlated random effects, this time according to the beta distribution with overdispersion vector $\bm{\rho} = (0.02, 0.01, 0.005, 0.015, 0.018)$. We fit the datasets using the integrated random effects MCMC algorithm provided in the original manuscript, and using the barrier effects model but sampling the random effects directly.

\subsection{Results}

The results for the size estimates corresponding to the \cite{zheng2006many} model is shown in Supplementary Figure \ref{fig:zheng_sims}. For both the Poisson and negative binomial parameterizations, the size estimates are unbiased when the random effects are uncorrelated. However, for correlated random effects, the Poisson parameterizations produces biased size estimates, while the negative binomial model estimates stay unbiased. However, the negative binomial model produces very conservative uncertainty intervals, with the 95\% coverage probabilities of the 5 groups being 100, 100, 100, 97, and 97. 

\begin{figure}[!t]
    \centerline{\includegraphics[width=0.75\textwidth]{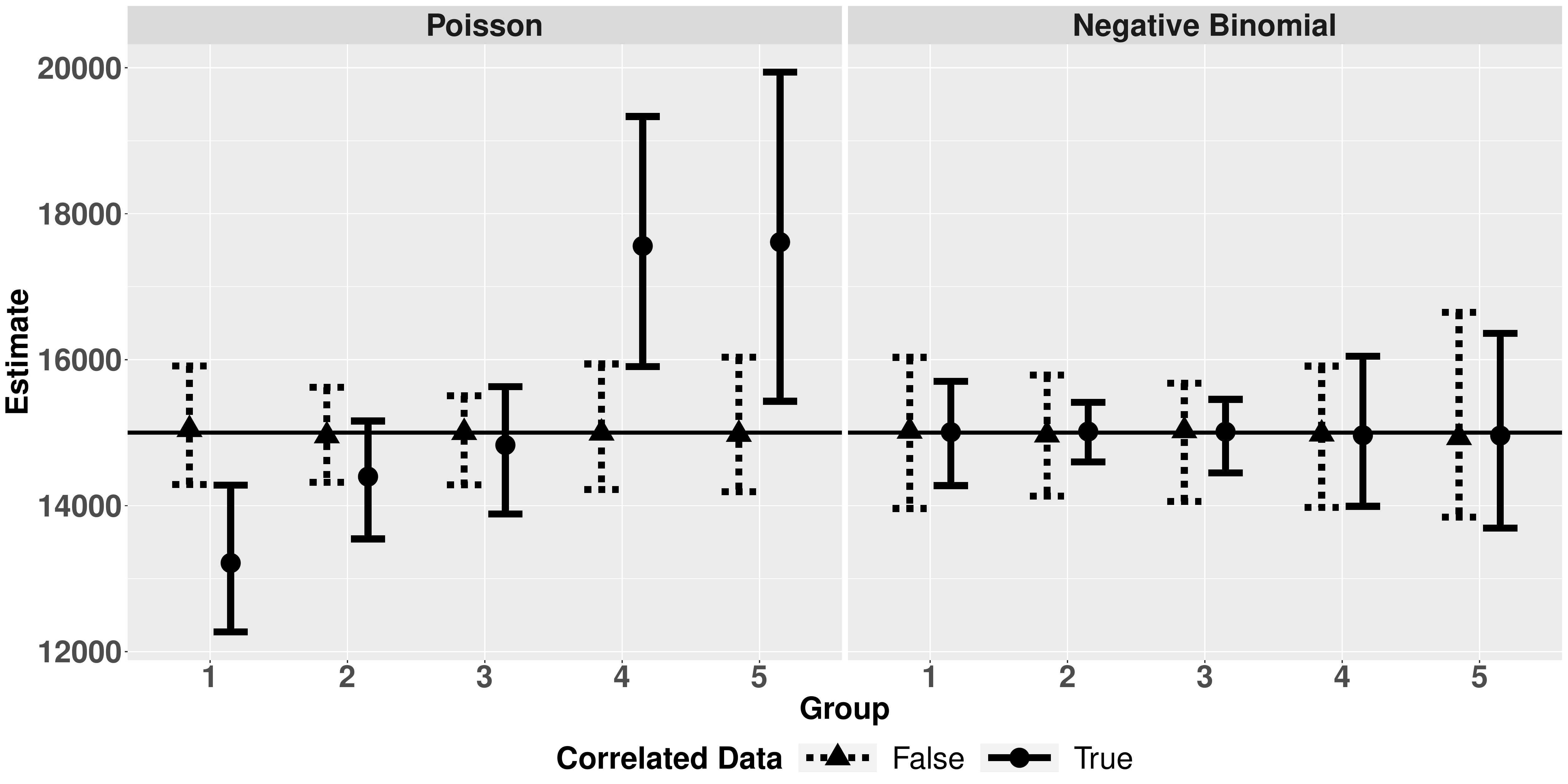}}
    \caption{95\% interval of posterior means of size estimates across 100 simulations for the Poisson and negative binomial \cite{zheng2006many} models. The true size is represented by the horizontal black line.}
    \label{fig:zheng_sims}
\end{figure}

Regarding the correlation matrix, we present the estimated correlation matrices for the correlated data rounded to the second decimal place in Equation \eqref{eq:zheng_poisson_corr} for the Poisson model and Equation \eqref{eq:zheng_nb_corr} for the negative binomial model. We calculate this matrix for the Poisson model by first calculating the residual matrix for simulation $sim$ as $r^{m,sim}_{ik} = \sqrt{y^{sim}_{ik}} - \sqrt{exp\{\delta^{m,sim}_{i} + \rho^{m,sim}_{k}\}\gamma^{m,sim}_{ik}}$, where the superscript ${m,sim}$ refers to the $m^{th}$ posterior sample for simulation $sim$. For the negative binomial model, we calculate the residual matrix for simulation $sim$ as $r^{m,sim}_{ik} = \sqrt{y^{sim}_{ik}} - \sqrt{exp\{\delta^{m,sim}_{i} + \rho^{m,sim}_{k}\}}$, following the procedure in \cite{zheng2006many}. Then, for each residual matrix corresponding to each posterior sample, $\bm{\rho}^{m,sim}$, we calculate the sample correlation matrix $\Omega^{m,sim}$. Finally, we take the average of these $\Omega^{m,sim}$ first across $m$, and then across sim, yielding our estimate $\hat{\Omega}$.

For both \cite{zheng2006many} models, the correlations between correlated populations is underestimated while the correlation between uncorrelated populations is overestimated. While the correlation matrices estimated from the residuals offer some insight into the true underlying correlation matrix, the model is unable to completely recover the true correlation through the residuals. Instead, the estimated correlation matrices are reflecting the correlation induced by the degrees. In more complicated datasets, further correlation will also be induced by the covariates.

\begin{align}
\label{eq:zheng_poisson_corr}
   \hat{\Omega}^{Poisson} =  \begin{pmatrix}
1 & 0.65 & 0.54 & 0.15 &  0.17\\
0.65 & 1 & 0.49 & 0.17 &  0.12\\
0.54 & 0.49 & 1 & 0.16 &  0.16\\
0.15 & 0.17 & 0.16 & 1 &  0.56\\
0.17 & 0.12 & 0.16 & 0.56 &  1
\end{pmatrix}
\end{align}

\begin{align}
\label{eq:zheng_nb_corr}
\hat{\Omega}^{NB} = \begin{pmatrix}
1 & 0.75 & 0.62 & 0.15 &  0.17\\
0.72 & 1 & 0.61 & 0.22 &  0.16\\
0.62 & 0.61 & 1 & 0.25 &  0.23\\
0.15 & 0.22 & 0.25 & 1 &  0.69\\
0.17 & 0.16 & 0.23 & 0.49 &  1
\end{pmatrix}
\end{align}

The results for the size estimates corresponding to the \cite{maltiel2015estimating} models are shown in Supplementary Figure \ref{fig:maltiel_sims}. Like for the \cite{zheng2006many} models, both \cite{maltiel2015estimating} models produce accurate size estimates when the underlying random effects are uncorrelated. Based on the findings of the \cite{zheng2006many} simulations, we first hypothesized that models that sample the random effects directly (e.g. the sampled barrier effects model) would produce biased estimates, while models that integrate out random effects (e.g. the integrated barrier effects model) would produce unbiased estimates. However, these do not seem to be the case. In this case, both the integrated and sampled barrier effects models produce biased size estimates for correlated data.

\begin{figure}[!t]
    \centerline{\includegraphics[width=0.75\textwidth]{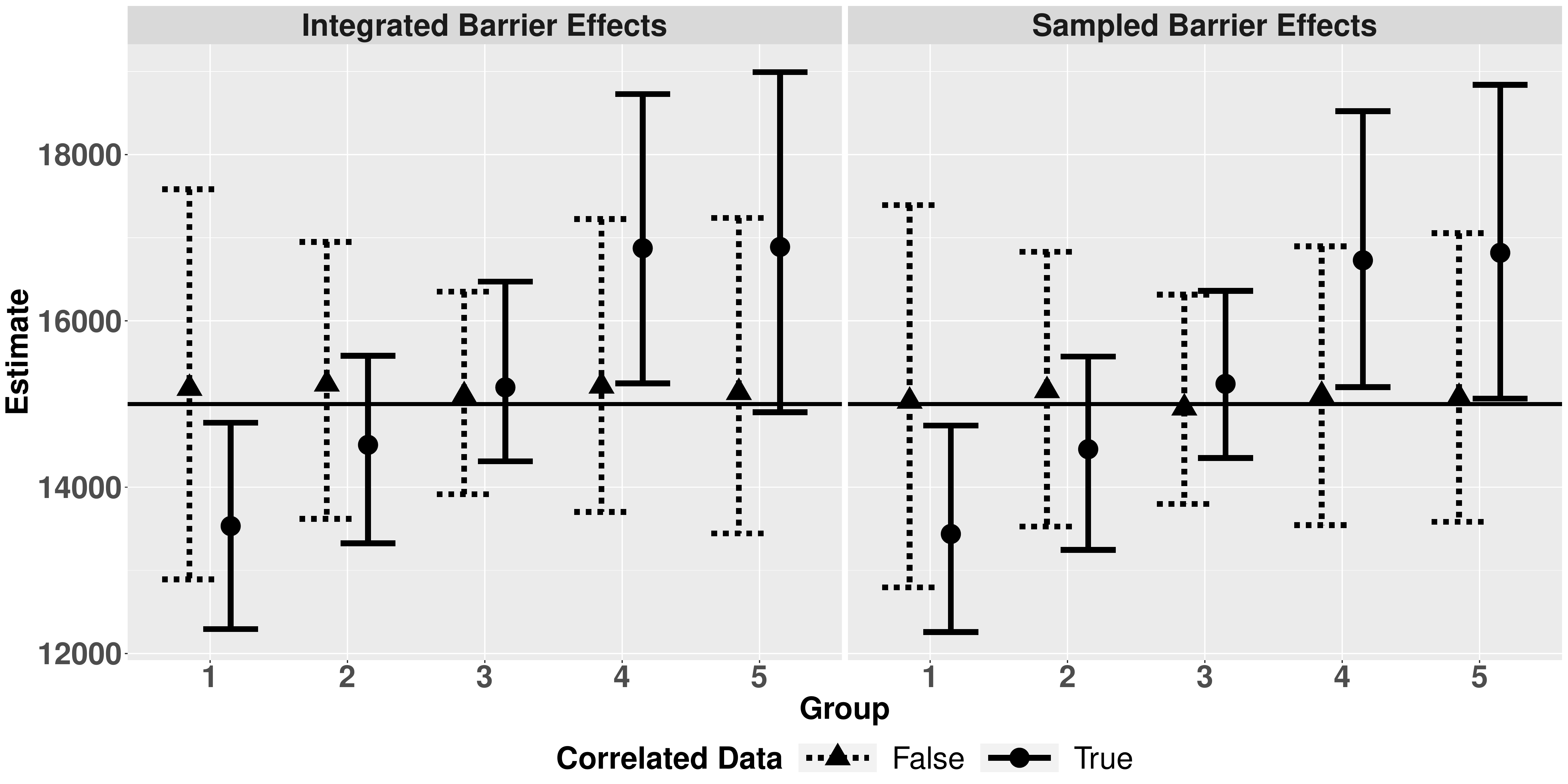}}
    \caption{95\% interval of posterior means of size estimates across 100 simulations for the integrated barrier effects and sampled barrier effects \cite{maltiel2015estimating} models. The true size is represented by the horizontal black line.}
    \label{fig:maltiel_sims}
\end{figure}

For the correlation matrix estimated from the residuals, we estimate the correlation matrix using the same approach, except now where $r^{m,sim}_{ik} = \sqrt{y^{sim}_{ik}} - \sqrt{d^{m,sim}_i m^{m,sim}_{k}}$ for the integrated barrier effects model and where $r^{m,sim}_{ik} = \sqrt{y^{sim}_{ik}} - \sqrt{d^{m,sim}_i q^{m,sim}_{ik}}$ for the sampled barrier effects model. The estimated correlation matrix for the integrated barrier effects model is in Equation \eqref{eq:maltiel_integrated_corr}, while the estimated correlation matrix for the sampled barrier effects model is in Equation \eqref{eq:maltiel_sampled_corr}. Given that the true subpopulation sizes are fixed for all but the unknown subpopulation for the \cite{maltiel2015estimating} models, these correlation matrices resemble neither the true underlying random effect correlation matrix nor the average sample correlation of the simulated datasets. These simulations show that under the true model, the underlying correlation matrix is not accurately estimated from the residuals.

\begin{align}
\label{eq:maltiel_integrated_corr}
   \hat{\Omega}^{Integrated} =  \begin{pmatrix}
1 & 0.38 & 0.14 & -0.29 &  -0.23\\
0.38 & 1 & 0.08 & -0.19 &  -0.26\\
0.14 & 0.08 & 1 & -0.14 &  -0.13\\
-0.29 & -0.19 & -0.14 & 1 &  0.62\\
-0.23 & -0.26 & -0.13 & 0.62 &  1
\end{pmatrix}
\end{align}

\begin{align}
\label{eq:maltiel_sampled_corr}
\hat{\Omega}^{Sampled} = \begin{pmatrix}
1 & 0.22 & 0.10 & -0.10 &  -0.07\\
0.22 & 1 & 0.07 & -0.07 &  -0.11\\
0.10 & 0.07 & 1 & -0.06 &  -0.06\\
-0.10 & -0.07 & -0.06 & 1 &  0.21\\
-0.07 & -0.11 & -0.06 & 0.21 &  1
\end{pmatrix}
\end{align}


Based on the findings, we conjecture that models which have separate parameters to estimate the mean of the data and the overdispersion can produce unbiased estimates when data are correlated, while models that have parameters that influence both the mean and the variance simultaneously may lead to biased size estimates. In general, we believe that it is important to model the correlation directly, both for obtaining reliable inference results and for understanding the network structure.

\FloatBarrier
\section{Surrogate Residuals}
\label{sec:diagnostics}

Diagnostics to evaluate how well the assumed model fits the data have been largely underdeveloped for models involving ARD. Leave-one-out (LOO) subpopulation size estimates have been considered in almost all NSUM literature. In addition to LOO, \cite{zheng2006many} used standardized residuals from their model (defined as $r_{ik} = \sqrt{y_{ik}} - \sqrt{a_i b_k}$) to study the residual correlation between subpopulations and the relationship between the residuals and the individual-level predictors. \cite{mccormick2012latent} and \cite{maltiel2015estimating} also used absolute relative error of the $N_k$ estimates to evaluate the performance of models under different assumptions, although this line of thinking has limited applicability to real data because the true $N_k$ is unknown for the hard-to-reach populations. Here we propose new Bayesian surrogate residuals to evaluate goodness of fit.

Residuals for discrete data often have limited use because of the unusual visual appearance of the residuals and the lack of any theoretical distribution for the residuals under the correctly specified model. \cite{liu2018residuals} introduced the surrogate residuals, which solve these problems by generating continuous residuals that follow a known distribution under the null distribution, allowing the user to interpret the residuals in a familiar way. For ARD models, surrogate residuals are most useful because they can be used to detect anomalies between the observed data and the assumed distribution, inform variable selection, and diagnose the relationship between the response and the covariates. While our model is not in a class of cumulative link regression models as was the primary focus of \cite{liu2018residuals}, the authors provided two methods to calculate surrogate residuals for general models. Specifically, the authors showed  under their jittering strategy (B) that given an outcome $Y \sim F_a(y; \bm{X}, \bm{\beta})$, jittered surrogate variables $S | Y = y \sim U(F_a(y-1), F_a(y))$, and residuals $R = S - E_0\{S|X\}$, the residuals $R$ have properties $E\{R|X\}=0$ and $R|\bm{X} \sim U(-1/2, 1/2)$ under the null hypothesis $F_a = F_0$.

Calculating the surrogate residuals in our setting is not obvious. \cite{liu2018residuals} proposed the surrogate residuals under a frequentist paradigm, while our model is inherently Bayesian. To the best of our knowledge, the surrogate residuals have only been applied to one Bayesian model \citep{park2021bayesian}. In this case the model was ordinal, and the authors provided little detail on how to calculate the residuals. Thus, we propose a method to calculate Bayesian surrogate residuals. Actually calculating the residuals is difficult since $E_0\{S|\bm{X}\}$ is typically analytically intractable. However, it is straightforward to estimate $E_0\{S|\bm{X}\}$ using Monte Carlo approximations. In Algorithm 2 we outline our approach for calculating the residuals $R$, which relies on samples from the posterior predictive distribution. Note that the algorithm is given for a single $y_{ik}$ and the process should be repeated for all combinations of $i$ and $k$.

\begin{algorithm}[!t]
\SetAlgoLined
\KwResult{Surrogate residuals $R$}
    Set $N_{mc}$ equal to the number of posterior samples\\
 \For{each ind in 1:$N_{mc}$}{
    Calculate $F_{a,cond}^{ind}(y_{ik} - 1 | \hat{\bm{\theta}}^{ind})$\;
    Calculate $F_{a,cond}^{ind}(y_{ik} | \hat{\bm{\theta}}^{ind})$\;
    Simulate $y^{ind} \sim Poisson(\hat{\bm{\theta}}^{ind})$\;
    Calculate $F_a^{ind}(y^{ind} - 1 | \hat{\bm{\theta}}^{ind})$\;
    Calculate $F_a^{ind}(y^{ind} | \hat{\bm{\theta}}^{ind})$\;
    Simulate $S^{ind} \sim U(F_a^{ind}(y^{ind} - 1 | \hat{\bm{\theta}}^{ind}), F_a^{ind}(y^{ind} | \hat{\bm{\theta}}^{ind}))$\;
    Simulate $S_{cond}^{ind} \sim U(F_{a,cond}^{ind}(y_{ik} - 1 | \hat{\bm{\theta}}^{ind}), F_{a,cond}^{ind}(y_{ik} | \hat{\bm{\theta}}^{ind}))$\;
 }
 Calculate $\hat{E}_0(S^{ind} | \bm{X}) = \frac{1}{N_{mc}} \sum_{ind=1}^{N_{mc}} S^{ind}$\;
 Calculate $R^{ind} = S_{cond}^{ind} - \hat{E}_0(S^{ind} | \bm{X})$\;
 \caption{Bayesian Surrogate Residuals}
\end{algorithm}
In our case, $F_a(y^{ind})$ ($F_{a,cond}(y_{ik})$) is the CDF of the Poisson distribution evaluated at the simulated responses (observed responses). Note that under the Bayesian paradigm, for each observation $y_{ik}$, there is actually a vector of surrogate residuals $R^{ind}$. Thus, we can study the posterior distribution of the surrogate residuals, and we recommend looking at the behavior of the surrogate residuals across multiple posterior samples. Alternatively, we could use a point estimate for $\hat{\bm{\theta}}$ (e.g. posterior mean) to calculate a single vector of surrogate residuals. This approach is more similar to the original frequentist approach, but doesn't capture the uncertainty of the parameter estimates.

\FloatBarrier
\newpage
\section{Ukraine Analysis}

\subsection{Posterior predictive check descriptions}
Posterior predictive checks are a staple of Bayesian model diagnostics, but have remained largely unexplored in models for ARD. \cite{zheng2006many} looked at posterior predictive checks to evaluate the performance of the model in accurately estimating the correct proportions of responses for $y_{ik}$ equal to each non-negative integer. They concluded their model underestimated the proportion of respondents who know exactly one person in each group of the McCarty dataset, and that data was ``heaped'' around nice numbers, like 10. We develop the use of posterior predictive checks and present here several more informative checks for ARD models and recommend their use in evaluating model fit for future studies.

\textit{Probability mass function}: First, we recommend replicating the probability mass function diagnostic from \cite{zheng2006many}. We especially recommend looking at $P(y_{ik} = 0)$ and $P(y_{ik} = 1)$, as these values don't typically suffer from rounding, but can inform the user about unexpected behavior in the data (e.g. fewer people know 0 doctors than expected from the number of doctors because only respondents who don't have a family doctor will report knowing 0 doctors).

\textit{Mean and standard deviation}: Second, we recommend plotting two forms of mean and standard deviation: (1) the mean of the responses against the standard deviation of the responses and (2) the mean of the positive responses against the standard deviation of the positive responses, grouped by each subpopulation (i.e. the columns of $Y$). The conditioned plots offer an additional comparisons of the summary statistics between empirical data and the estimates, which is especially needed when several subpopulations have more than 99\% of the responses equal to 0. In this way, we are studying two separate but important properties of the data. The conditionally positive mean $\mu^+$ and conditionally positive standard deviation $\sigma^+$ for subpopulation $k$ are calculated by
\begin{equation*}
    \mu^+ = \frac{1}{n^+} \sum_{i=1}^n y_{ik} I(y_{ik} > 0) \qquad \sigma^+ = \sqrt{\frac{\sum_{i=1}^n (y_{ik} - \mu^+)^2 I(y_{ik} > 0)}{n^+}},
\end{equation*}
where $n^+$ is the number of positive responses in group $k$, i.e. $n^+ = \sum_{i=1}^n I(y_{ik} > 0)$.

\textit{Correlation}: Posterior predictive checks can also be used to measure the ability to capture the correlation between a respondent's responses across subpopulations. For each posterior sample, the sample correlation is calculated, producing pairwise correlations between each subpopulation. The distribution of pairwise correlations across the posterior samples are then compared to the pairwise correlations of the original Ukraine data. For two given subpopulations $k_1$ and $k_2$, the pairwise correlation is calculated by
\begin{equation*}
    r = \frac{\sum_{i=1}^n (y_{ik_1} - \bar{y}_{ik_1})(y_{ik_2} - \bar{y}_{ik_2})}{\sqrt{\sum_{i=1}^n (y_{ik_1} - \bar{y}_{ik_1})^2 \sum_{i=1}^n (y_{ik_2} - \bar{y}_{ik_2})^2}}
\end{equation*}

\subsection{Leave-one-out}
A leave-one-out procedure can be implemented to evaluate the subpopulation size estimates. For models where the known subpopulation sizes are fixed in the estimation procedure (e.g. the Killworth MLE estimator), the known sizes are assumed to be unknown one at a time and estimated using the remaining known subpopulations. Thus, the model needs to be refit for each known subpopulation. For models in which all subpopulation sizes are estimands (e.g. the overdispersed model and our correlated model), the LOO procedure does not require re-estimating the parameters. Instead, since the subpopulation sizes are scaled using known sizes, the known size of the left out subpopulation is excluded when scaling the size estimates. The estimates are then compared to the truth, either visually or using some metric.

\subsection{Diagnostic Results}
\textit{Mean and standard deviation}:
Next, we consider the joint distribution of (positive) mean and positive standard deviation. The results are shown in Supplementary Figures \ref{fig:uncond_scatter_ppp} and \ref{fig:cond_scatter_ppp}. There is no clear lack of fit for either of the models, indicating reasonable modeling of the ARD. We slightly underestimate the mean of several of the groups, potentially indicating that our model is not properly capturing one of the tail behaviors. However, the underestimation is mild, and no attempts to better capture the mean proved successful.

\textit{Probability mass function}:
For the probability mass function, we consider 9 different probabilities in Supplementary Figure \ref{fig:pmf_ppp}. where the true proportions are plotted against the simulated proportions with 95\% credible intervals included. We denote the known subpopulations as circles and the unknown subpopulations as triangles. We also observe the heaping pointed out by \cite{zheng2006many}, especially for $P(Y = 2)$ and $P(Y = 10)$. Perhaps surprisingly, $P(Y=9)$ is estimated fairly accurately while $P(Y=11)$ is estimated poorly.

\textit{Correlation}:
While we do include correlated random effects in our model, we do not explicitly model the correlation of the responses. Thus, we still consider the posterior predictive p-values of the correlation of the responses, as shown in Supplementary Figure \ref{fig:corr_ppp} for both the in-sample and out-of-sample simulations. Under the true model, we typically expect the posterior predictive p-values to be concentrated near 0.5. In both cases, the posterior predictive p-values are reasonably distributed, although the values for the in-sample checks are much closer to 0.5, as expected.

\textit{Leave-one-out}:
Next, we show the accuracy and validity of our subpopulation size estimates using LOO. The estimates are shown in Supplementary Figure \ref{fig:ukr_LOO}. The subpopulations are ordered by largest to smallest, from largest on the left to smallest on the right. In general, subpopulations are estimated reasonably well, given the difficulty of estimating subpopulation sizes. It is important to note that the size estimates are much more accurate for our weighted scaling when there are corresponding populations with high correlation, like men aged 15-17 and female aged 15-17 when using all populations to scale. Because these subpopulations are more likely to have similar biases, our correlated scaling approach brings these estimates much closer to the truth.

Many authors have noted that respondents typically overestimate the number of people they know in large subpopulations and underestimate the number of people they know in small subpopulations \citep{zheng2006many, mccormick2007adjusting, maltiel2015estimating}. For the correlated model with the Ukraine dataset, this does not seem to be the case.

\textit{Surrogate residuals}:
Finally, we review some results from the surrogate residuals. While the surrogate residuals can be used in many ways, in this section we show their ability to detect missing covariates, as is common in ordinary linear regression. Figure \ref{fig:surrogates} shows one realization of the surrogate residuals corresponding to a single randomly chosen posterior sample against the standardized age covariate for the subpopulations people who died in 2007 and FSW. Since there is a vector of surrogate residuals for each posterior sample, to visualize the variability in the residuals across these samples, loess curves for 25 random samples were are added for reference. In general, there is very little variability across the posterior samples, except in areas with very few observations (respondents aged older than 80). For people who died in 2007, the uncorrelated basic model residuals are highly correlated with age, resulting in a linear loess curve. On the other hand, the residuals for the correlated model are mostly uncorrelated with age, indicating that the regression structure in the correlated model does accurately capture the relationship between age and the responses. There is no indication that higher order terms are needed except for age$^2$.

For FSW, we see that the relationship between the residuals and age is much weaker, and even the loess curve for the uncorrelated basic model is relatively flat with a slight downward trend. For the correlated model the loess curve is almost absolutely flat. In this case, age seems to play a larger role in the responses for people who died in 2007 than it does for FSW, which is expected.

\begin{figure}[!t]
    \centering
    \begin{subfigure}{0.7\textwidth}
        \includegraphics[width=\textwidth]{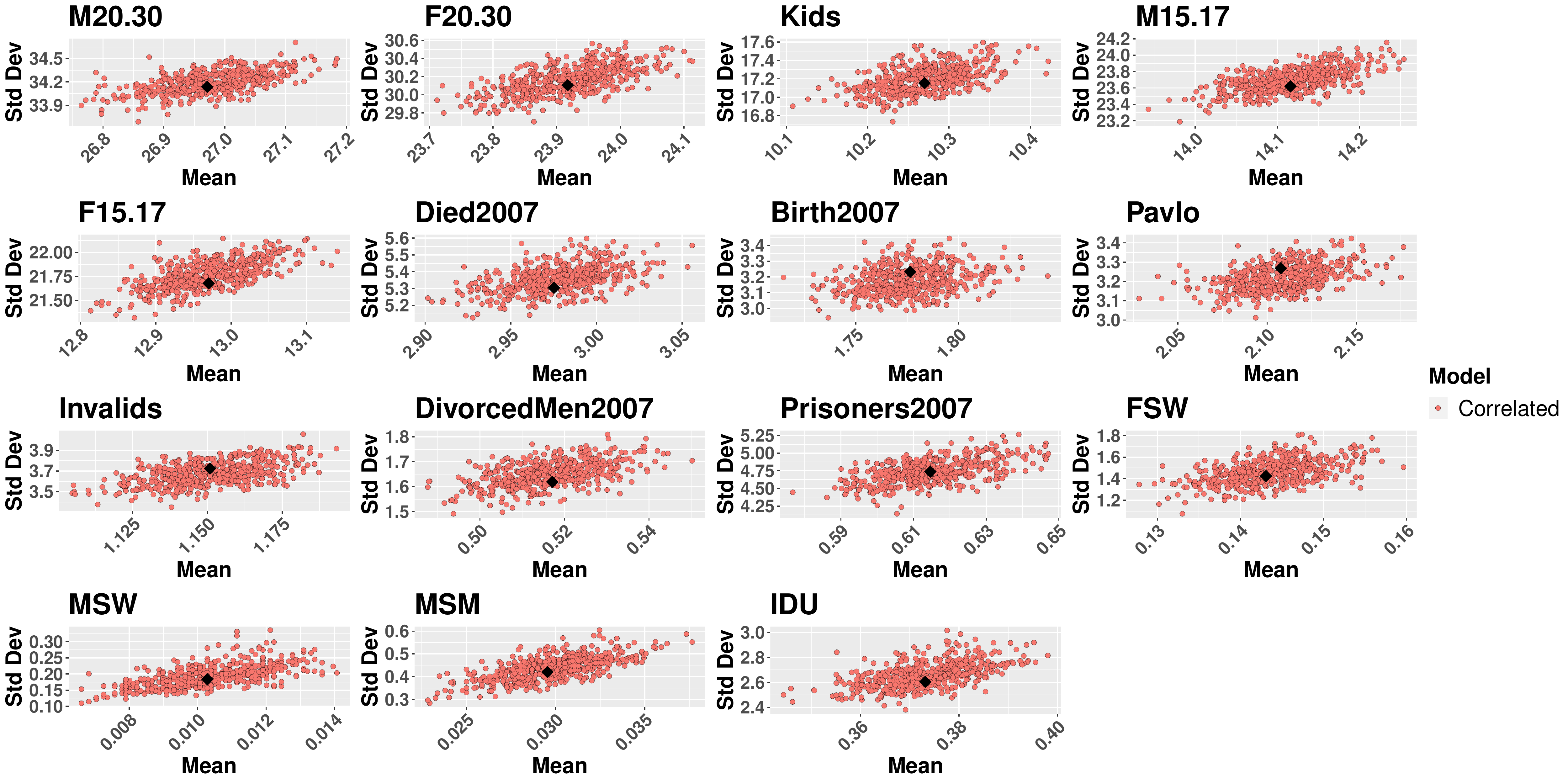}
        \caption{}
    \end{subfigure}
    \\
    \begin{subfigure}{0.7\textwidth}
        \includegraphics[width=\textwidth]{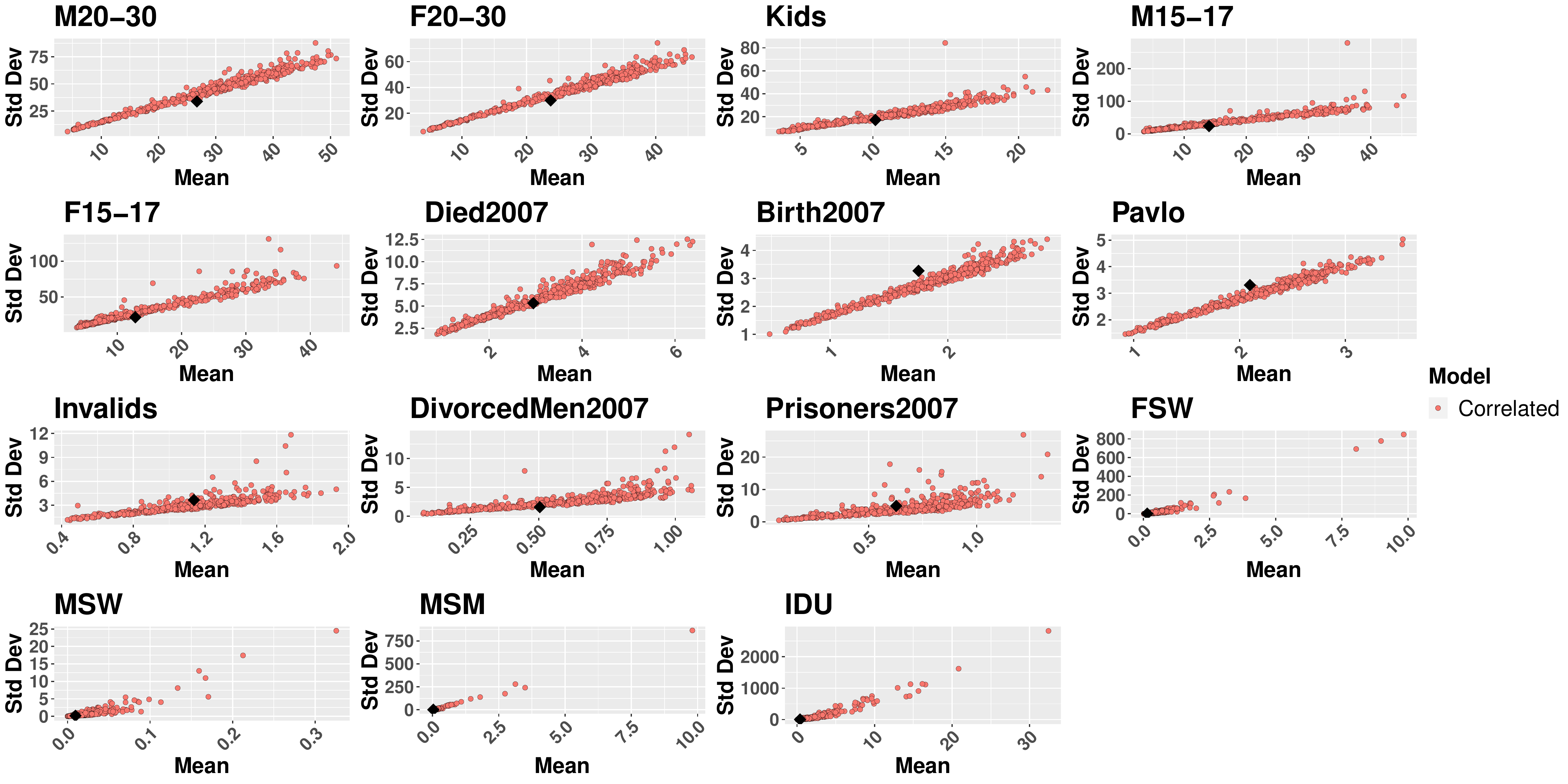}
        \caption{}
    \end{subfigure}
    \caption{Posterior predictive checks for the mean and standard deviation for in-sample simulations (a) and out-of-sample simulations (b). The mean and standard deviation of posterior samples is shown across subpopulation for our correlated model (pink) and for the observed data (black square). For visualizations, only 500 posterior samples are shown for each model.}
    \label{fig:uncond_scatter_ppp}
\end{figure}

\begin{figure}[!t]
    \centering
    \begin{subfigure}{0.7\textwidth}
        \includegraphics[width=\textwidth]{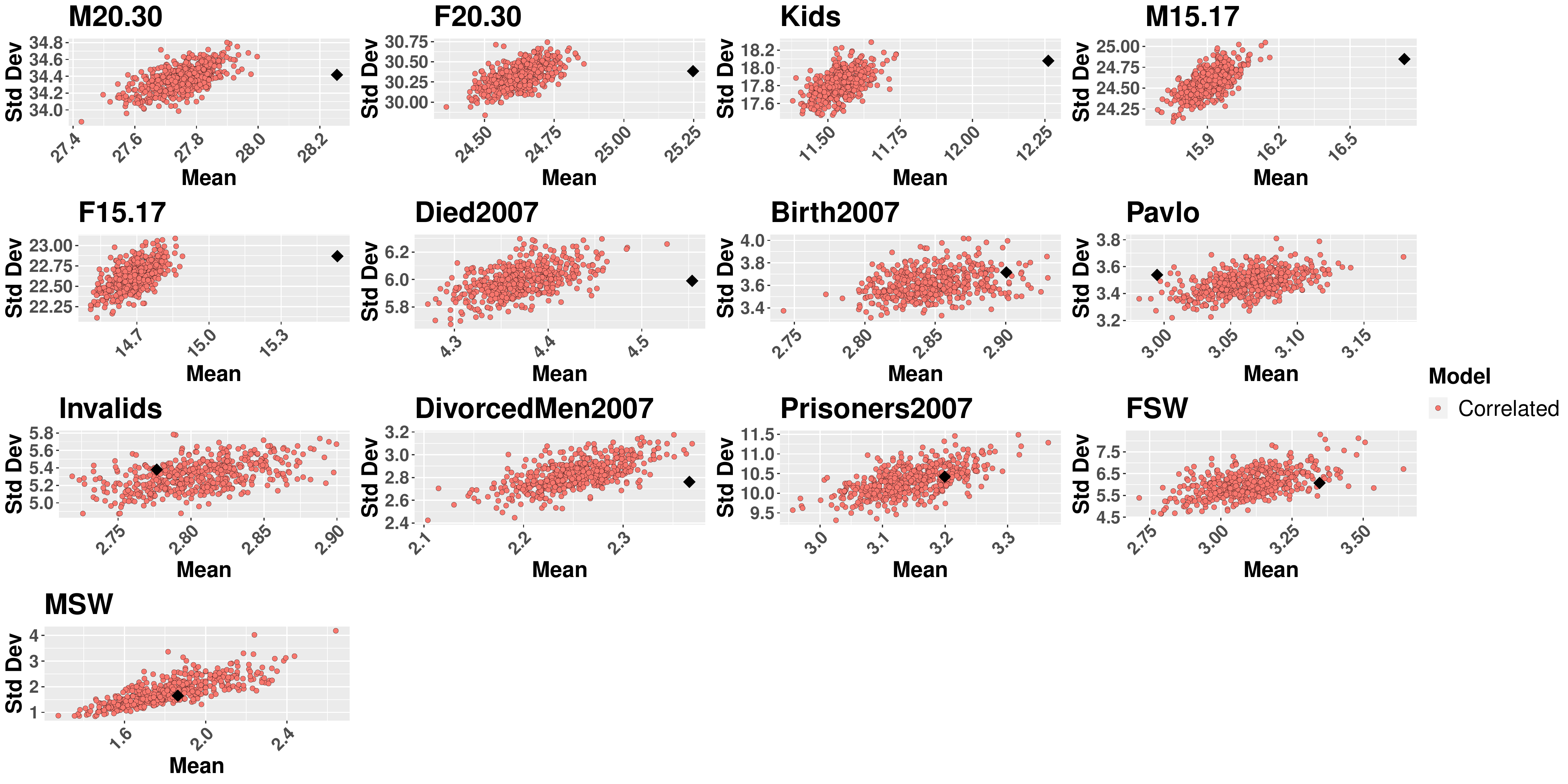}
        \caption{}
    \end{subfigure}
    \\
    \begin{subfigure}{0.7\textwidth}
        \includegraphics[width=\textwidth]{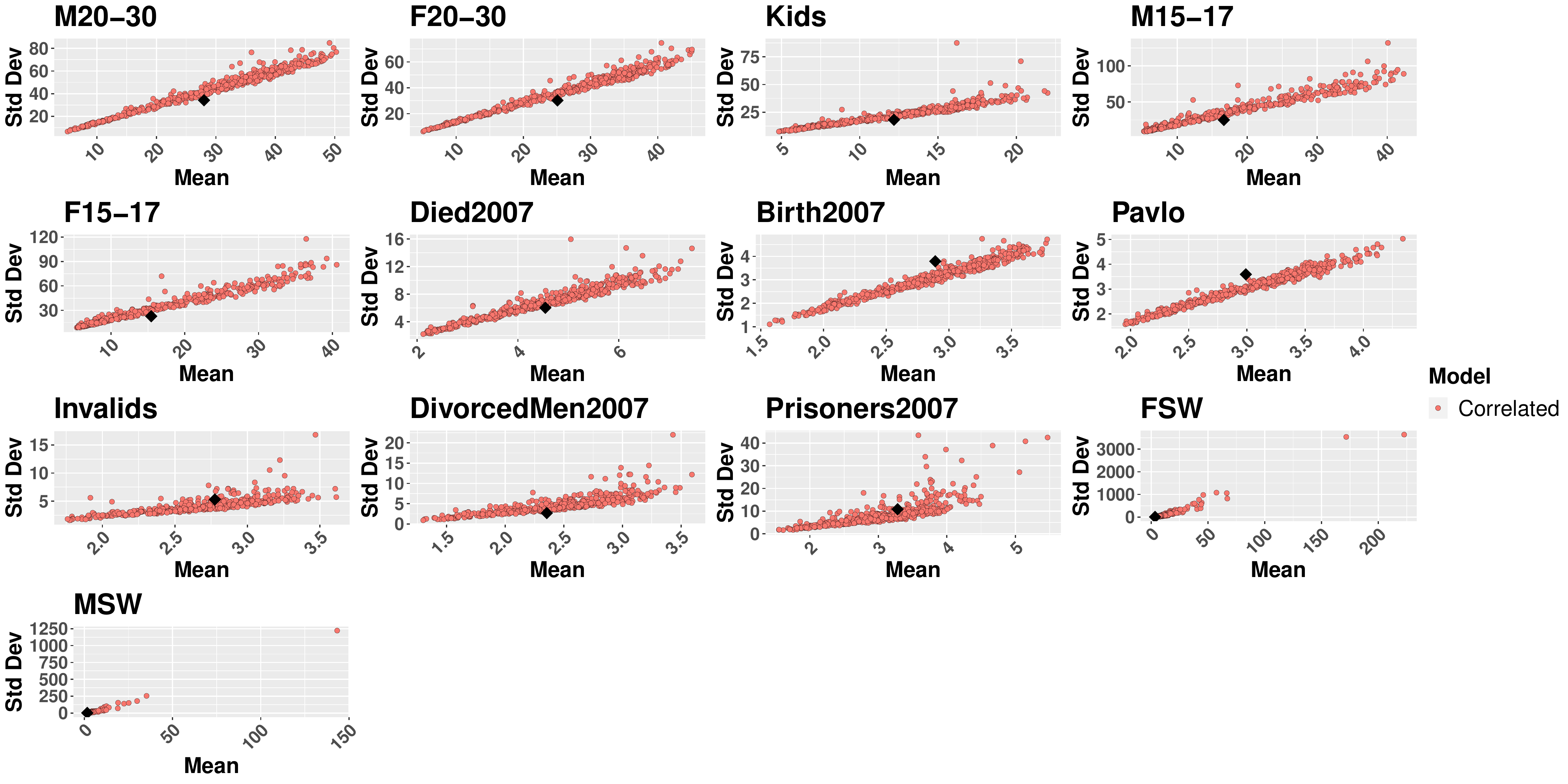}
        \caption{}
    \end{subfigure}
    \caption{Posterior predictive checks for conditionally-positive mean and standard deviation for in-sample simulations (a) and out-of-sample simulations (b). The mean and standard deviation of posterior samples is shown across subpopulation for our correlated model (pink) and for the observed data (black square). For visualizations, only 500 posterior samples are shown for each model.}
    \label{fig:cond_scatter_ppp}
\end{figure}

\begin{figure}[!t]
    \centering
    \begin{subfigure}{0.7\textwidth}
        \includegraphics[width=\textwidth]{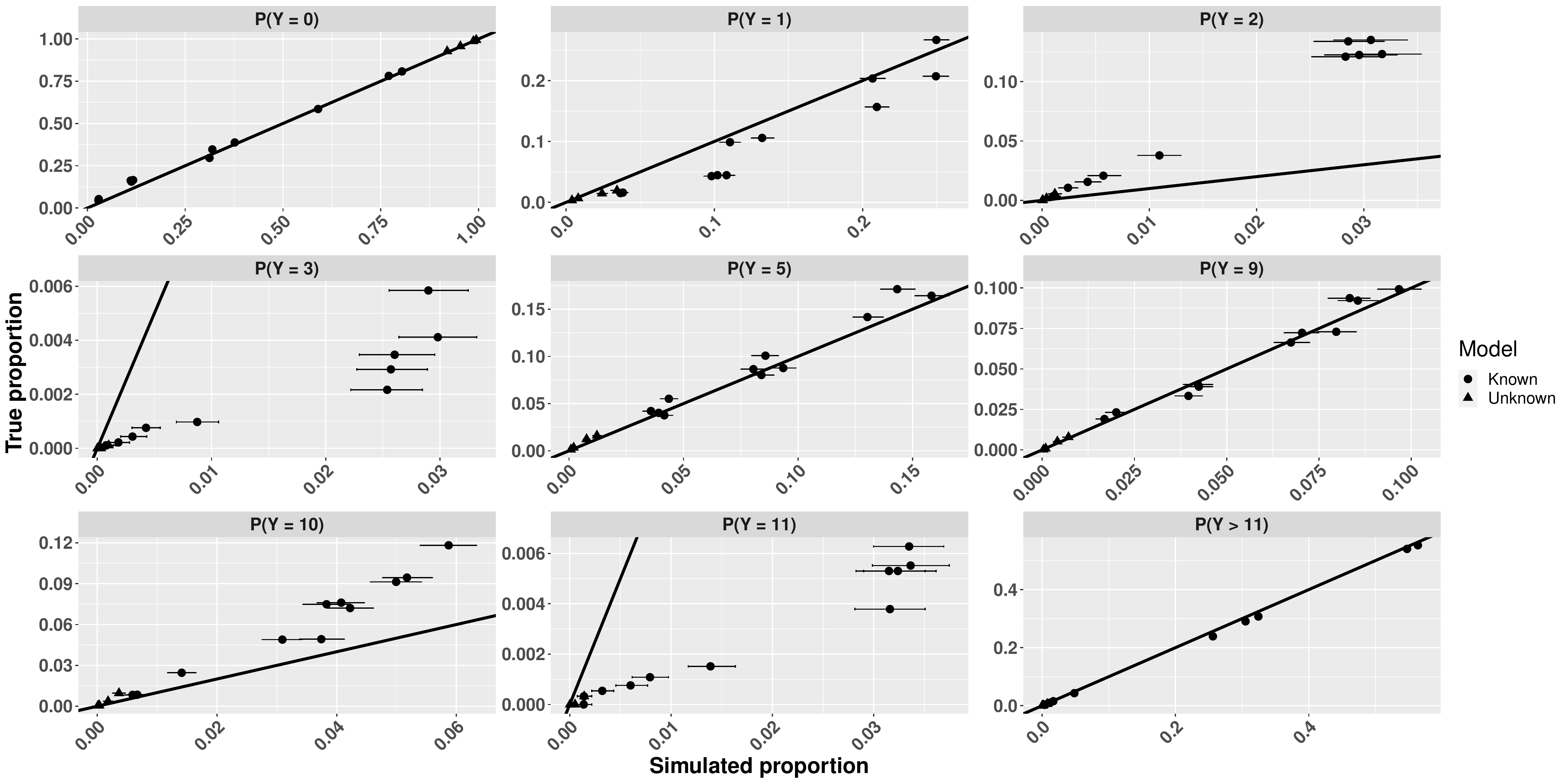}
        \caption{}
    \end{subfigure}
    \\
    \begin{subfigure}{0.7\textwidth}
        \includegraphics[width=\textwidth]{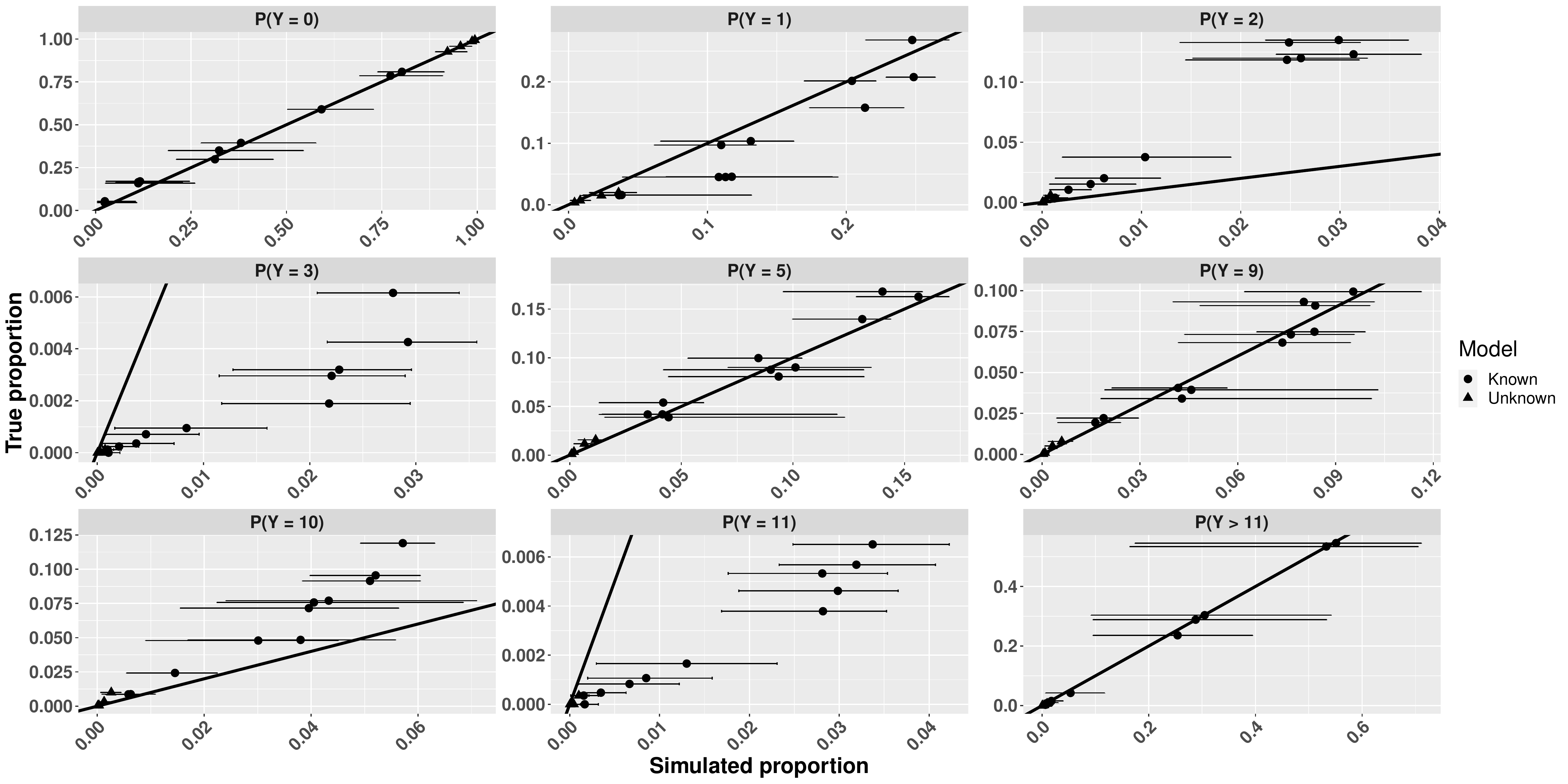}
        \caption{}
    \end{subfigure}
    \caption{Posterior predictive checks for the probability mass function and associated 95\% posterior credible intervals for the in-sample posterior simulations (a) and the out-of-sample simulations (b). The x-axis and y-axis represents the proportion of simulated responses and observed responses that are equal to the integer values considered, respectively. The mean and 95\% credible interval of the proportions over the posterior samples is shown using the shape and error bars. The diagonal line represents true proportion equal to simulated proportion.}
    \label{fig:pmf_ppp}
\end{figure}

\begin{figure}[!t]
    \centering
    \begin{subfigure}{0.7\textwidth}
        \includegraphics[width=\textwidth]{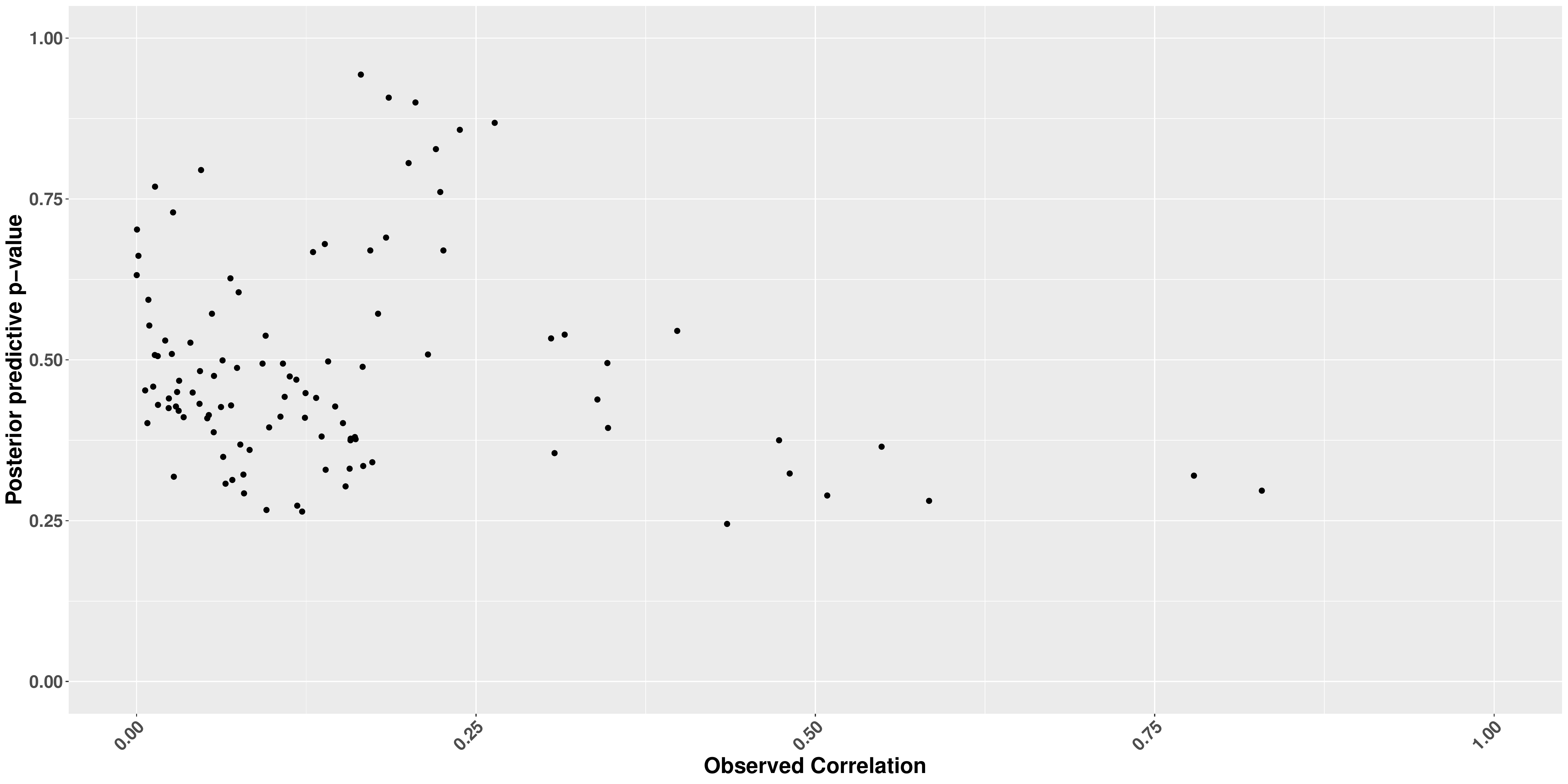}
        \caption{}
    \end{subfigure}
    \\
    \begin{subfigure}{0.7\textwidth}
        \includegraphics[width=\textwidth]{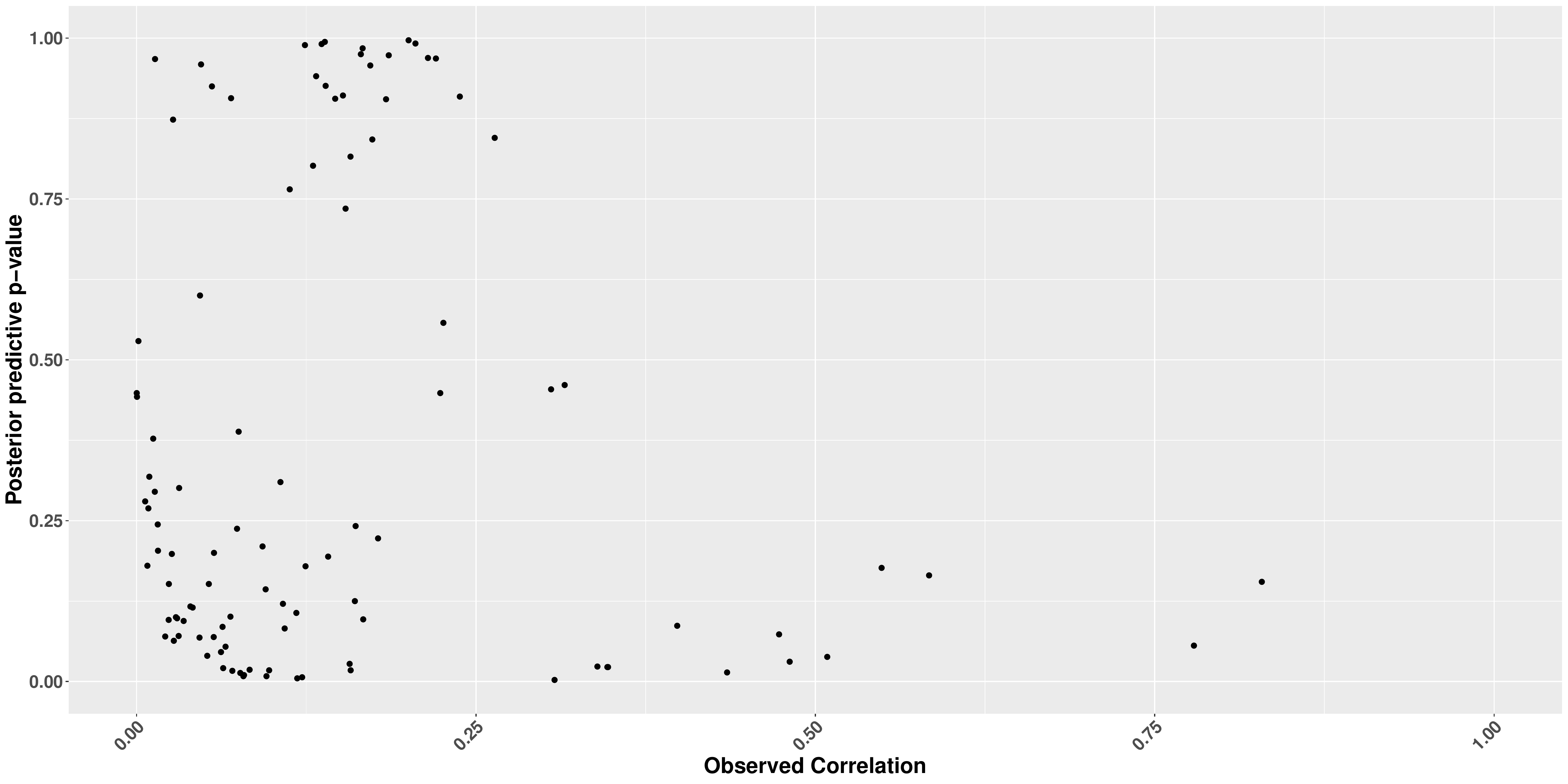}
        \caption{}
    \end{subfigure}
    \caption{Posterior predictive checks for subpopulation correlations for the in-sample simulations (a) and the out-of-sample simulations (b). The posterior predictive p-values are plotted against the observed correlations of the ARD.}
    \label{fig:corr_ppp_vs_observed}
\end{figure}

\begin{figure}[!t]
    \centering
    \begin{subfigure}{0.7\textwidth}
        \includegraphics[width=\textwidth]{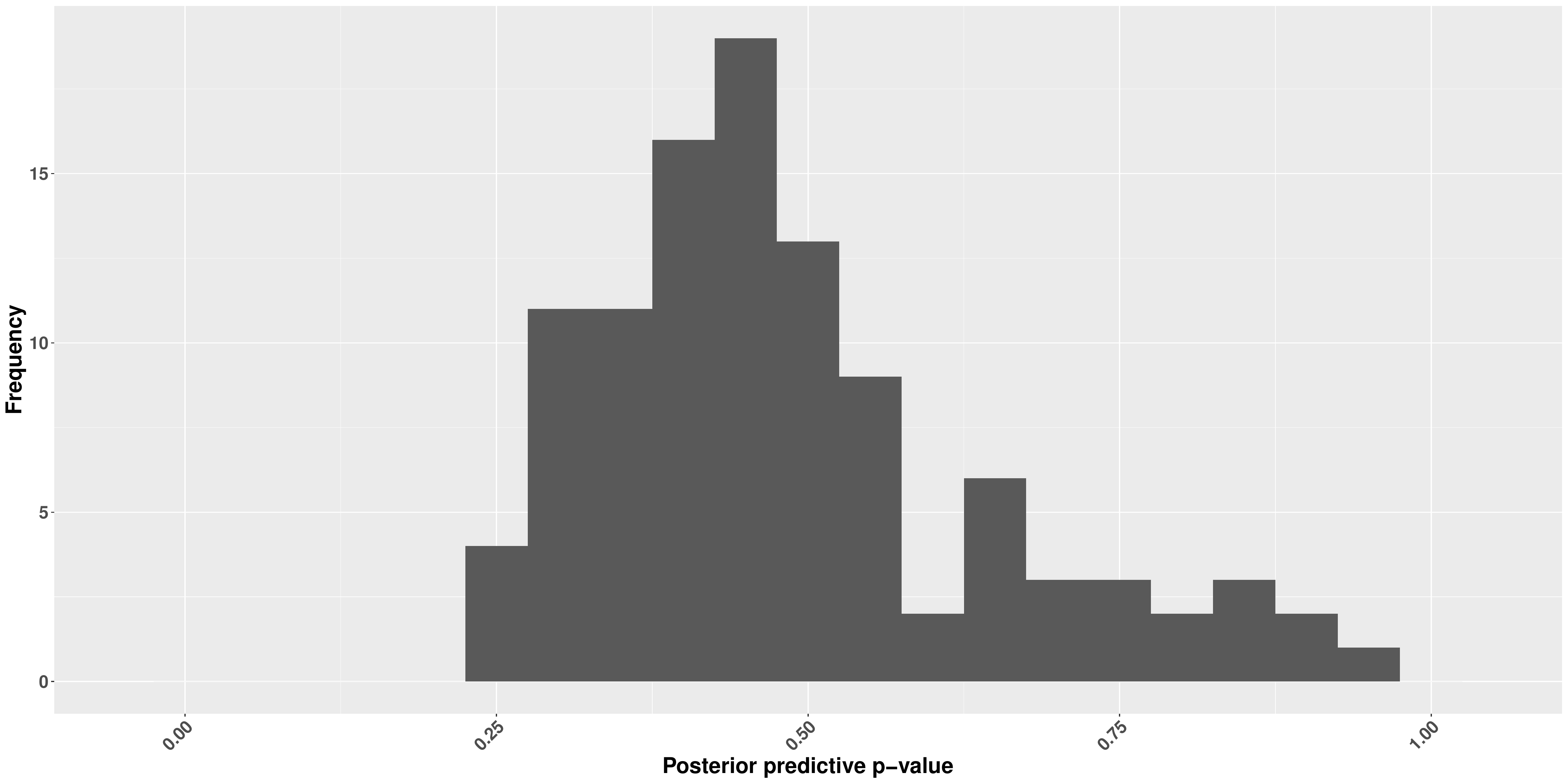}
        \caption{}
    \end{subfigure}
    \\
    \begin{subfigure}{0.7\textwidth}
        \includegraphics[width=\textwidth]{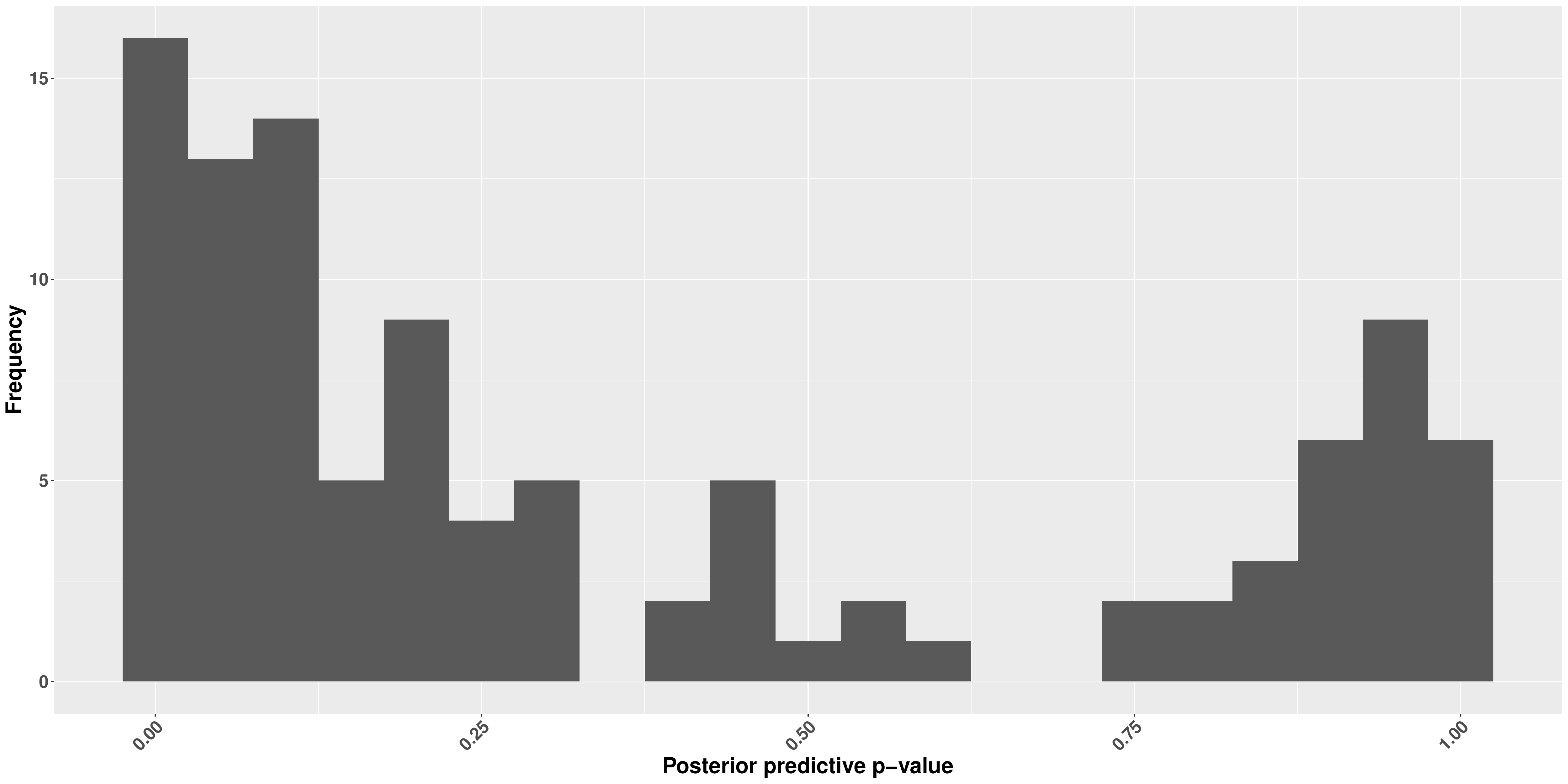}
        \caption{}
    \end{subfigure}
    \caption{Posterior predictive checks for subpopulation correlations for the in-sample simulations (a) and the out-of-sample simulations (b).}
    \label{fig:corr_ppp}
\end{figure}

\begin{figure}[!t]
    \centerline{\includegraphics[width=\textwidth]{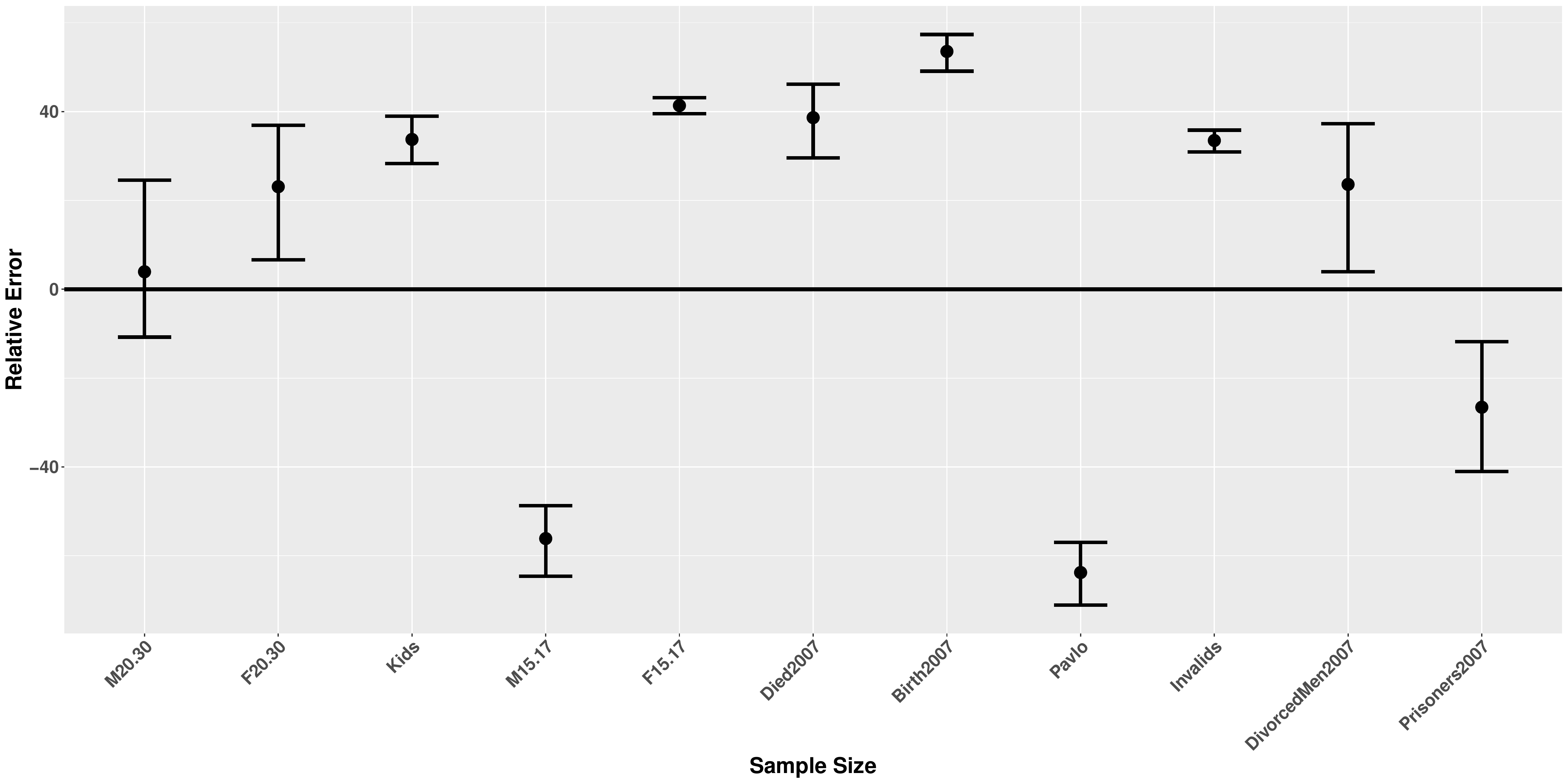}}
    \caption{Subpopulation leave-one-out estimates. The subpopulations are ordered from largest to smallest size, and the relative error represents $100 * (Truth - Predicted) / Truth$. The estimates scaled using all subpopulations is shown with a dotted line and our correlated scaling results with a solid line.}
    \label{fig:ukr_LOO}
\end{figure}

\begin{figure}[!t]
    \centering
    \begin{subfigure}{0.35\textwidth}
        \includegraphics[width=\textwidth]{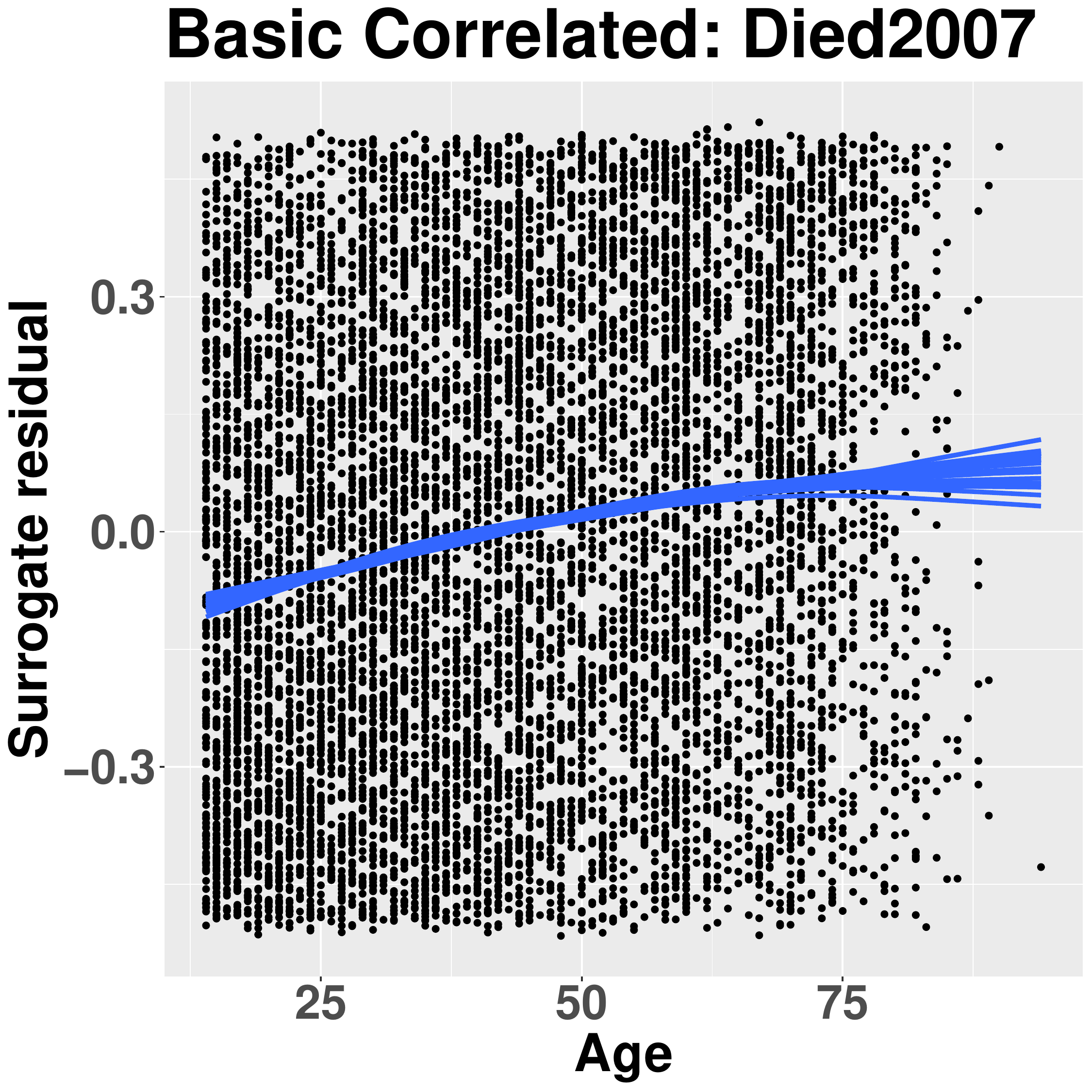}
        \caption{}
    \end{subfigure}
    \begin{subfigure}{0.35\textwidth}
        \includegraphics[width=\textwidth]{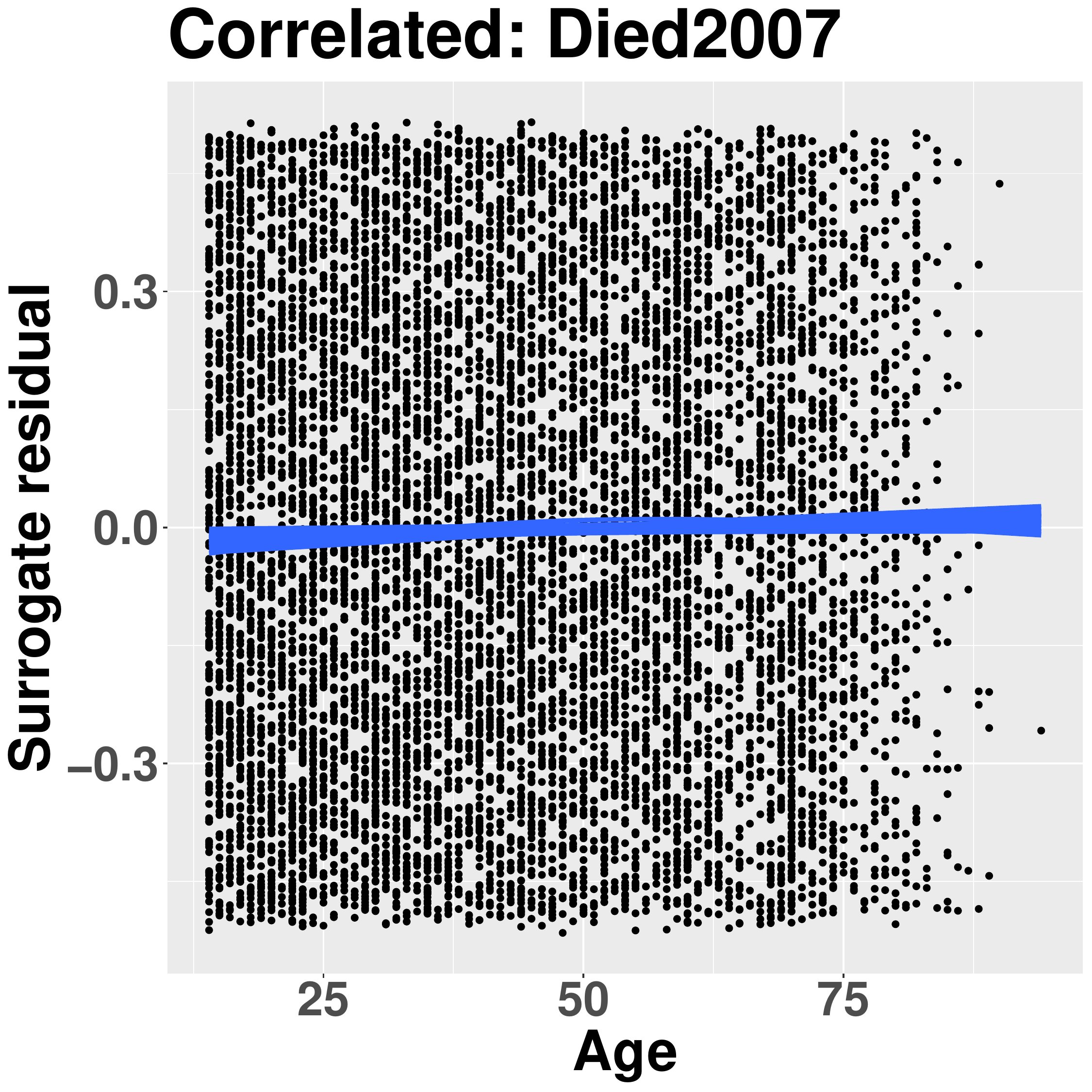}
        \caption{}
    \end{subfigure}
    \\
    \begin{subfigure}{0.35\textwidth}
        \includegraphics[width=\textwidth]{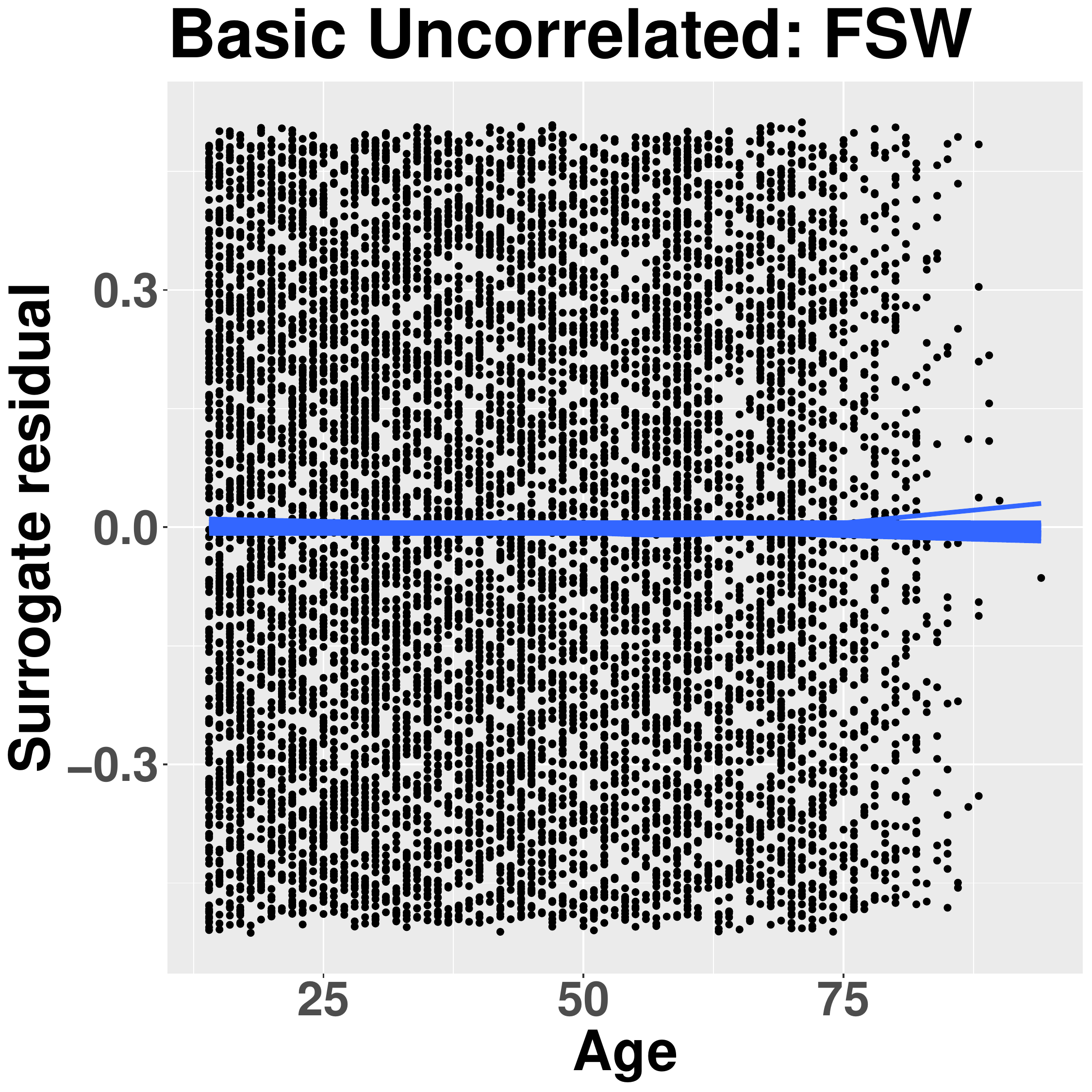}
        \caption{}
    \end{subfigure}
    \begin{subfigure}{0.35\textwidth}
        \includegraphics[width=\textwidth]{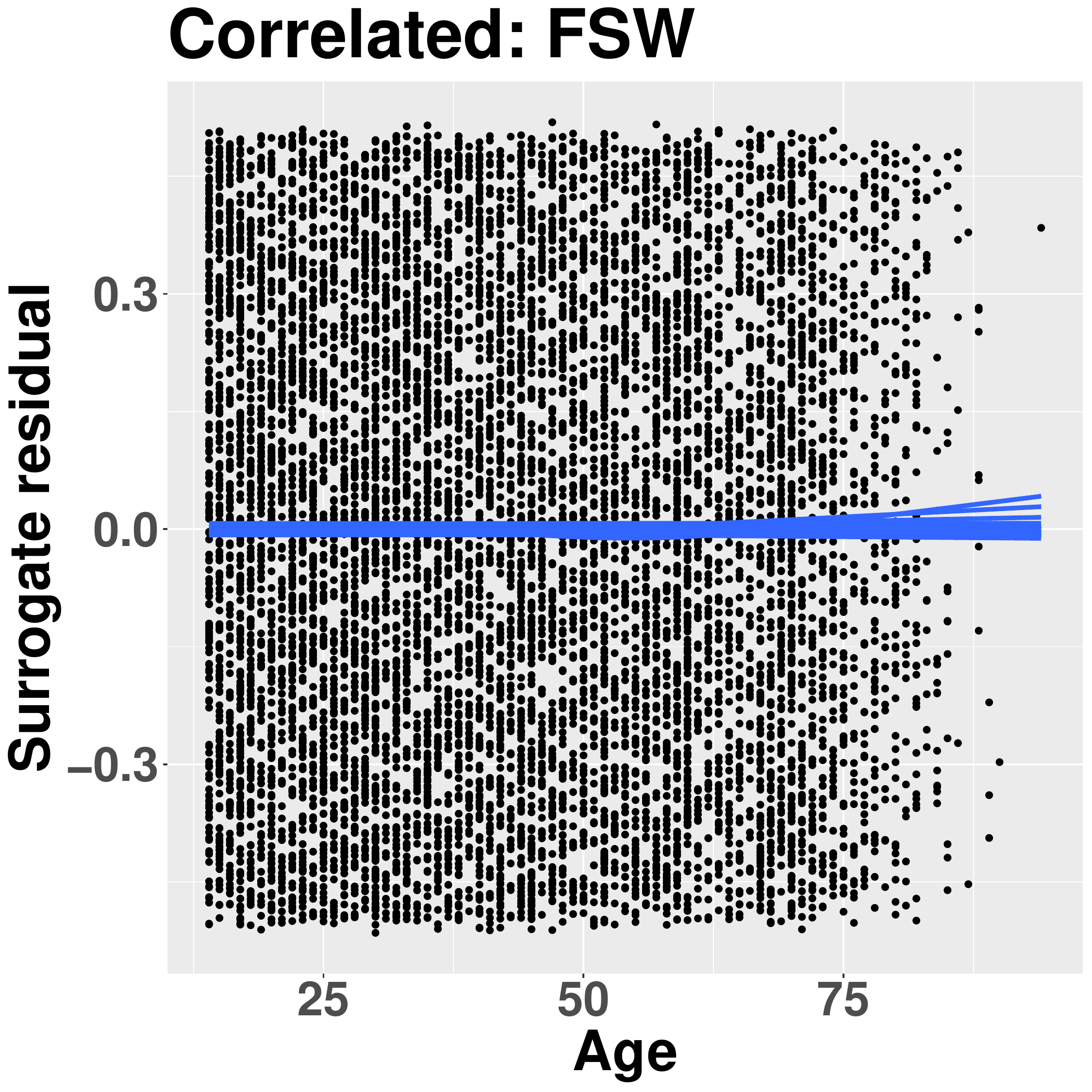}
        \caption{}
    \end{subfigure}
    \caption{Surrogate residuals from the uncorrelated basic model for subpopulations people who died in 2007 (a) and FSW (c) and from the correlated model for subpopulations people who died in 2007 (b) and FSW (d). Loess smoothing curves are plotted over residuals for reference.}
    \label{fig:surrogates}
\end{figure}

\FloatBarrier
\newpage
\section{Sensitivity Analysis}
In this section, we perform sensitivity analysis to assess to robustness of our model to different choices of priors. Specifically, keep all other priors the same, and change the following five priors, one at a time:\\
\textbf{Flat $\rho_k$}: To assess how influential the shared Normal prior is on the prevalence parameters, we instead assume $\rho_k \propto 1$.\\
\textbf{Cauchy slope parameters}: We assume an even more diffuse prior for all slope parameters simultaneously, so assess whether our original priors shrink the slope parameters closer to 0. We assuming the following Cauchy priors:
\begin{align*}
    \alpha_k &\sim Cauchy(0, 2.5) \\
    \beta^{global}_{j} &\sim Cauchy(0, 2.5) \\
    \beta^{group}_{k,j} &\sim Cauchy(0, 2.5)
\end{align*}
\textbf{Correlation parameters}: We also examine the effect that our hyper-parameters have on the correlation estimates by testing both large and small values of the $LkjCholesky$ prior. Larger hyper-parameters correspond to smaller correlations, while smaller hyper-parameters correspond to larger correlations. For reference, a hyper-parameter of 1 corresponds to the uniform density over all correlation matrices.
\begin{equation*}
    \Omega^{1/2} \sim LkjCholesky(0.1)
\end{equation*}
\begin{equation*}
    \Omega^{1/2} \sim LkjCholesky(10)
\end{equation*}
\textbf{Standard deviation parameter}: Finally, we consider the sensitivity of our parameter estimates to the standard deviation of the random effects. There is a long history about what priors to use for variance and standard deviations. In the main manuscript, we chose a diffuse half-Cauchy prior, as suggested by \cite{gelman2006prior}. We also consider a fairly strong Gamma prior which will shrink the $\tau_{N,K}$ parameters towards 0, favoring a smaller standard deviation. This is an extreme alternative to the prior used in the main manuscript.
\begin{equation*}
    \tau_{N,k} \sim Gamma(0.01, 0.01)
\end{equation*}

\subsection{Results}
\begin{figure}[!t]
    \centerline{\includegraphics[width=0.8\textwidth]{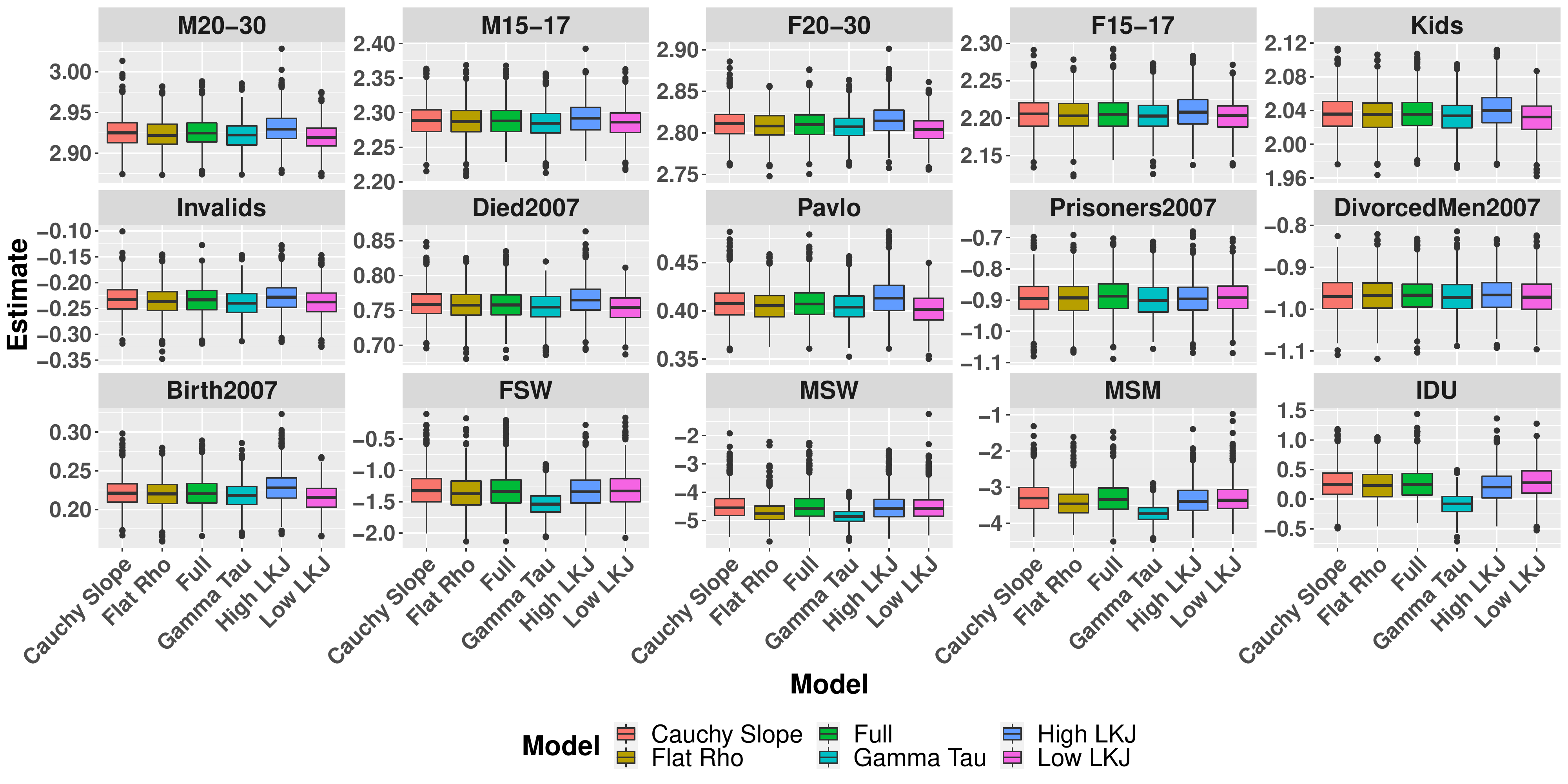}}
    \caption{Estimates of $\rho$ under different priors.}
    \label{fig:sens_rho}
\end{figure}

\begin{figure}[!t]
    \centerline{\includegraphics[width=0.8\textwidth]{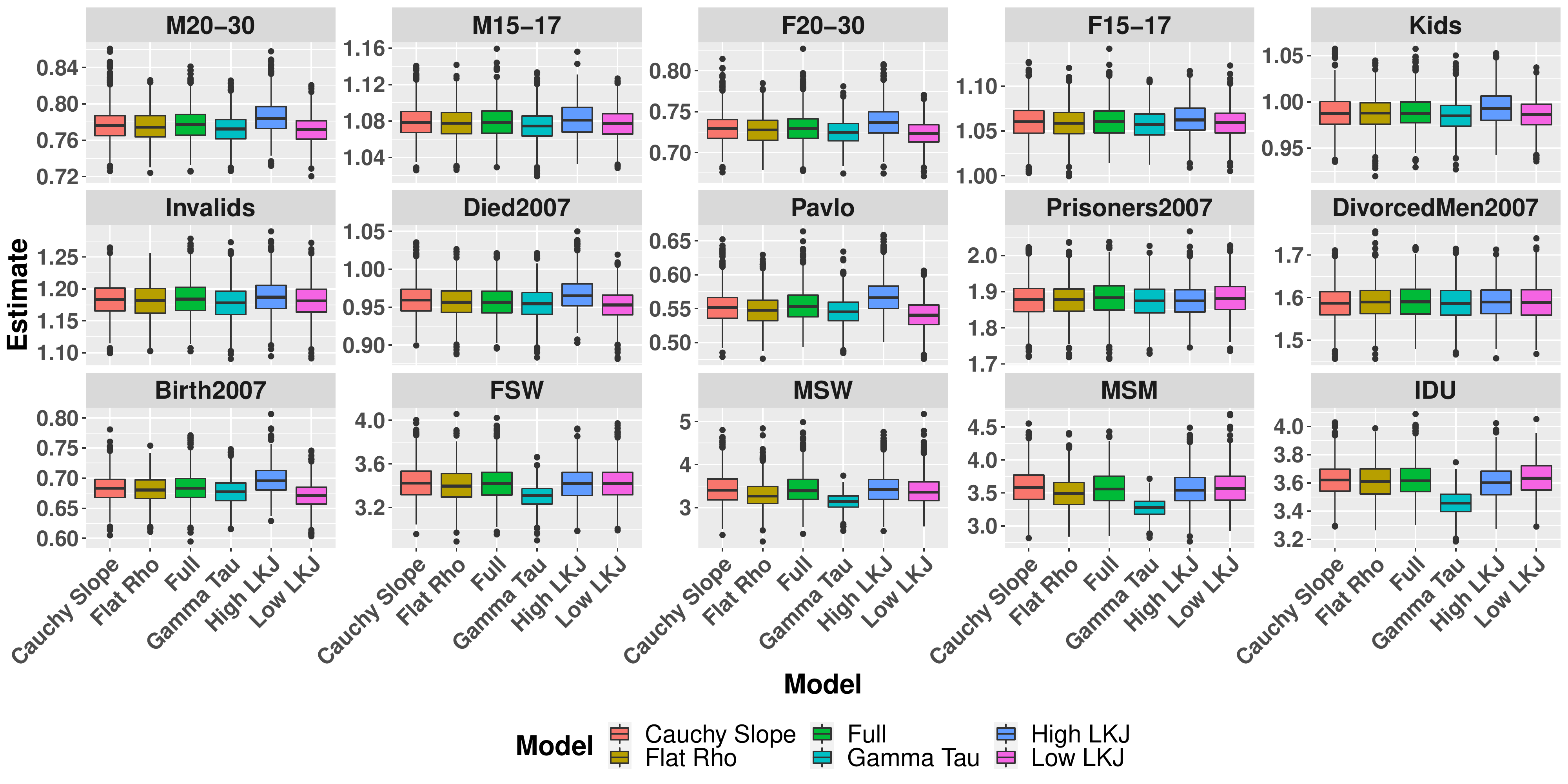}}
    \caption{Estimates of $\tau_k$ under different priors.}
    \label{fig:sens_tau}
\end{figure}

\begin{figure}[!t]
    \centerline{\includegraphics[width=0.8\textwidth]{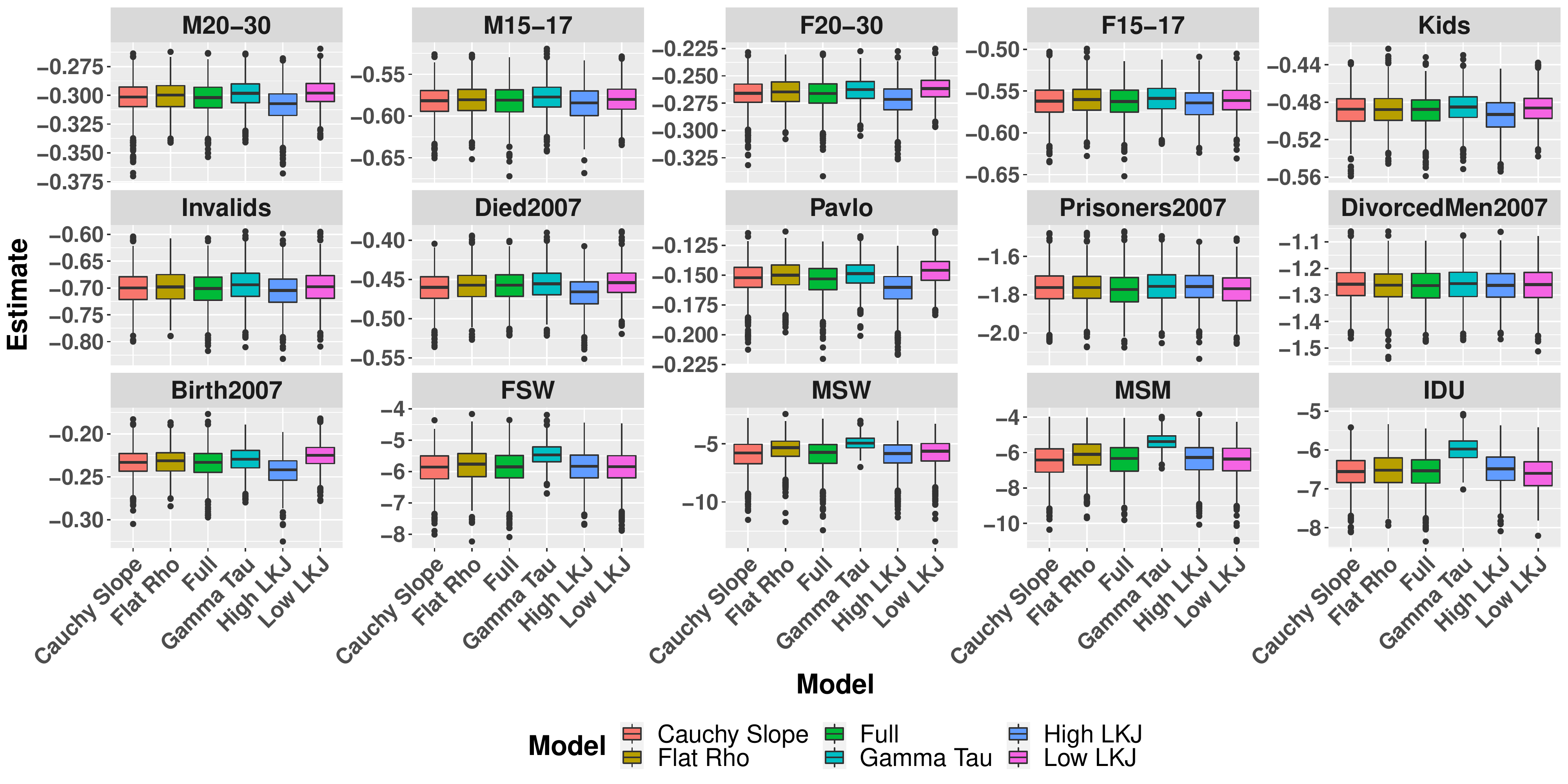}}
    \caption{Estimates of $\mu_k$ under different priors.}
    \label{fig:sens_mu}
\end{figure}

\begin{figure}[!t]
    \centerline{\includegraphics[width=0.8\textwidth]{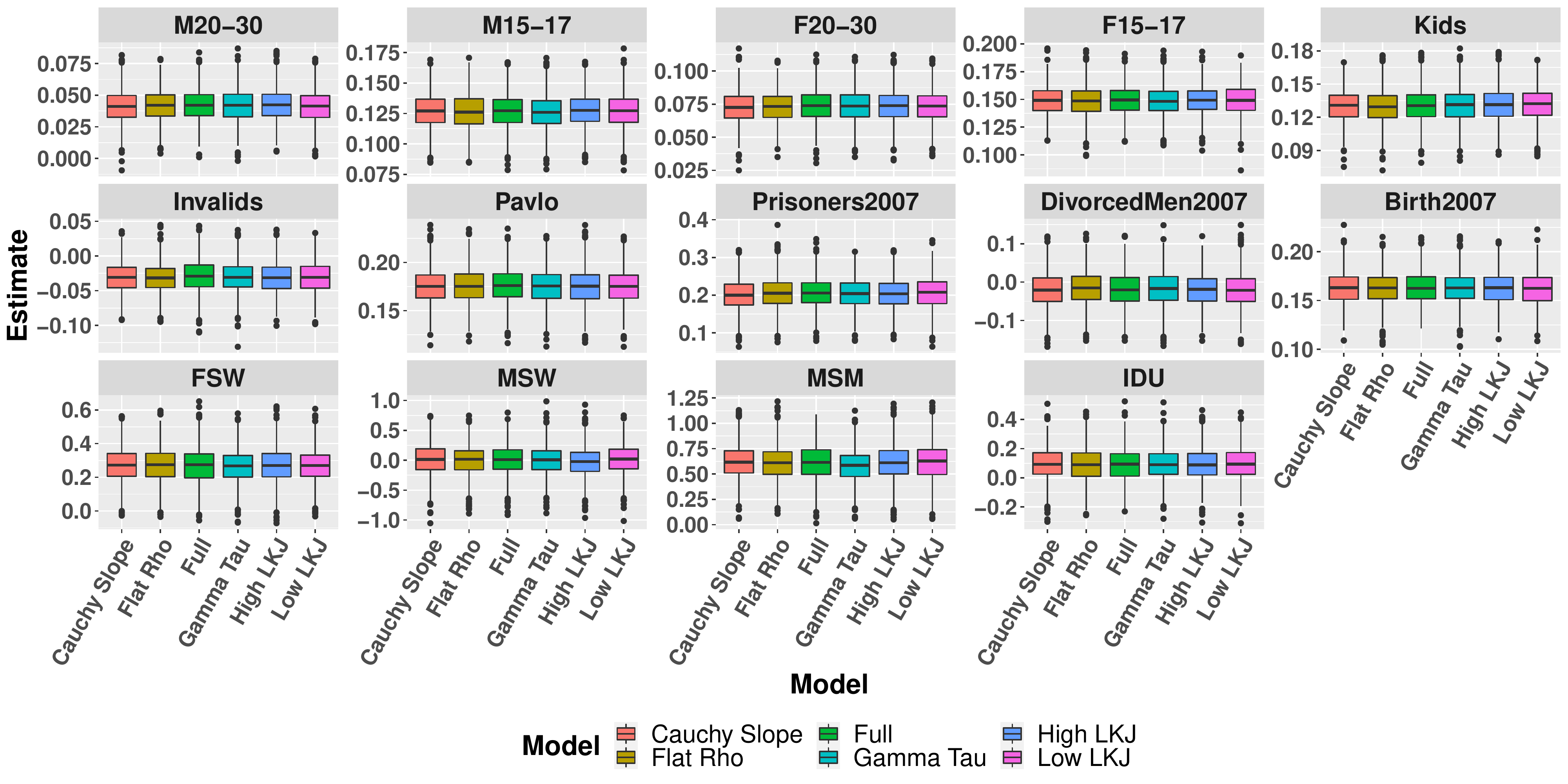}}
    \caption{Estimates of $\bm{\alpha}$ under different priors.}
    \label{fig:sens_alpha}
\end{figure}

\begin{figure}[!t]
    \centerline{\includegraphics[width=0.8\textwidth]{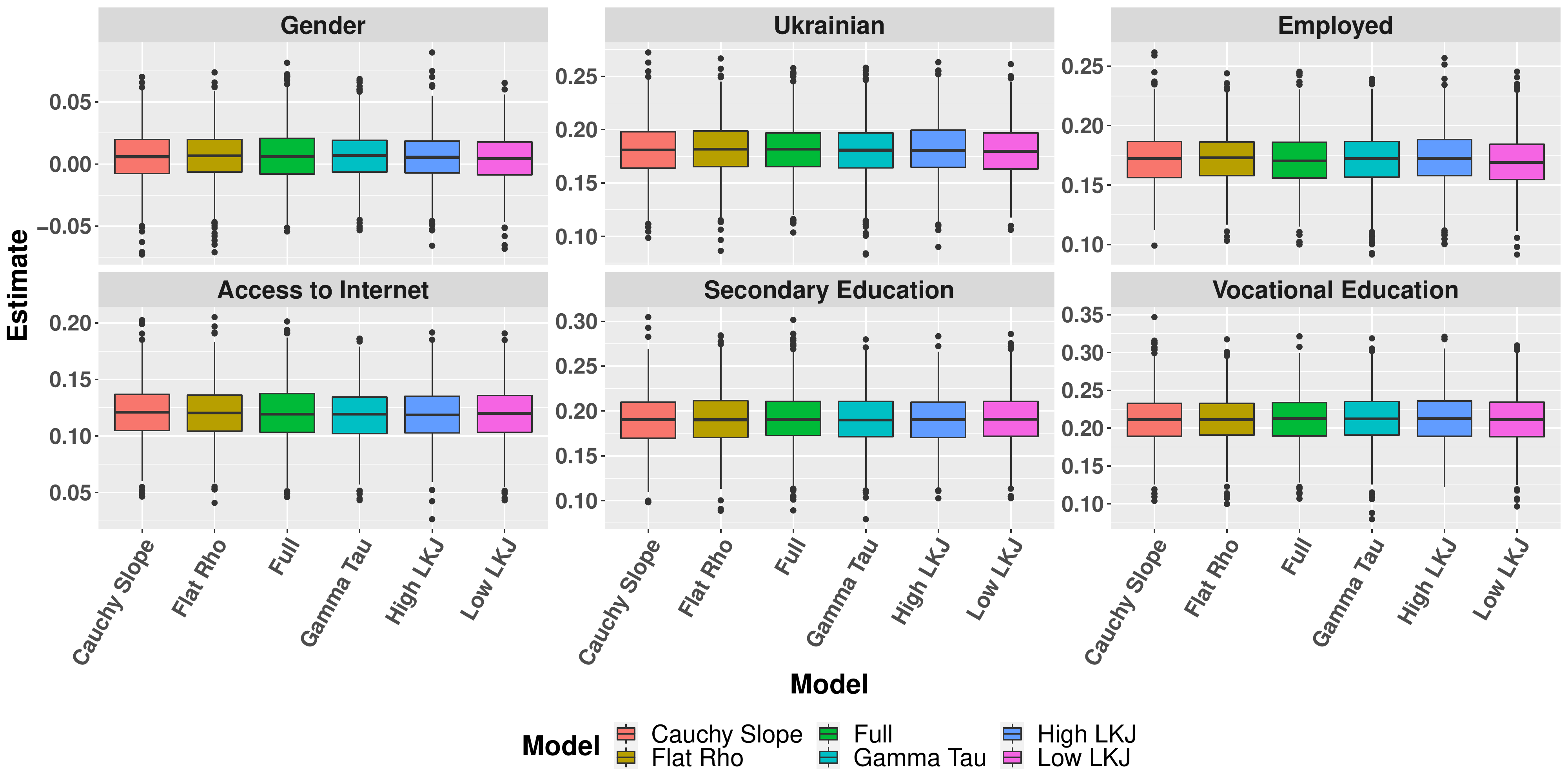}}
    \caption{Estimates of $\bm{\beta}^{global}$ under different priors.}
    \label{fig:sens_global}
\end{figure}

\begin{figure}[!t]
    \centerline{\includegraphics[width=0.8\textwidth]{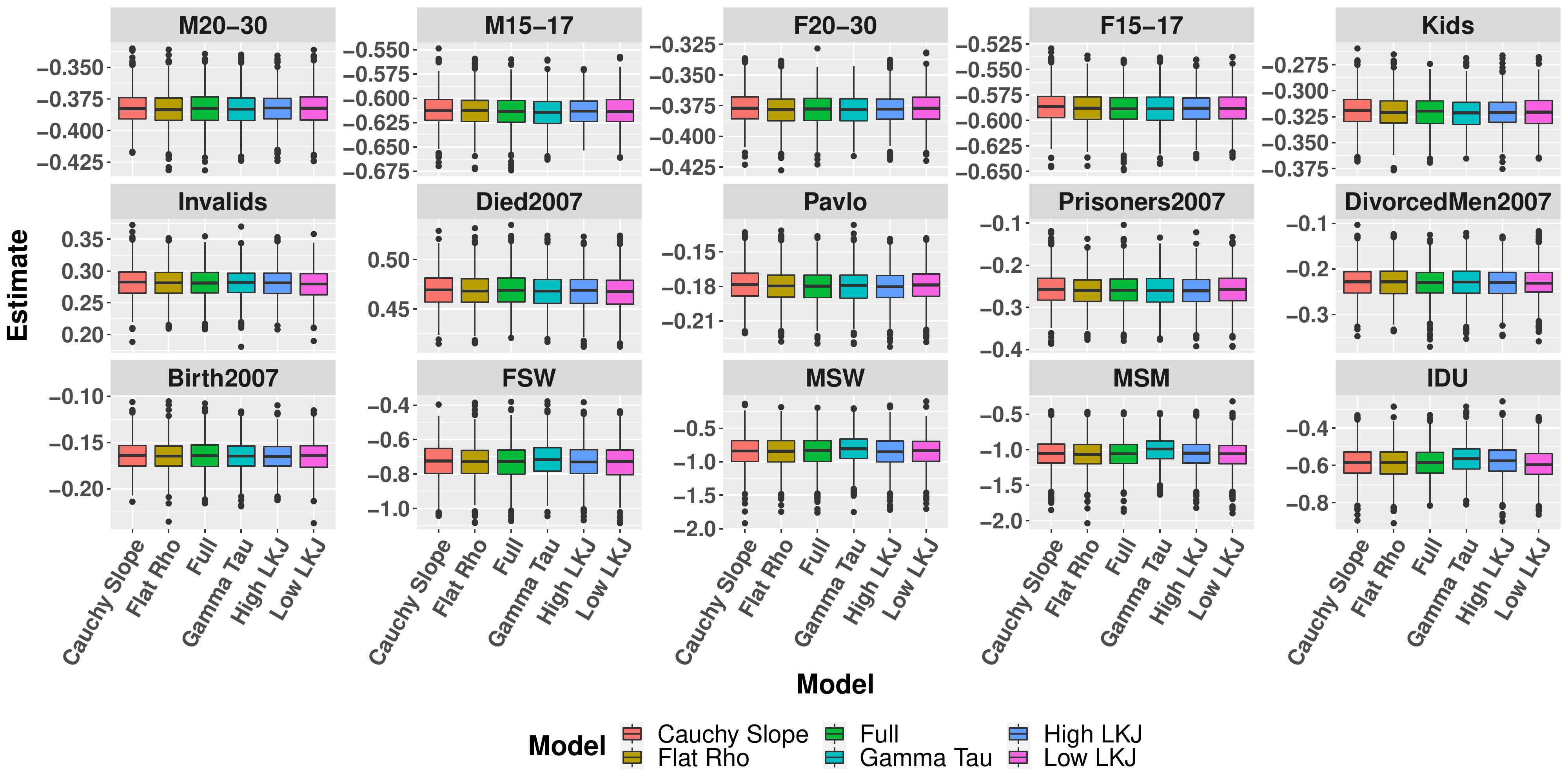}}
    \caption{Estimates of $\bm{\beta}_{k}^{group}$ for age under different priors.}
    \label{fig:sens_age}
\end{figure}

\begin{figure}[!t]
    \centerline{\includegraphics[width=0.8\textwidth]{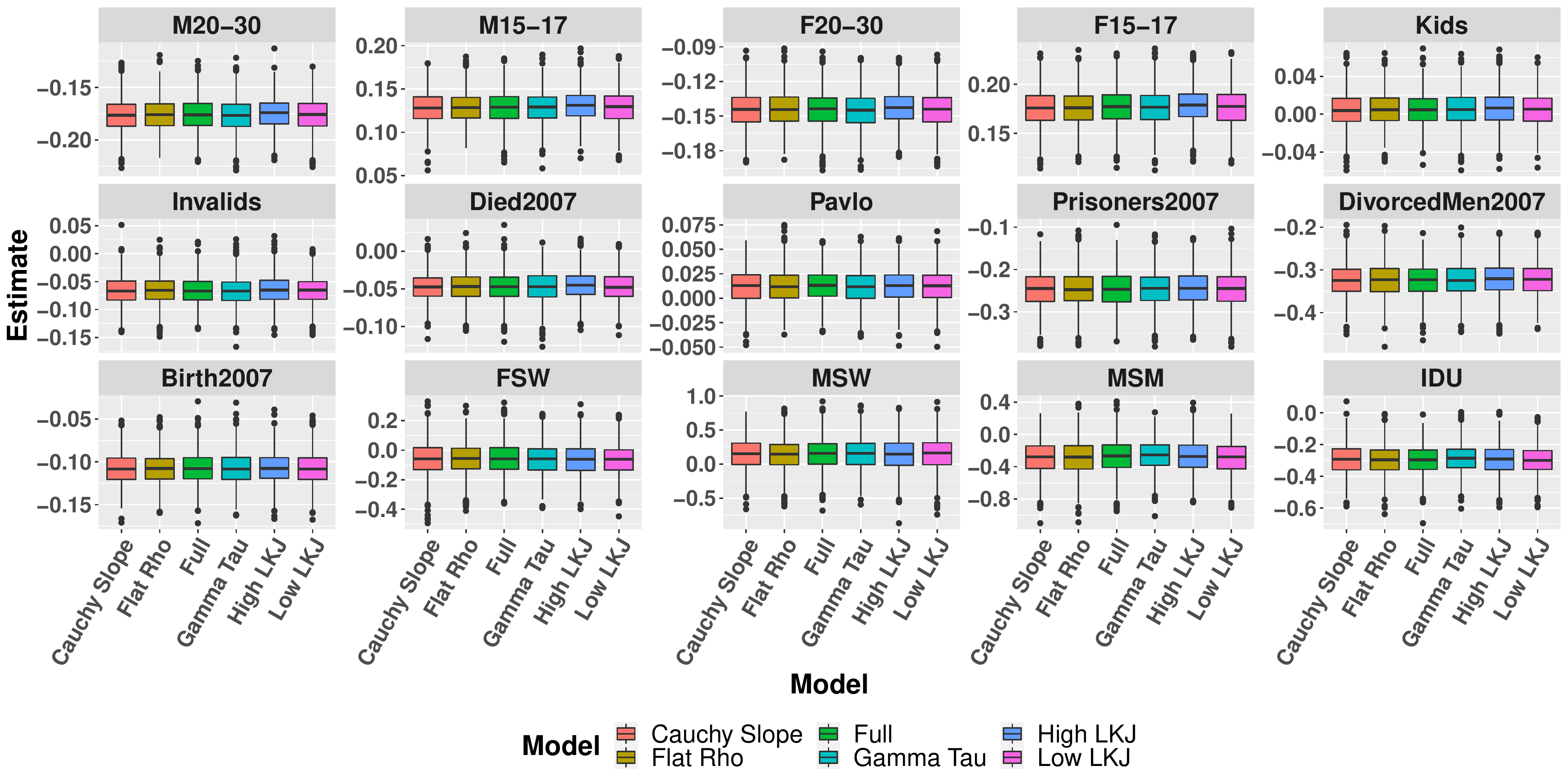}}
    \caption{Estimates of $\bm{\beta}_{k}^{group}$ for age$^2$ under different priors.}
    \label{fig:sens_age2}
\end{figure}

\begin{figure}[!t]
    \centerline{\includegraphics[width=0.8\textwidth]{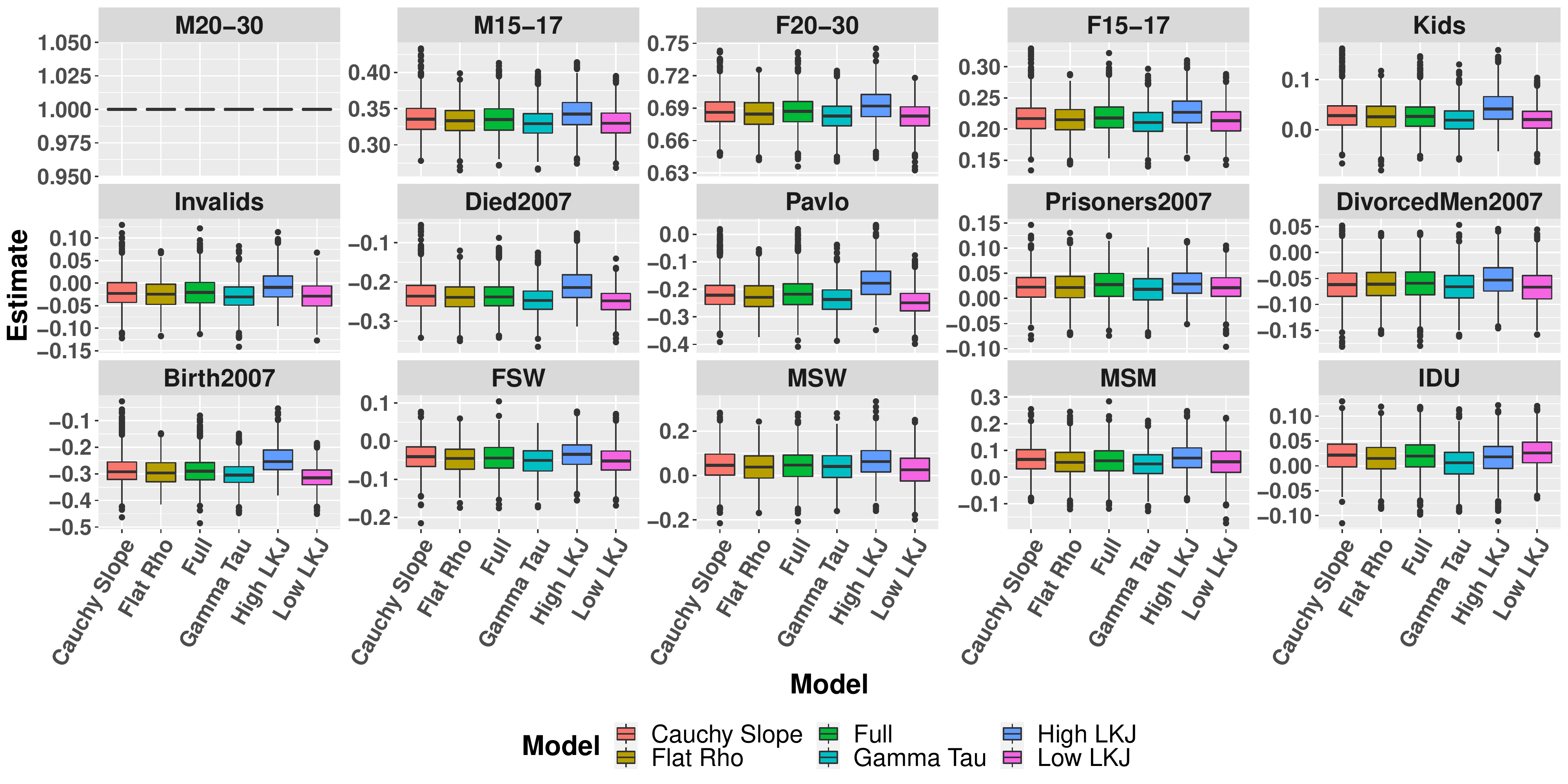}}
    \caption{Estimates of correlations with men 20-30 under different priors.}
    \label{fig:sens_m2030}
\end{figure}

\begin{figure}[!t]
    \centerline{\includegraphics[width=0.8\textwidth]{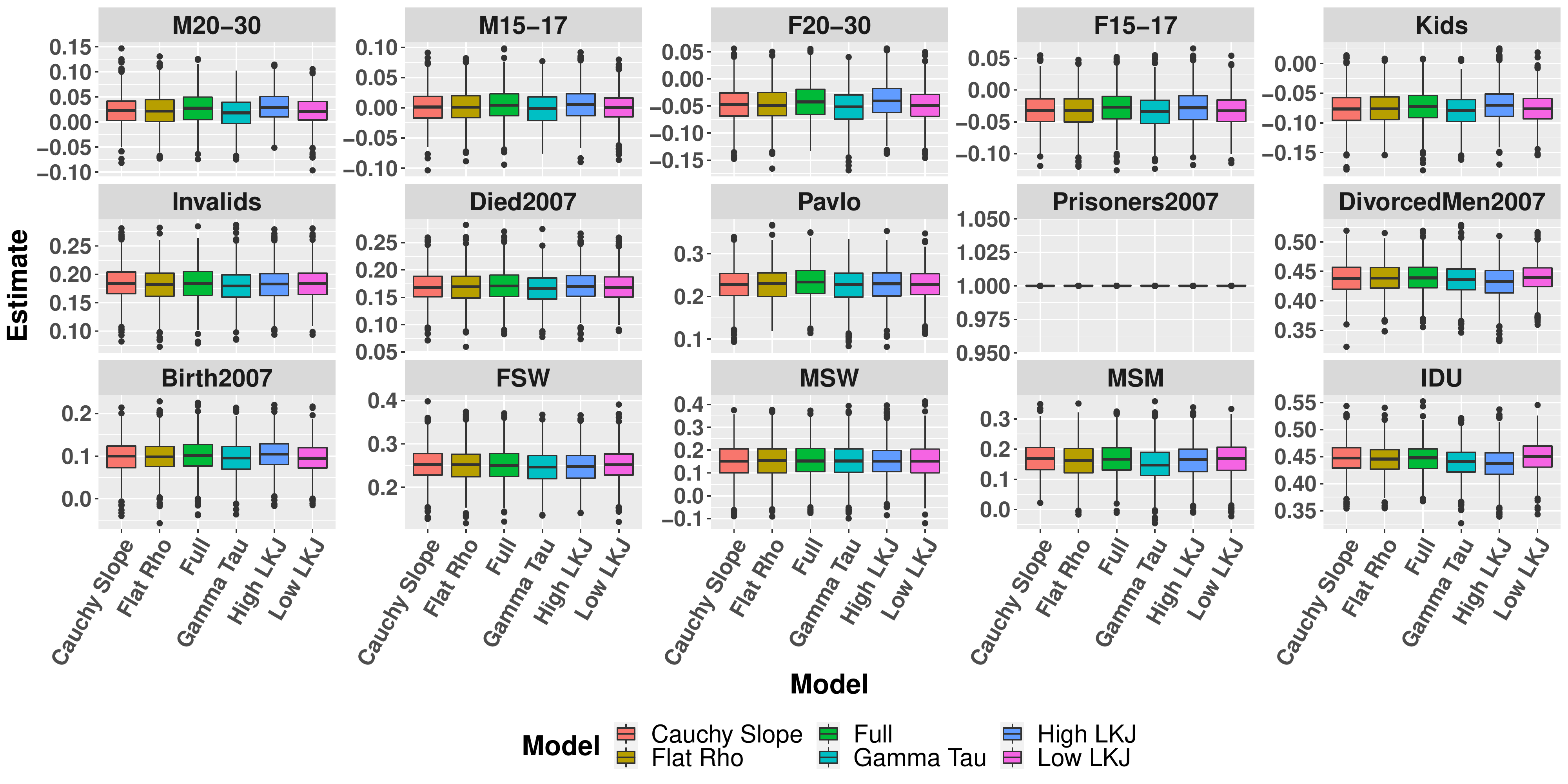}}
    \caption{Estimates of correlations with prisoners in 2007 under different priors.}
    \label{fig:sens_prison}
\end{figure}

\begin{figure}[!t]
    \centerline{\includegraphics[width=0.8\textwidth]{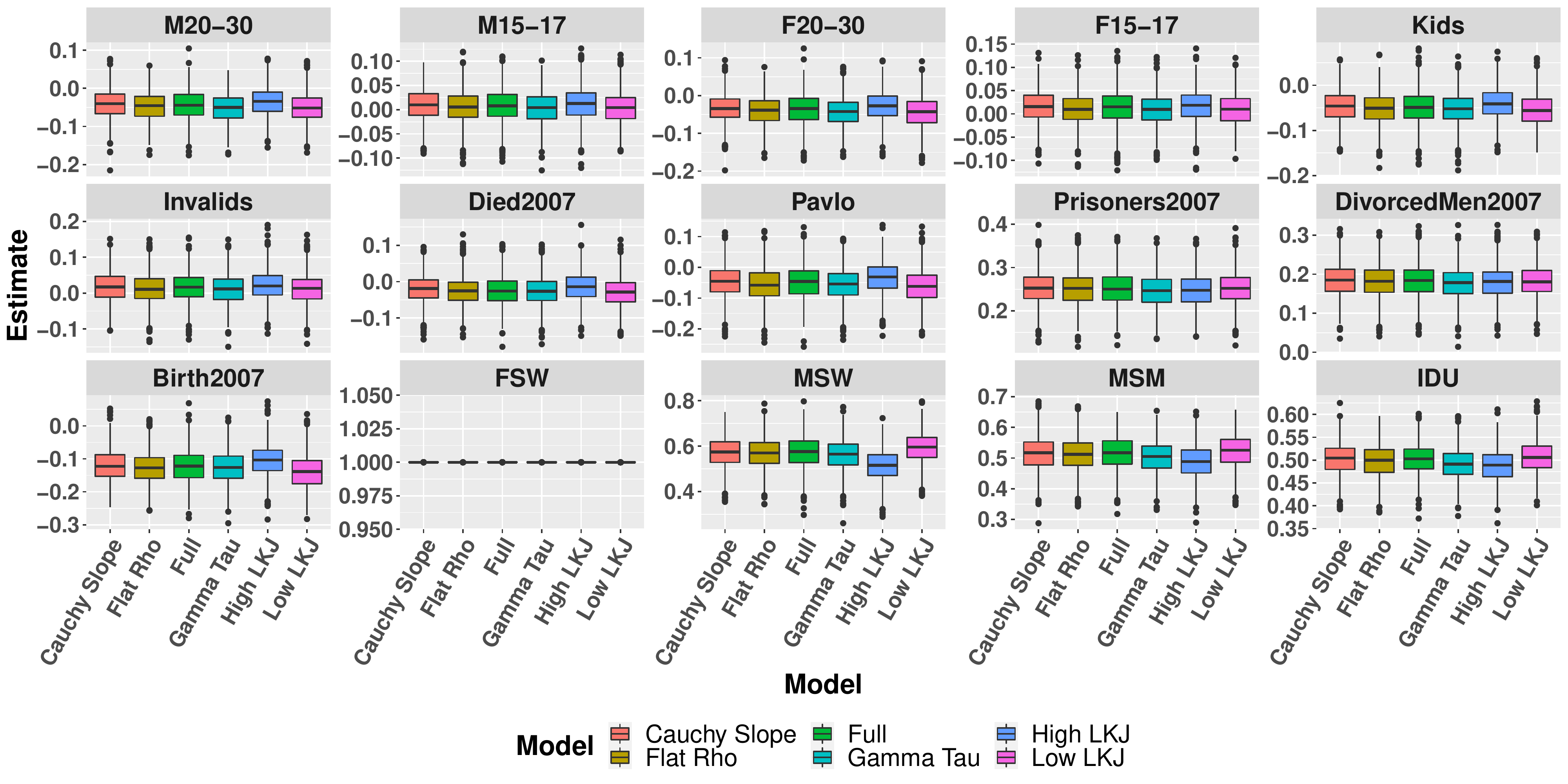}}
    \caption{Estimates of correlations with female sex workers under different priors.}
    \label{fig:sens_fsw}
\end{figure}

We examine the result by comparing the boxplots of the posterior samples for different parameters in the model. We examine the results for all $\rho_k$ (Supplementary Figure \ref{fig:sens_rho}), $\tau_{k}$ (Supplementary Figure \ref{fig:sens_tau}), $\mu_{k}$ (Supplementary Figure \ref{fig:sens_mu}), $\bm{\alpha}$ (Supplementary Figure \ref{fig:sens_alpha}), $\bm{\beta}_{global}$ (Supplementary Figure \ref{fig:sens_global}), $\bm{\beta}_{k,group}$ for age (Supplementary Figure \ref{fig:sens_age}), $\bm{\beta}_{k,group}$ for age$^2$ (Supplementary Figure \ref{fig:sens_age2}), and correlations with groups Men 20-30 (Supplementary Figure \ref{fig:sens_m2030}), Prisoners (Supplementary Figure \ref{fig:sens_prison}), and female sex workers (FSW) (Supplementary Figure \ref{fig:sens_fsw}). Note that we look at the estimates for $\tau_{k}$ and $\mu_{k}$ instead of $\tau_{N,k}$, since the distribution of $\tau_{N,k}$ is very right-skewed, making it difficult to visualization. Also, there is no level-of-respect response for people who died in 2007, so this subpopulation is removed from Supplementary Figure \ref{fig:sens_alpha} corresponding to $\bm{\alpha}$. In all figures, the ``Full'' model describes the original prior setup described in the main manuscript.

We find that in general, the results are very stable, even for strong priors like the Gamma prior for $\tau_{N,k}$. For the prevalence parameters $\rho_k$, we find that the Normal pooling prior does not change the results compared to a flat prior. However, all three priors on the random errors (standard deviation and correlation) do influence the results. We find that a Gamma prior of $\tau_{N,k}$ has the largest effect, which reduces the size of the unknown subpopulations, but has minimal influence on the estimates for the known subpopulations. This is because compared to the original prior specification, the Gamma prior shrinks the estimates much closer to 0, as seen in Supplementary Figures \ref{fig:sens_tau} and \ref{fig:sens_mu}. We believe that the Gamma prior is too influential and does not correspond to our prior beliefs. The results for $\tau_k$ and $\mu_k$ are similar to the results for $\rho_k$. This result is expected given the connection between the random effects and the prevalence parameters.

Across all priors, the slope estimates are stable.

Finally, for the correlation estimates in Supplementary Figures \ref{fig:sens_m2030}, \ref{fig:sens_prison}, and \ref{fig:sens_fsw}, we find that the correlation priors have the largest influence on the parameter estimates. In particular, a larger hyper-parameter results in smaller correlation estimates while a smaller hyper-parameter results in larger correlation estimates. This is expected and consistent with the prior specification. However, the correlation estimates are still fairly stable. In the most extreme case here for the correlation between men aged 20-30 and men named Pavlo, the posterior mean from the large hyper-parameter is -0.172 while the posterior mean from the small hyper-parameter is -0.246. Considering the extreme difference between these two priors, this difference is reasonably small and does not change the interpretation of the results.

\FloatBarrier
\newpage
\section{MCMC Diagnostics}
\FloatBarrier
\begin{figure}[!h]
    \centerline{\includegraphics[width=0.8\textwidth]{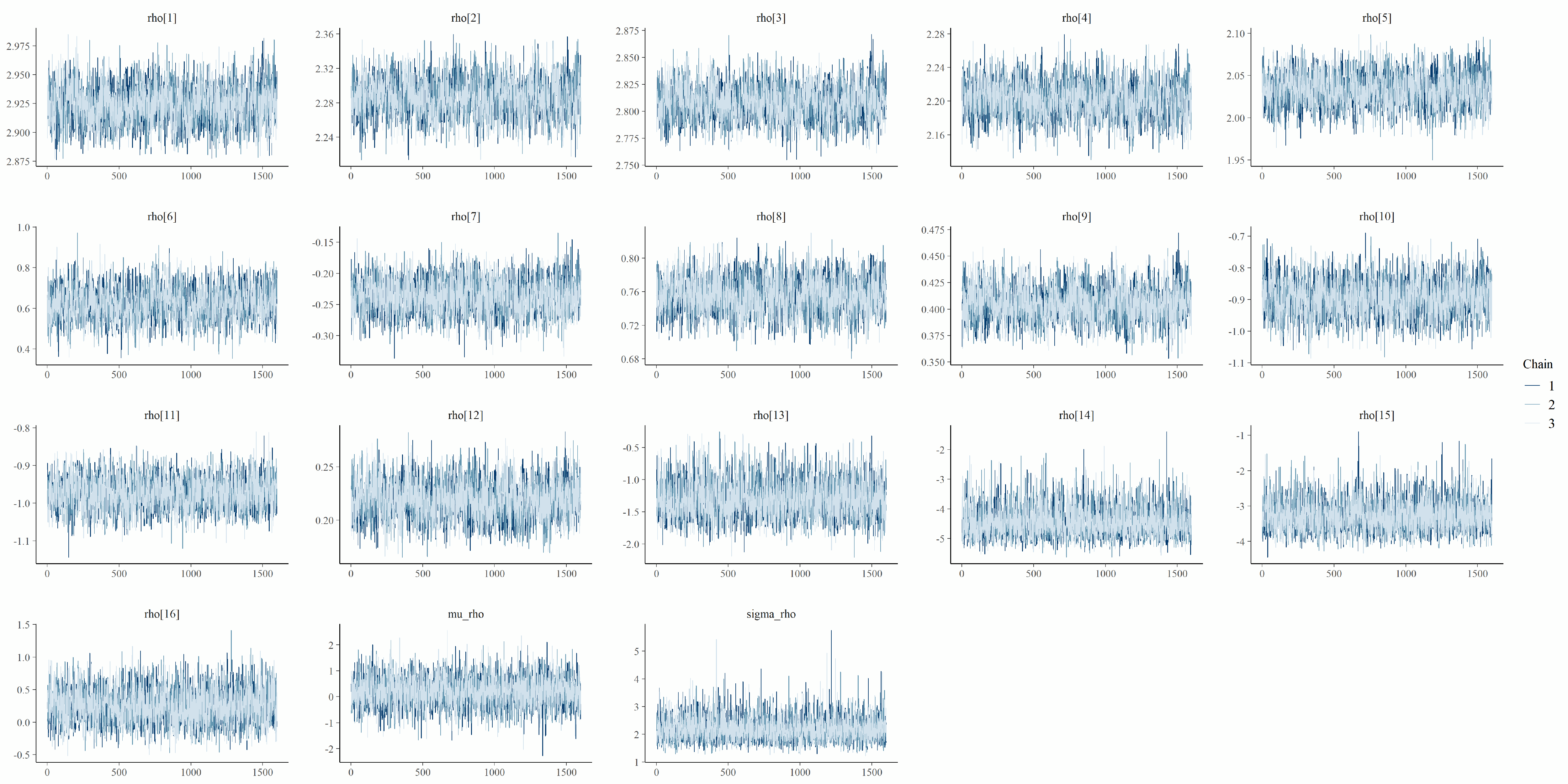}}
    \caption{Trace plots for parameters related to $\rho$.}
\end{figure}

\begin{figure}[!h]
    \centerline{\includegraphics[width=0.8\textwidth]{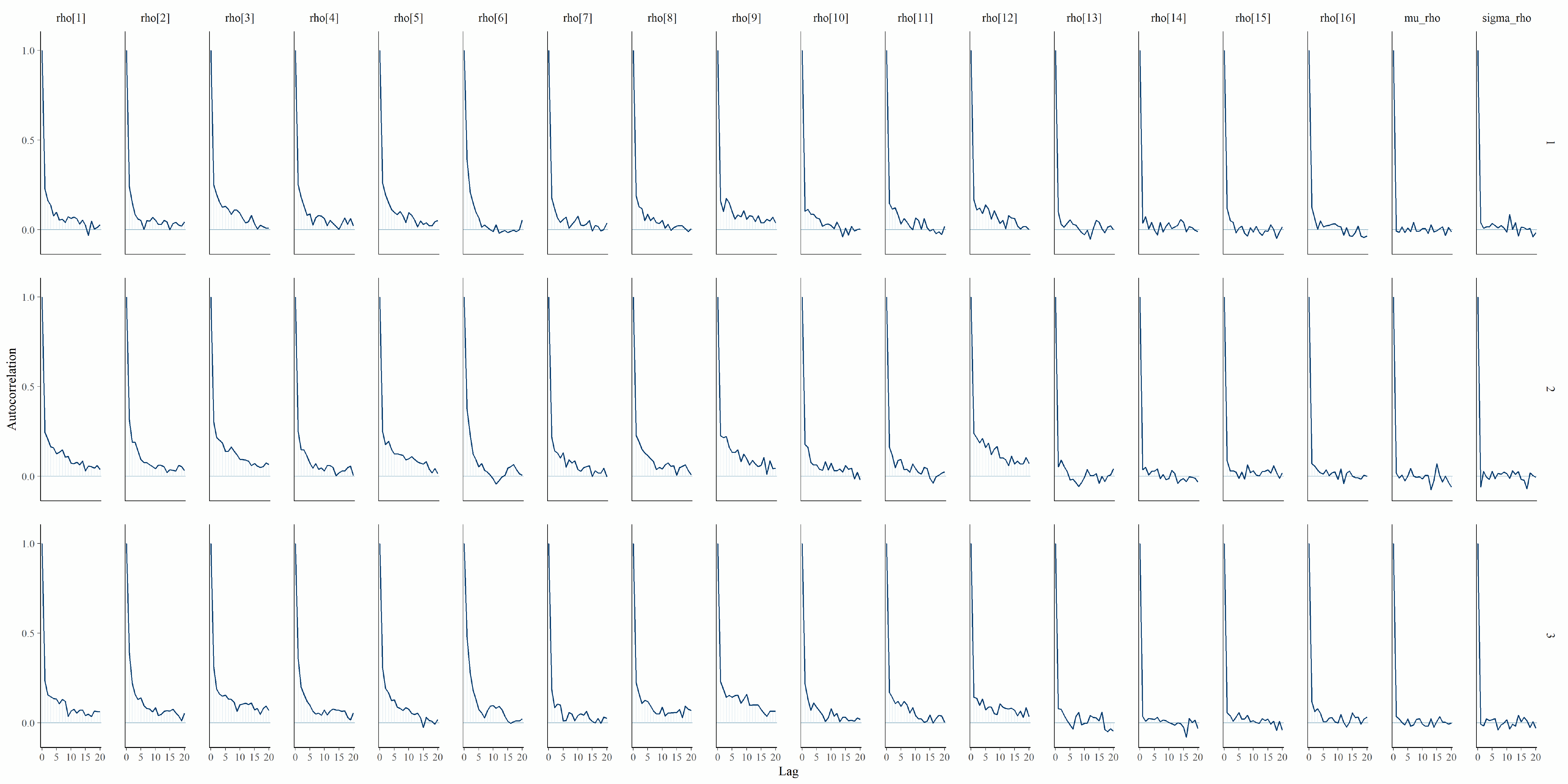}}
    \caption{ACF plots for parameters related to $\rho$.}
\end{figure}

\begin{figure}[!t]
    \centerline{\includegraphics[width=0.8\textwidth]{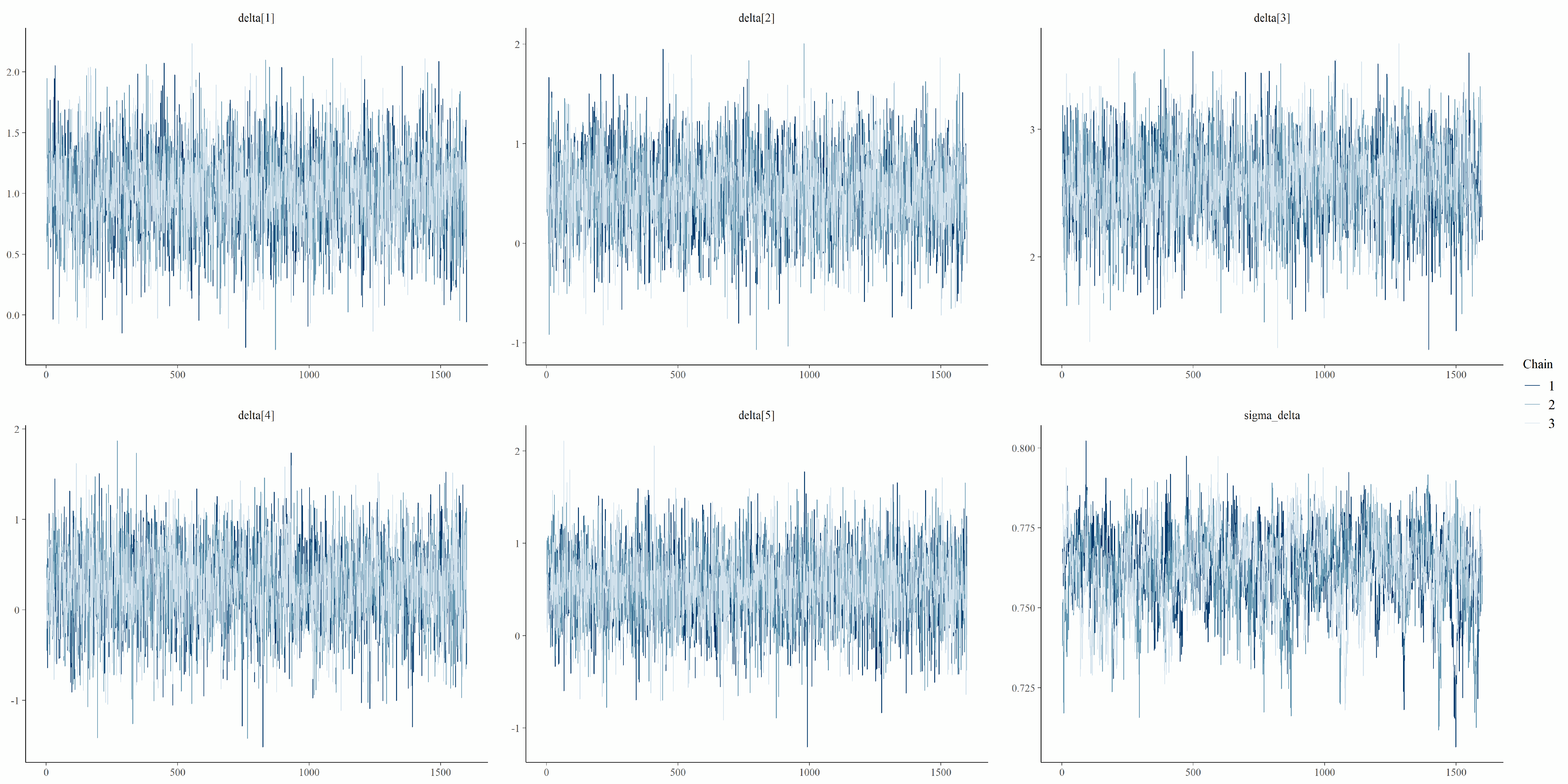}}
    \caption{Trace plots for parameters related to $\delta$.}
\end{figure}

\begin{figure}[!t]
    \centerline{\includegraphics[width=0.8\textwidth]{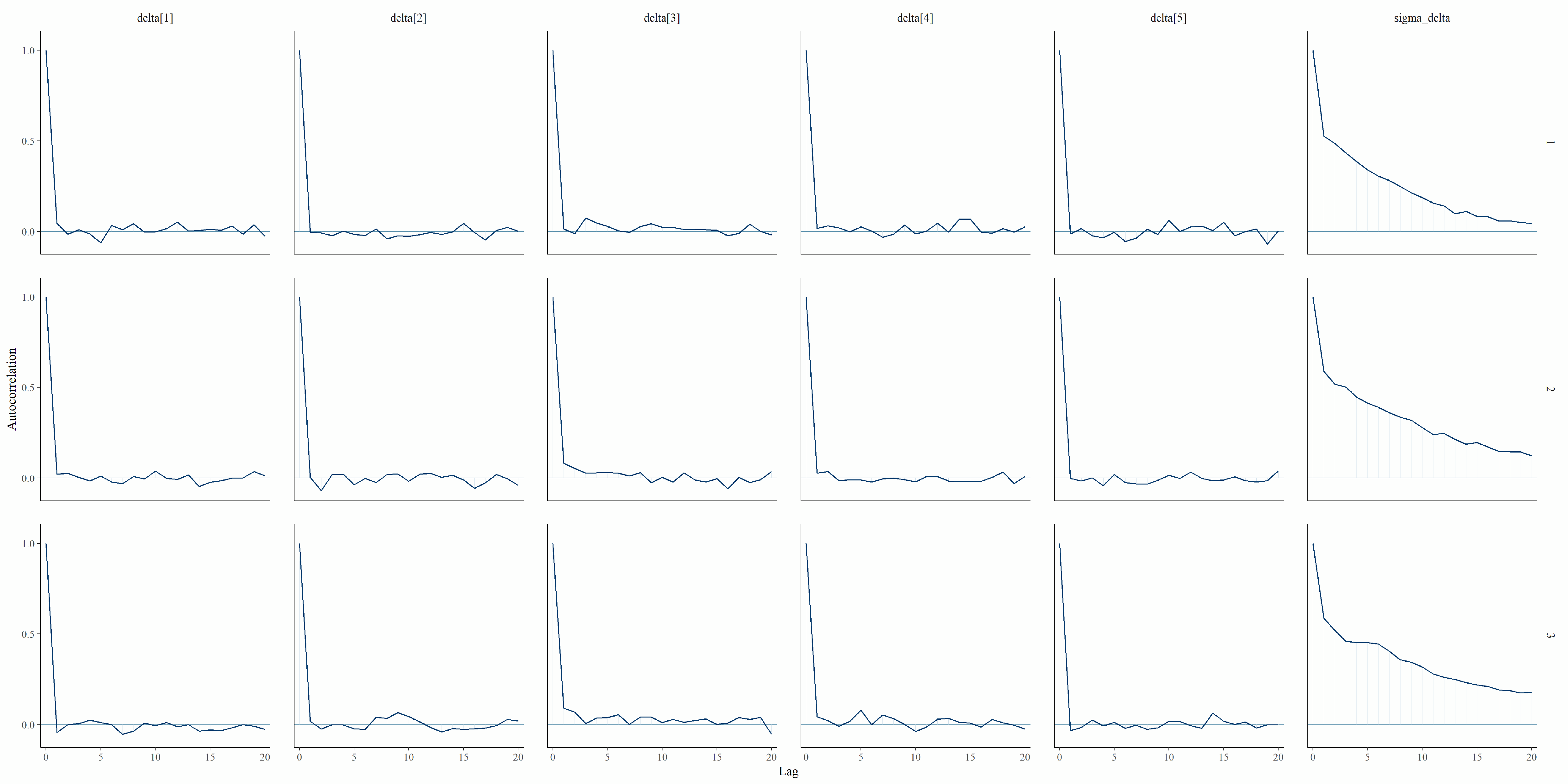}}
    \caption{ACF plots for parameters related to $\delta$.}
\end{figure}

\begin{figure}[!t]
    \centerline{\includegraphics[width=0.8\textwidth]{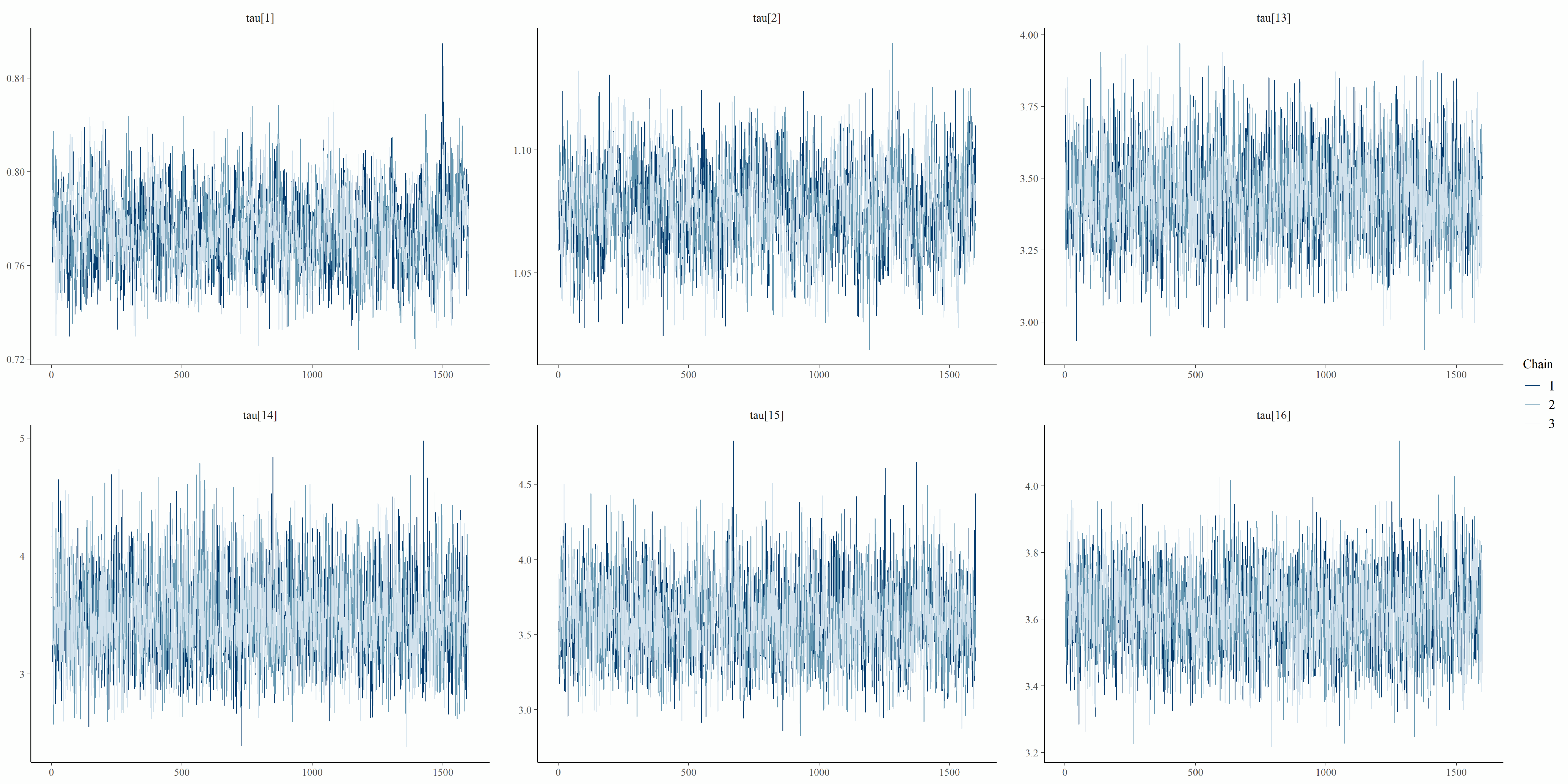}}
    \caption{Trace plots for parameters related to $\tau$.}
\end{figure}

\begin{figure}[!t]
    \centerline{\includegraphics[width=0.8\textwidth]{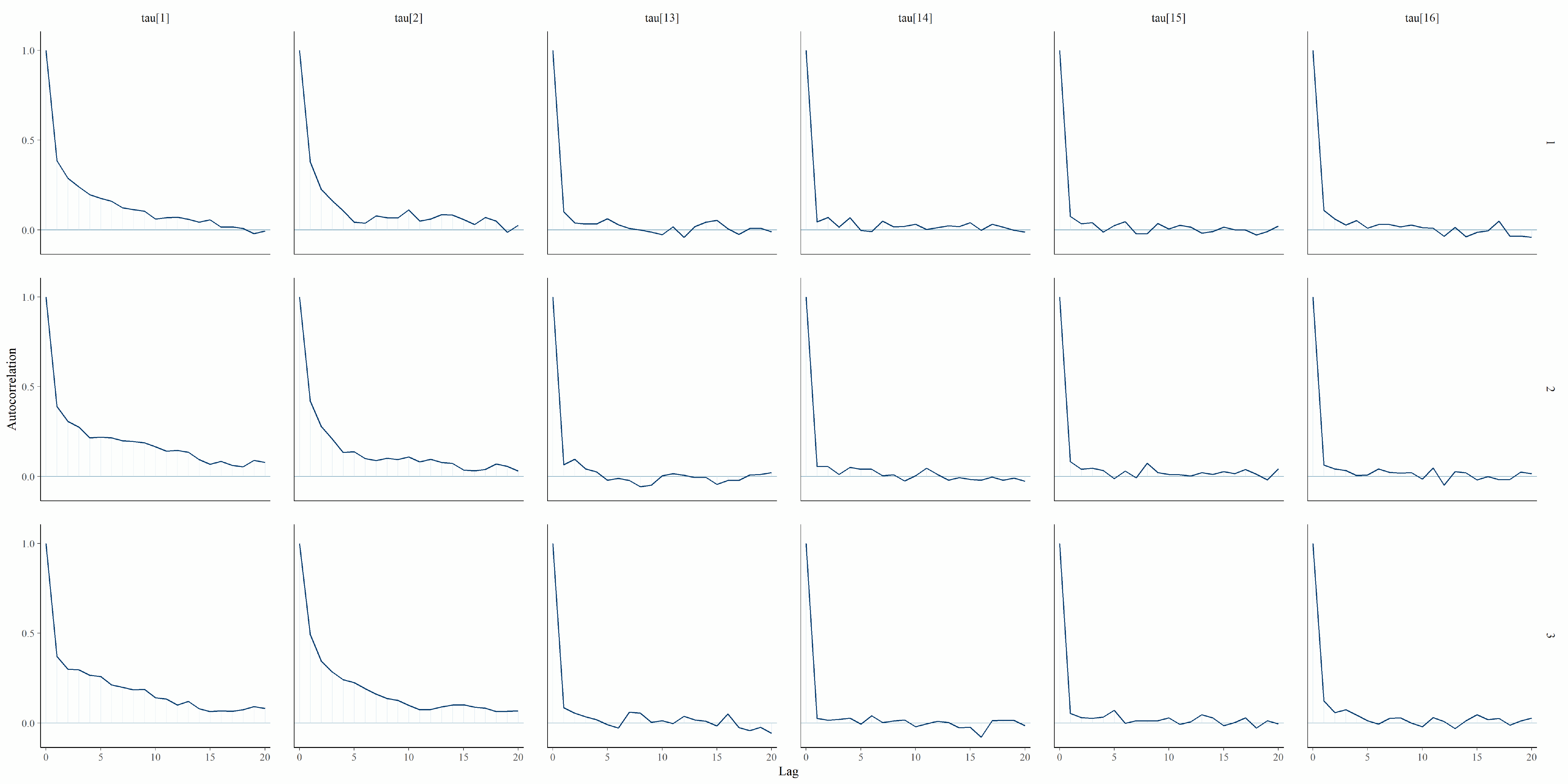}}
    \caption{ACF plots for parameters related to $\tau$.}
\end{figure}

\begin{figure}[!t]
    \centerline{\includegraphics[width=0.8\textwidth]{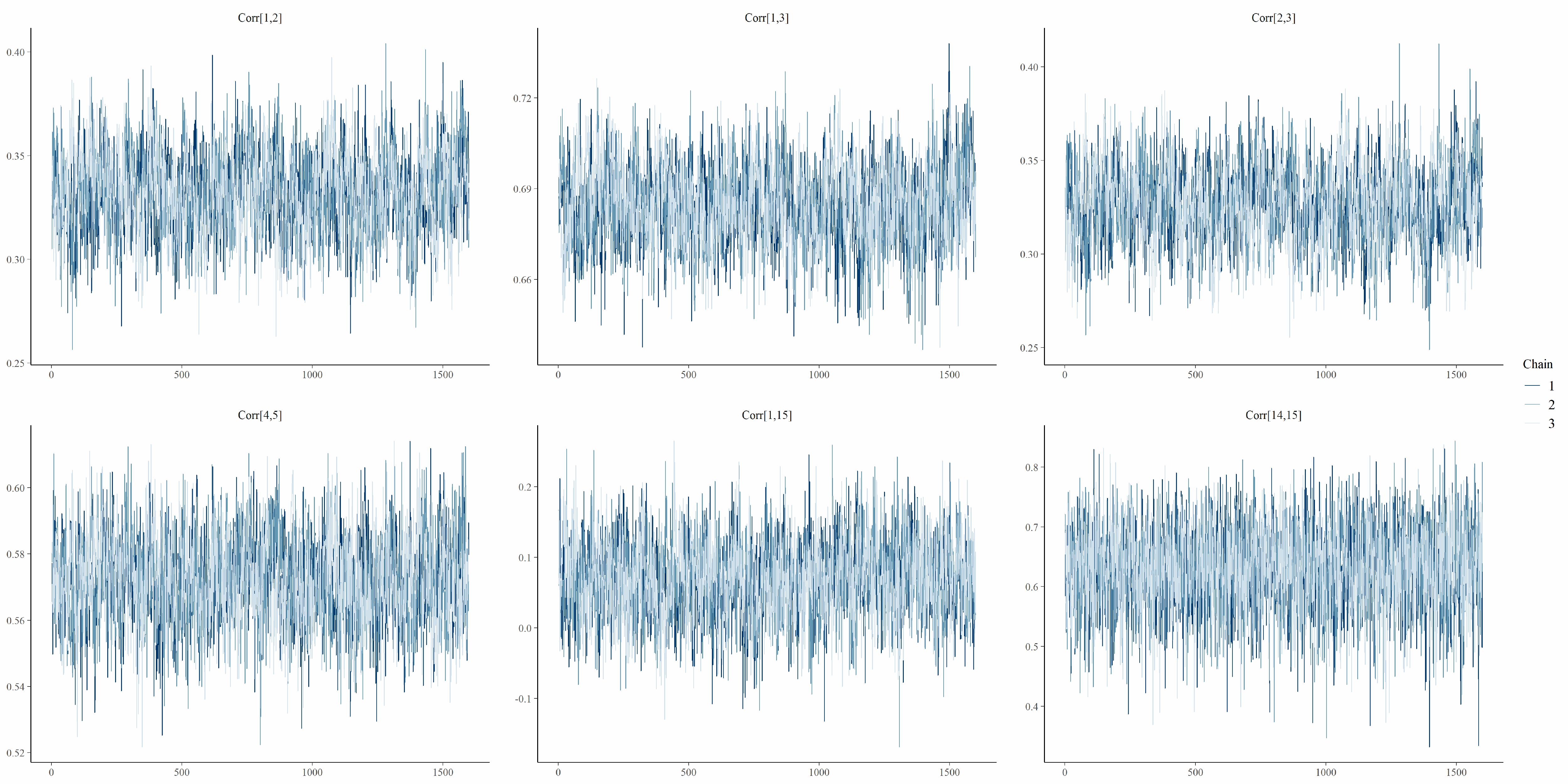}}
    \caption{Trace plots for parameters related to $\Omega$.}
\end{figure}

\begin{figure}[!t]
    \centerline{\includegraphics[width=0.8\textwidth]{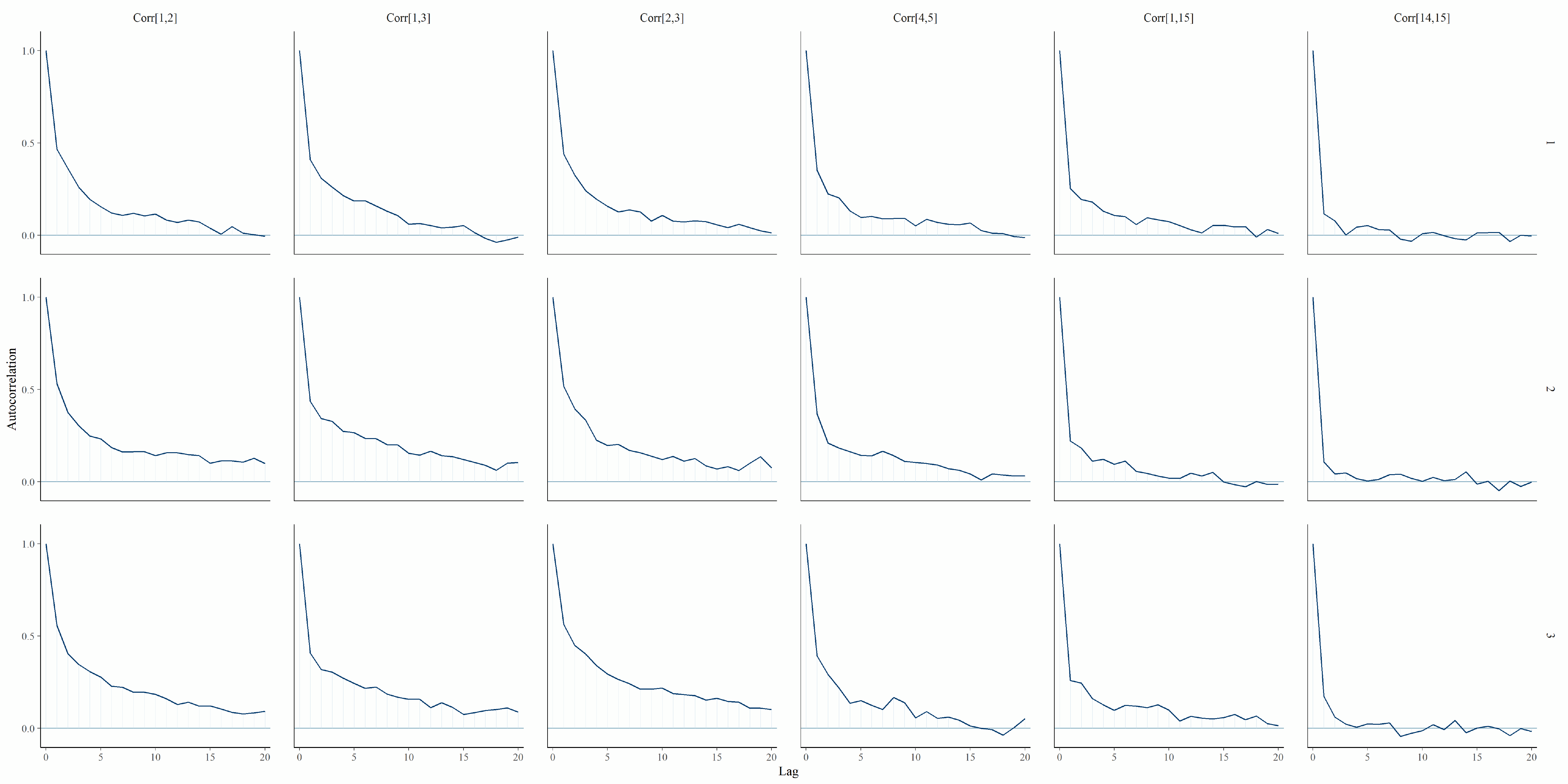}}
    \caption{ACF plots for parameters related to $\Omega$.}
\end{figure}

\begin{figure}[!t]
    \centerline{\includegraphics[width=0.8\textwidth]{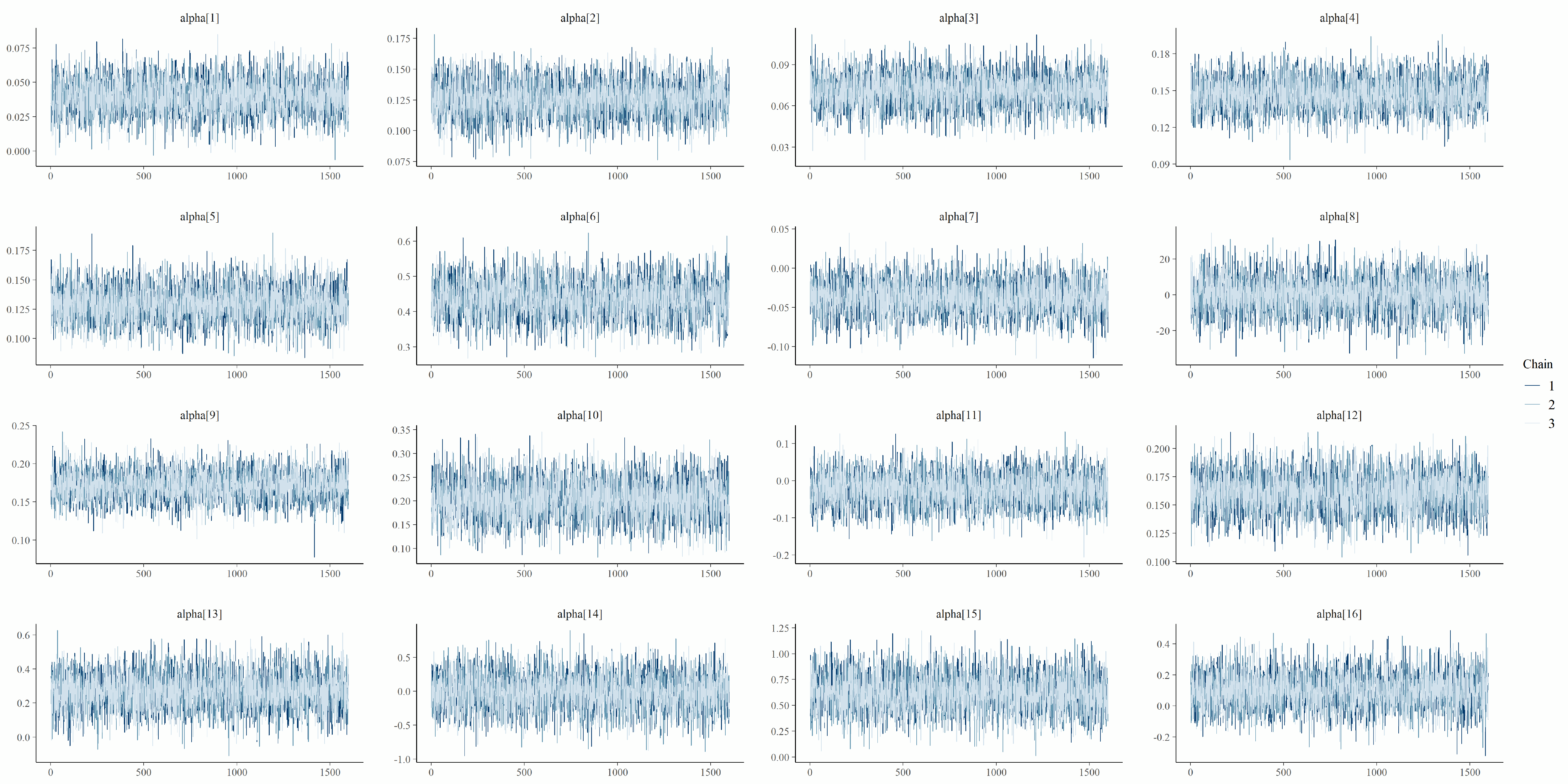}}
    \caption{Trace plots for parameters related to $\alpha$.}
\end{figure}

\begin{figure}[!t]
    \centerline{\includegraphics[width=0.8\textwidth]{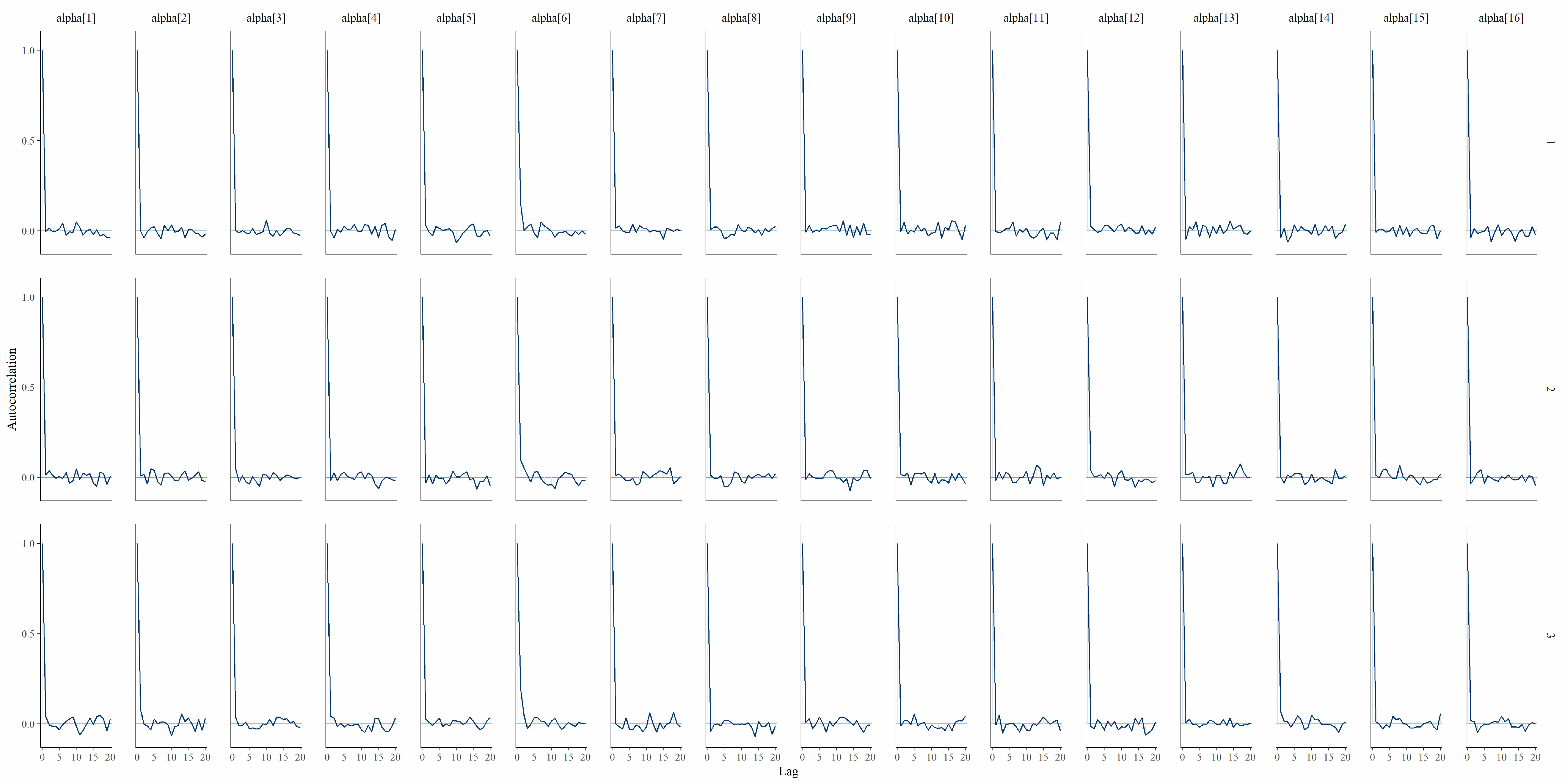}}
    \caption{ACF plots for parameters related to $\alpha$.}
\end{figure}

\begin{figure}[!t]
    \centerline{\includegraphics[width=0.8\textwidth]{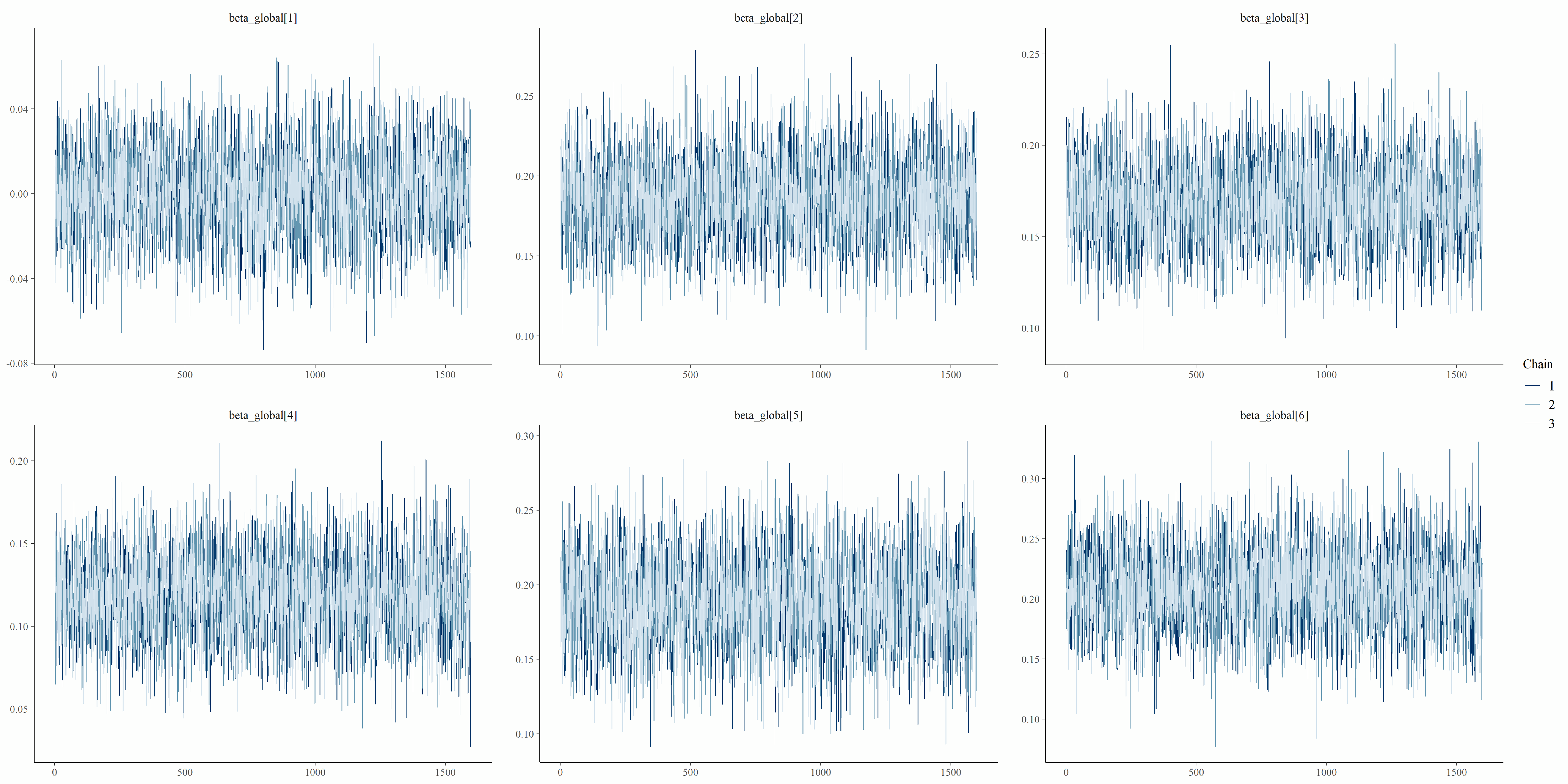}}
    \caption{Trace plots for parameters related to $\beta^{global}$.}
\end{figure}

\begin{figure}[!t]
    \centerline{\includegraphics[width=0.8\textwidth]{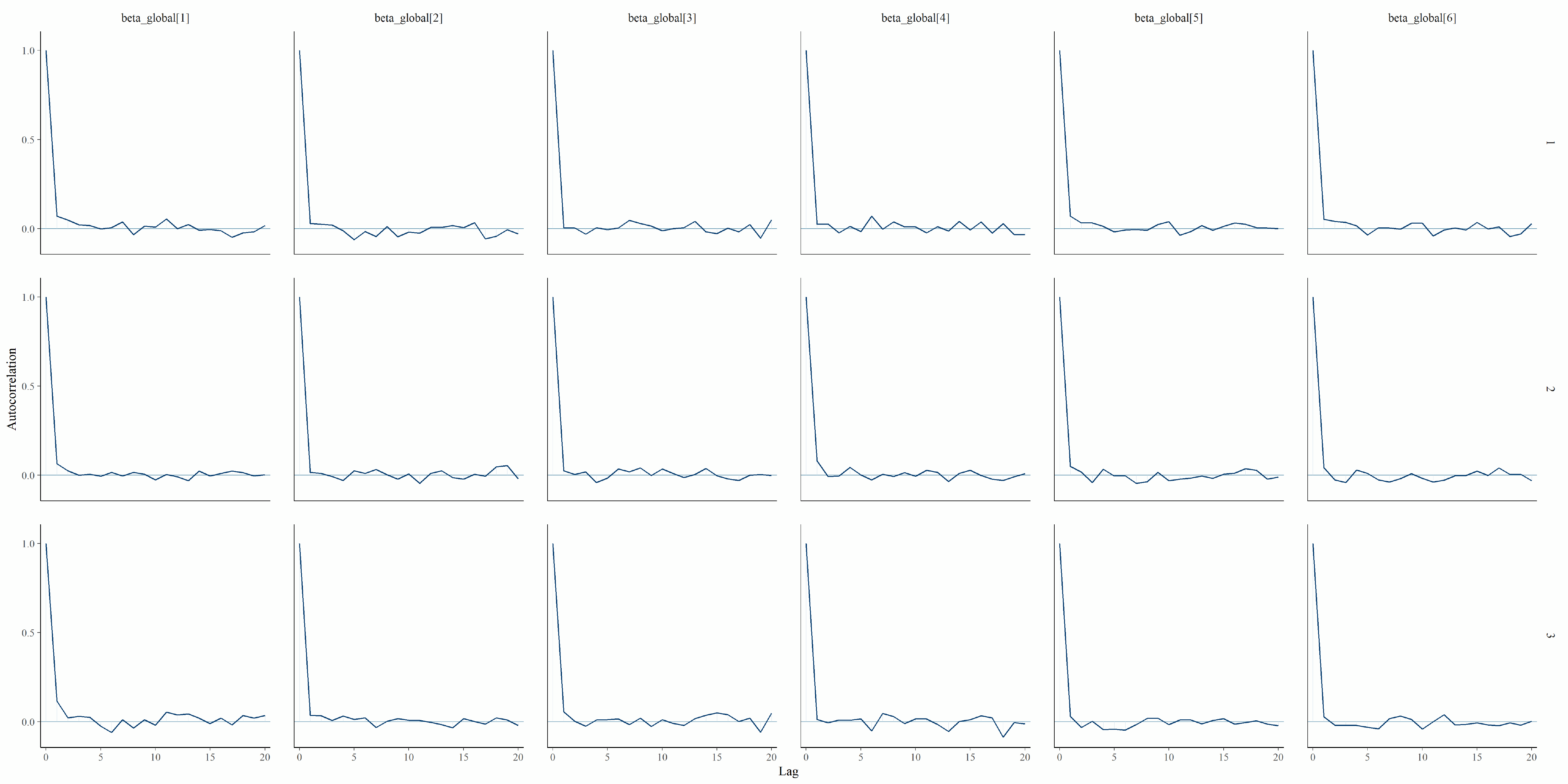}}
    \caption{ACF plots for parameters related to $\beta^{global}$.}
\end{figure}

\begin{figure}[!t]
    \centerline{\includegraphics[width=0.8\textwidth]{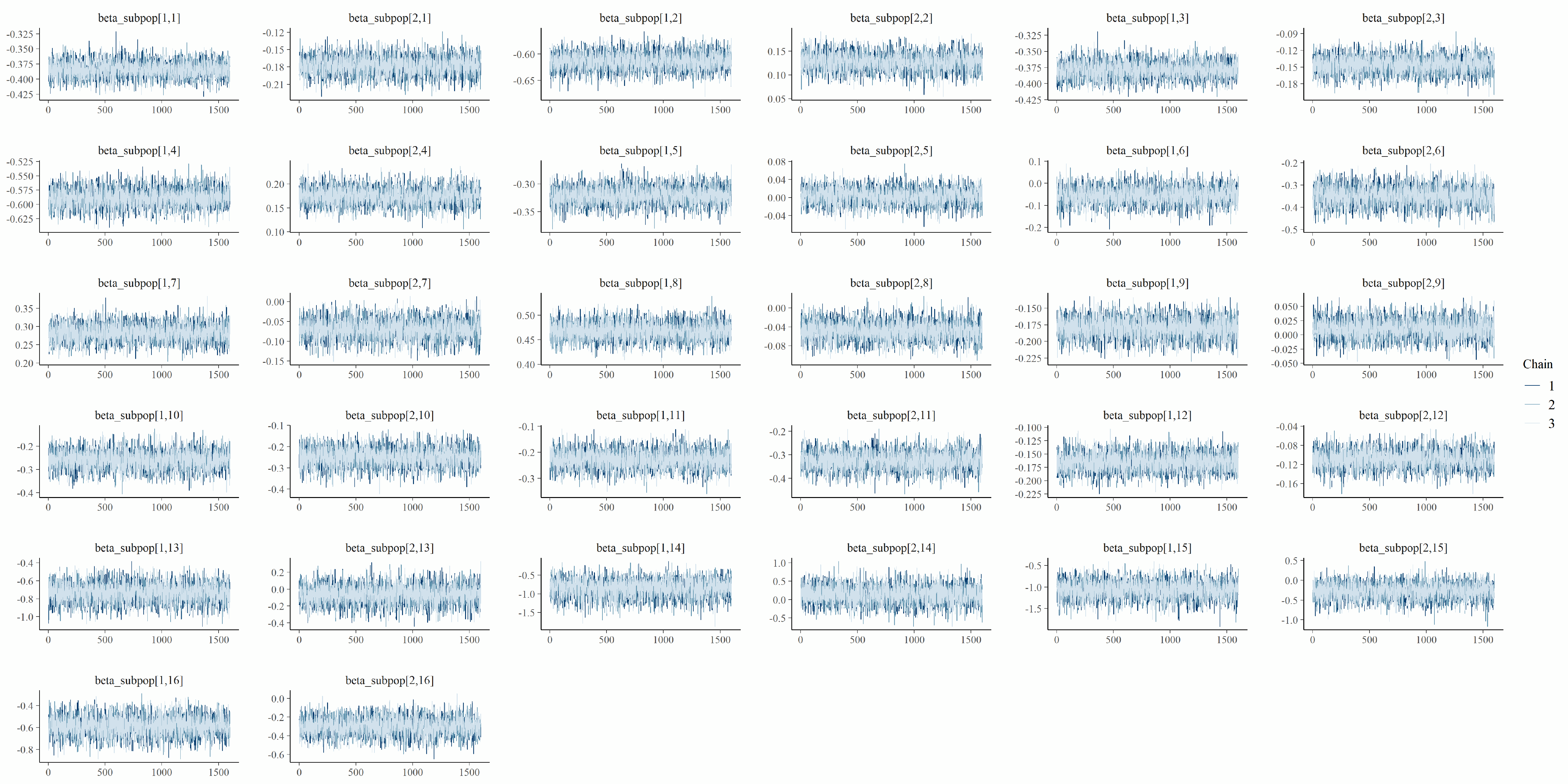}}
    \caption{Trace plots for parameters related to $\beta^{group}$.}
\end{figure}

\begin{figure}[!t]
    \centerline{\includegraphics[width=0.8\textwidth]{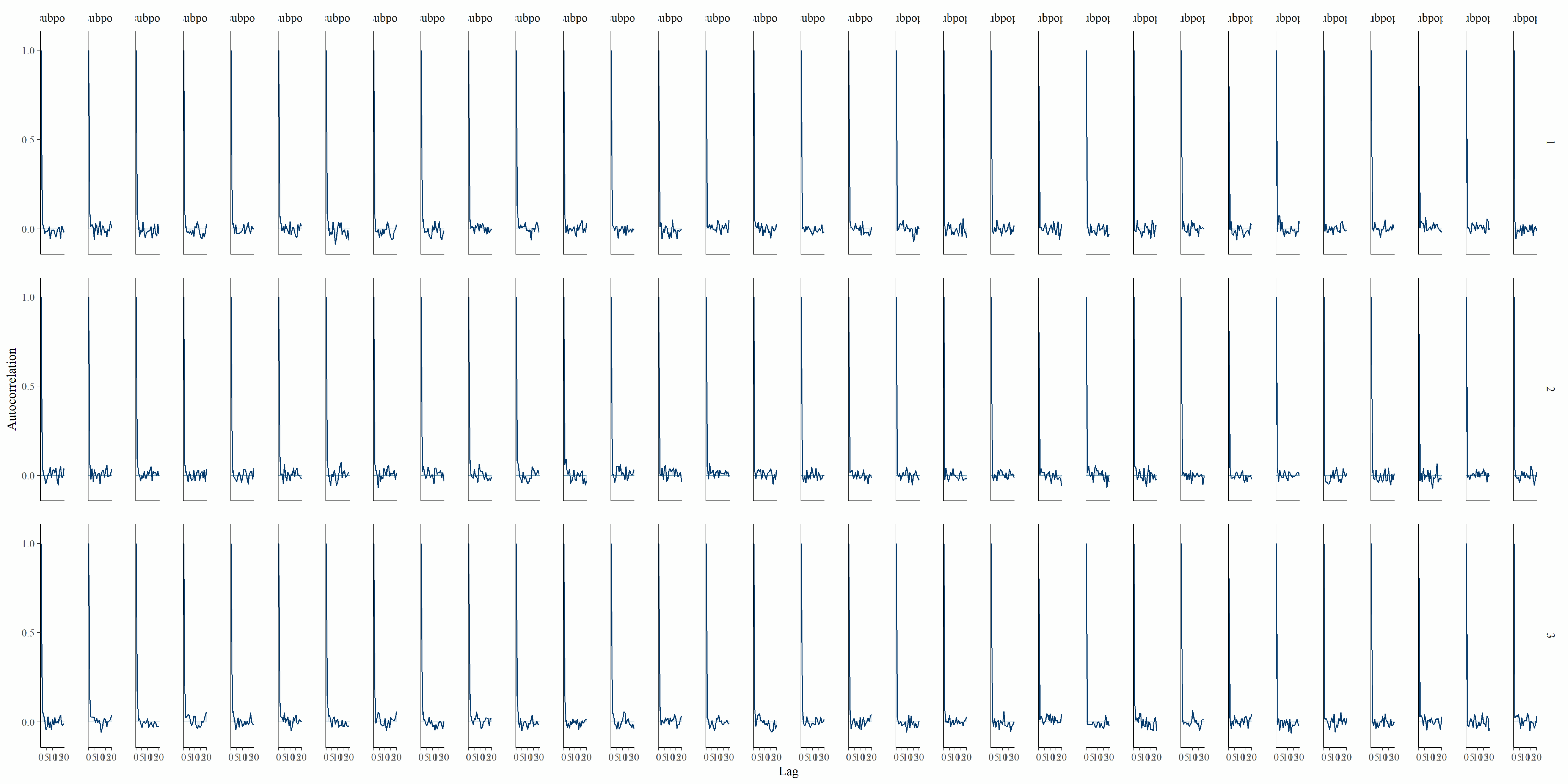}}
    \caption{ACF plots for parameters related to $\beta^{group}$.}
\end{figure}

\begin{figure}[!t]
    \centerline{\includegraphics[width=0.8\textwidth]{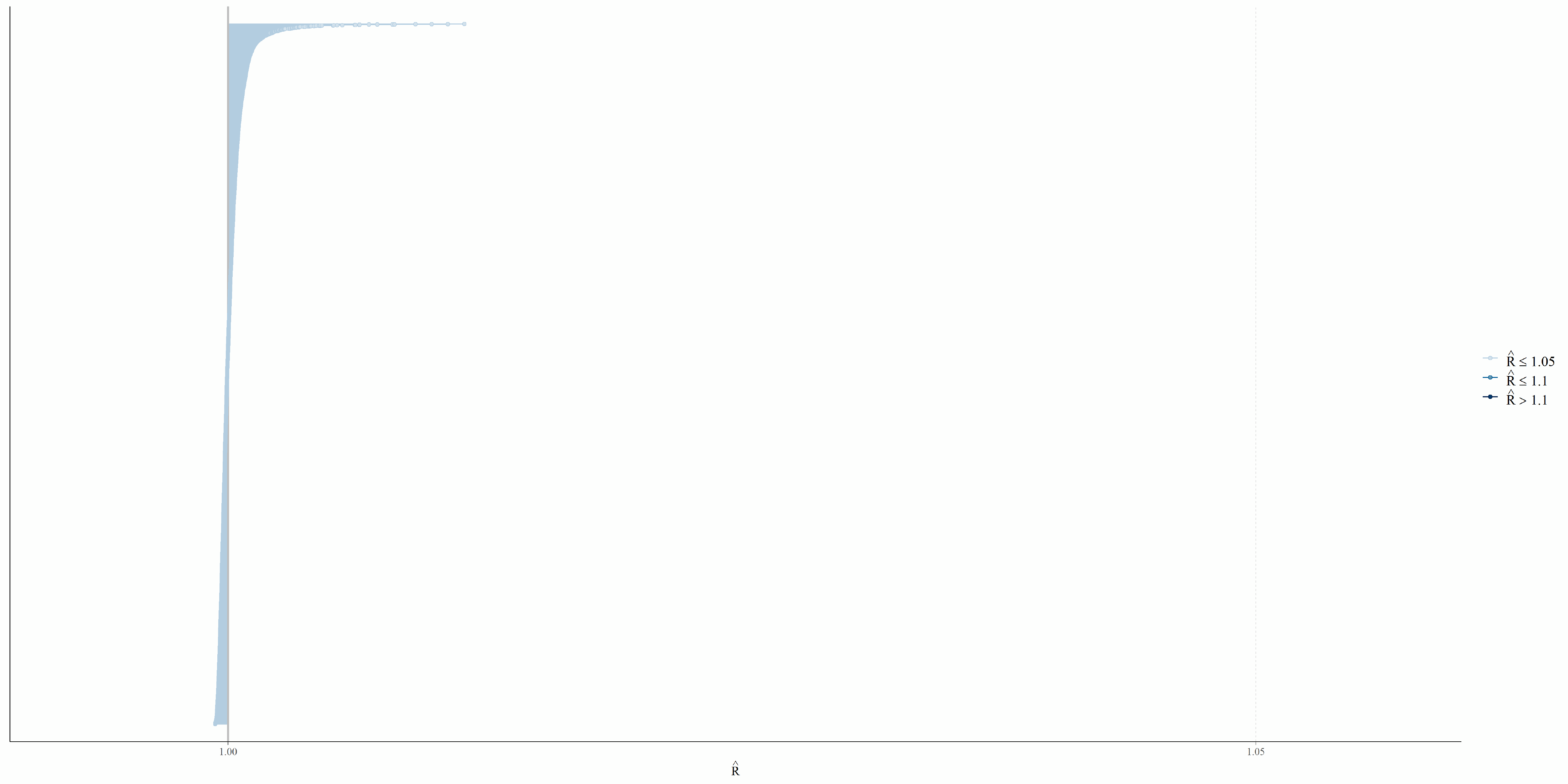}}
    \caption{$\hat{R}$ values.}
\end{figure}

\FloatBarrier

\bibliographystyle{apalike}
\bibliography{NSUM_bib}

\end{document}